\documentclass[fleqn,usenatbib]{mnras}

\usepackage{newtxtext,newtxmath}
\usepackage{array}
\usepackage{multicol}
\usepackage{multirow}
\usepackage{graphicx}
\usepackage{subcaption}
\usepackage{stfloats}
\usepackage[dvipsnames]{xcolor}

\usepackage[T1]{fontenc}

\DeclareRobustCommand{\VAN}[3]{#2}
\let\VANthebibliography\thebibliography
\def\thebibliography{\DeclareRobustCommand{\VAN}[3]{##3}\VANthebibliography}

\defcitealias{PC2025}{PC25}

\usepackage{graphicx}	
\usepackage{amsmath}	
\usepackage[normalem]{ulem}

\title[HE-$\nu$ by H-rich interacting SNe: SN 2023ixf]{High-Energy Neutrinos by Hydrogen-rich Supernovae interacting with low-massive Circumstellar Medium: The Case of SN 2023ixf} 

\author[S.P. Cosentino, M.L. Pumo \& S. Cherubini]{
Stefano P. Cosentino$^{1,2}$\thanks{Contact e-mail: \href{mailto:stefano.cosentino@dfa.unict.it}{stefano.cosentino@dfa.unict.it}}, Maria L. Pumo$^{1,2,3}$\thanks{Contact e-mail: \href{mailto: marialetizia.pumo@unict.it}{marialetizia.pumo@unict.it}} \& Silvio Cherubini$^{1,3}$\\
$^{1}$ Università degli Studi di Catania, Dip. di Fisica e Astronomia "Ettore Majorana", Catania, Italy\\
$^2$ INAF - Osservatorio Astrofisico di Catania, Catania, Italy\\
$^3$ Laboratori Nazionali del Sud-INFN, Catania, Italy
}
\date{Accepted 2025 May 22. Received 2025 May 9; in original form 2024 August 23}

\pubyear{2024}

\begin{document}
\label{firstpage}
\pagerange{\pageref{firstpage}--\pageref{lastpage}}
\maketitle

\begin{abstract}
In hydrogen-rich (H-rich) Supernova  (SN) events, the collision between the H-rich ejecta and the Circum-Stellar Medium (CSM) can accelerate particles and produce high-energy neutrinos (HE-$\nu$, TeV-PeV) through proton-proton inelastic scattering. Despite understanding the production mechanism of these neutrinos, the lack of direct observations raises questions about particle acceleration efficiency and the involved astrophysical conditions. This study focuses on neutrino emission from H-rich SNe with low-mass CSM, such as SN 2023ixf. We developed a semi-analytical model to characterize the progenitor and CSM at the explosion time, allowing us to infer the expected neutrino flux at Earth during the SN's interaction phase.
Our model shows that neutrino emission depends not only on shock velocity and CSM mass but also on the spatial matter distribution of the CSM. By analysing the bolometric light curve of SN 2023ixf beyond 100 days post-explosion, we find that its ejecta, consisting of $9\,\text{M}_{\rm \odot}$ (including $0.07\,\text{M}_{\rm \odot}$ of radioactive $^{56}$Ni) and having a kinetic energy of $1.8\,\text{foe}$, collides with a low-mass CSM of $0.06\,\text{M}_{\rm \odot}$ distributed according to a power-law density profile with an exponent of $s=2.9$. Through these parameters, we estimate that up to $4\pm1\times 10^{-2}$ muon (anti-)neutrino events could be detected by IceCube within 50 days post-explosion. Although the predicted flux ($\lesssim 3\times 10^{-9}\,\text{GeV} \, \text{cm}^{-2} \, \text{s}^{-1}$) is below current IceCube sensitivity, future telescopes like IceCube-Gen2 and KM3NeT could detect HE-$\nu$ from similar SN events.
\end{abstract}

\begin{keywords}
 supernovae: general -- circumstellar matter -- shock waves -- acceleration of particles -- neutrinos -- supernovae: individual: SN 2023ixf 
\end{keywords}



\section{Introduction}
The last evolutionary stage of the majority of massive stars, i.e. those with zero-age main-sequence masses ($M_{\rm ZAMS}$) greater than about $8\, \text{M}_{\rm \odot}$ \citep[see, e.g.,][and references therein]{2009ApJ...705L.138P}, occurs when their core collapses, leading to core-collapse (CC) supernova (SN) events \citep[see, e.g.,][]{RevModPhys.74.1015}.
During the explosive phase, CC supernovae (CC-SNe) release a significant amount of energy in the form of MeV neutrinos, typically around $10^{53}$ erg (equivalent to $10^2$ foe). Of this energy, only one hundredth is transferred to the ejected material (from here on called ejecta) as thermal and kinetic energy \citep[see, e.g.,][]{Janka_12}. In the case of CC-SNe with hydrogen-rich (H-rich) ejecta, also known as type II SNe (SNe II), if there is a Circum-Stellar Medium (CSM) surrounding the SN stellar progenitor, the ejecta-CSM collision can be responsible for a secondary neutrino emission in the High-Energy (HE) range \citep[TeV-PeV; see, e.g.,][]{Murase11,Fang2020}{}{}.\par
Whereas the neutrino-driven CC-SNe scenario has been widely studied, besides being confirmed by the explosion neutrinos of SN 1987A \citep{AnnRev87A}, there are still unsolved issues about the mechanism and the astrophysical conditions that lead to the production of High-Energy neutrinos \citep[HE-$\nu$; see e.g.][]{Sarmah_2022,2023MNRAS.524.3366P}.
Specifically, these neutrinos can be generated by the shock-interaction between the rapidly expanding ejecta and its CSM. Indeed, when the faster ejecta collides with the H-rich CSM, two shock wave-fronts start to propagate inside them, in reverse and forward way respectively \citep[][]{2017hsn..book..875C}{}{}. Both of them contribute to the acceleration of particles, mainly protons, which are swept by the two shock fronts. However, the energy of these protons efficiently increases only when the forward shock propagates inside the optically thin regions of the CSM \citep[][]{Suzuki_2020}. In this way, the collisions between accelerated protons of the shock-shell and the nuclei of the H-rich CSM lead to the production of HE-$\nu$ and gamma rays by proton-proton (pp) inelastic scattering \citep[e.g.,][]{Murase14,Fang2020}. 
While the gamma radiation could be stopped by the matter surrounding the shock, the neutrinos easily escape from the production region becoming potentially detectable on Earth by the large volume neutrino observatories \citep[i.e. IceCube and KM3NeT; see, e.g.,][]{Ahlers2018}{}{}.
According to this scenario, the HE-$\nu$ emission depends on the physical properties of SN event \citep[e.g.,][]{Murase18,Sarmah_2022}, which in turn varies according to the configuration of the SN progenitor and the CSM matter distribution at explosion. 
These configurations, linked also to the pre-SN evolution, even determine the extreme variety of the spectro-photometric features which characterises the post-explosive phases of SNe II \citep[see, e.g.,][]{PZ2011,KK_2023}{}{}. Therefore, valuable information about SN physical properties is yielded by analysing their post-explosive electromagnetic emission which, in turn, can provide insights into the features of HE-$\nu$ emission \citep[e.g.,][]{2023MNRAS.524.3366P,CosPumChe2024,Salmaso24}.\par
The case of the IC200530A event recorded by \citet{2020GCN.27865....1I}, believed to be the neutrino counterpart of the optical transient AT2019fdr \citep{2019TNSCR1016....1C}, demonstrates how the study of optical observations in the post-explosive phase can guide the research and identification of astrophysical neutrino sources \citep[see also][]{Pitik2022}. However, this type of survey has primarily focused on SNe with particularly massive CSM ($\gtrsim 1-2$ times the mass of their ejecta), because in these cases, the duration of the interaction phase is as long enough as to allow the neutrino telescopes to give a significant constrain on the observed HE-$\nu$ flux \citep[see, e.g.,][]{2023MNRAS.524.3366P}.\par 
From the electromagnetic point of view, the ejecta-CSM interaction can lead to the presence of narrow (n) high ionization emission lines, typical for type IIn SNe \citep[see, e.g.,][and references therein]{DessartHiller2016}{}{}, and/or an increasing of the SN bolometric luminosity like in the case of the enigmatic Superluminous SNe \citep[SLSNe; see, e.g.,][]{Inserra2019}{}{}. 
However, events such as SLSNe and SNe IIn constitute only about $18\%$ of H-rich SNe \citep{Perley_2020}. Most SNe II have a low-massive CSM that does not significantly alter the kinetic energy of the ejecta after expansion has begun \citep[see, e.g.,][]{Moriya2013}. For these low-interacting SNe, the optical features due to the interaction are rarely observed and are typically present only during the initial stages, approximately $1-10$ days after the first light peak \citep[see, e.g.,][]{2017NatPh..13..510Y,Tsvetkov_2024}. 
Moreover, their CSM matter distribution depends on the mass loss history of the progenitor in the final months to decades before the explosion \citep[see, e.g.,][]{MorzovaPiroValenti2018,Strotjohann_2021}{}{}, governed by several poorly understood processes in the context of massive star evolution theory, including pre-supernova outbursts, wind acceleration, and binary interactions \citep[see, e.g.,][]{Fuller2017,MYGB2017,Smith2017}{}{}.
In this case, hence, the neutrino emission can be much more sensitive to the characteristics of the SN progenitor system \citep[see, e.g.,][]{Sarmah_2022}.\par
The nearby SN 2023ixf is an optimal example for testing the detection capability for HE-$\nu$ from this type of SNe \citep[e.g.,][]{KheiMura_2023,2024JCAP...04..083S}, as well as demonstrating the link between HE-$\nu$ emission and the physical properties of SN explosion. Specifically, SN 2023ixf was discovered in the nearby galaxy Messier 101 (M101) by \citet[][]{2023TNSTR1158....1I}{}{}, approximately less than one day after the estimated explosion epoch
 \citep[$\text{MJD}=60082.74\pm 0.08$; ][]{Hiramatsu_2023}. Seeing the host galaxy's distance of just $6.9\pm0.1\,\text{Mpc}$ \citep[][]{Riess_2022}, it has been possible to study this SN by multiwavelength follow-up observations \citep[see, e.g.,][]{Teja_2023,Bostroem_2023,Jacobson-Galán_2023,Grefenstette_2023,2023ATel16075....1M, zimmerman_complex_2024,Singh_2024,li_shock_2024, Martinez_AA2024, Chandra_2024ApJ,2025MNRAS.538..659K} and to identify the RSG progenitor from pre-explosion images \citep[see, e.g.,][]{Jencson_2023,Qin_2023,Niu_2023}.
The pre-explosion scenario inferred from these observations suggests that a RSG went through an increase in mass loss rate approximately 20 years prior to the explosion \citep{xiang_dusty_2023}. Subsequently, roughly 3 years before the explosion, the star transitioned into a yellow hypergiant state \citep[see, e.g.,][]{2014ARA&A..52..487S} with a mass loss rate of about $ 6\times 10^{-4}\,\text{M}_{\rm \odot}\,\text{yr}^{-1}$ \citep{ZHANG20232548}, and without any notable eruptive events \citep{Flinner_2023RNAAS}. While the presence of a high mass companion star ($\gtrsim 7\,\text{M}_{\rm \odot}$) has been ruled out \citep{Qin_2023}, the complete stellar evolution of the progenitor and its initial mass remain subjects of ongoing debate, with an estimated $M_{\rm ZAMS}$ raging from $10\,\text{M}_{\rm \odot}$ to $20\,\text{M}_{\rm \odot}$ \citep[see, e.g.,][]{Soraisam_2023,VanDyk_2023,Niu_2023,Pledger_2023,Kilpatrick_2023,VanDyk_2023}.
In this context, the analysis of the bolometric light curve (LC) of SN 2023ixf and its modelling can provide important information on both the structure of the CSM and the amount of mass and energy ejected during the SN event \citep[][]{Bersten_AA2024,Moriya2024}. This information, however, besides being important for a better understanding of the progenitor's nature, is essential for properly simulating the temporal evolution of HE-$\nu$ flux.\par
In this work, with the aim of improving our knowledge on the HE-$\nu$ emission from H-rich SNe with low-mass CSM, we present a semi-analytical description of the mechanisms involved in the production of this HE-$\nu$ radiation, linking them to the astrophysical characteristics of the progenitor and CSM at the time of the explosion (see Section \ref{sec:HEnuSNe}). This approach enables us to identify the SN modelling parameters having the greatest influence on the characteristics of the neutrino energy spectrum. Additionally, based on the same assumptions used to simulate the neutrino emission, in Section \ref{sec:EM_emi} we introduce a new model capable of describing the LC behaviour for interacting SNe II during their shock-interaction phase. This model has been applied to SN 2023ixf, allowing us to infer the main parameters that describe the properties of the SN ejecta and its CSM. Finally, in Section \ref{sec:detect_pot} we evaluate the detection capability of the HE-$\nu$ from low-interacting SNe II, like SN 2023ixf, using the efficiency limits of current and future large-volume neutrino telescopes, such as IceCube and KM3NeT.

\section{High-energy neutrinos from Interacting SNe}\label{sec:HEnuSNe}
The frequent discoveries of astrophysical neutrinos with TeV-PeV energies by IceCube \citep[][]{2013Sci...342E...1I}{}{} have motivated the theoretic community to explore the possible mechanisms which lead to their production \citep[see, e.g.,][]{Fang2020}{}{}, and the astronomical one to search the correspondent electromagnetic sources \citep[see, e.g.,][]{2018Sci...361.1378I,Pitik2022}{}{}. 
Considering the potential interaction between their ejecta and CSM, SNe II can be acceleration sites for hadronic collisions and, consequently, sources of HE-$\nu$ \citep[see, e.g.,][]{2023MNRAS.524.3366P}{}{}.
In this section, we will introduce a semi-analytic model to describe the acceleration and cooling time-scales of protons resulting from the shock interaction between H-rich SN ejecta and its CSM  (Section \ref{Subsec:p_acc}). Additionally, we will derive the neutrino energy distribution arising from pp-collisions (Section \ref{Subsec:nu_flux}) and examine its dependency on SN modelling parameters (Section \ref{Subsec:par_mod}).
\subsection{Relativistic protons in ejecta-CSM shock interaction}\label{Subsec:p_acc}
\begin{figure}
	\includegraphics[width=\columnwidth]{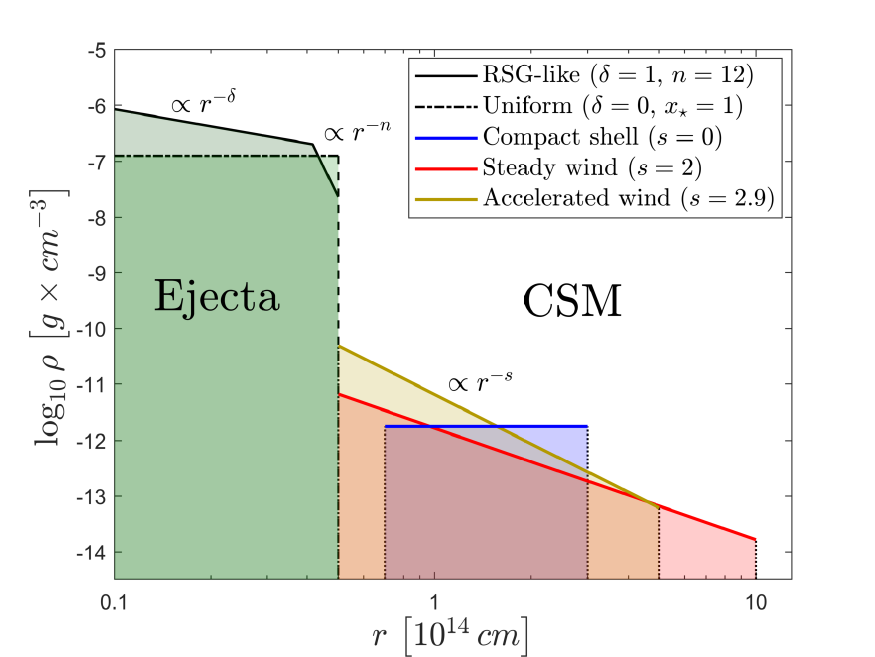}
    \caption{Initial density distribution for several realistic ejecta and CSM configurations. All density profiles share the same $M_{\rm ej}=10\,\text{M}_{\rm \odot}$, $R_{\rm \star}=5\times 10^{13}\,\text{cm}$ (dashed line),  $M_{\rm CSM}=0.1\,\text{M}_{\rm \odot}$ and $f_{\rm \Omega}=1$. The typical $n$ and $\delta$ density slopes for red supergiant (RSG), jointly to $x_{\rm \star}=1.2$, have been set according to the pre-SN models presented in \citet[][]{Moriya_11}, which are based on the results of \citet{RevModPhys.74.1015}. The colored curves show the CSM's density profiles for three remarkable scenarios \citep[see][and references therein for further details]{2022MNRAS.517.1483D}. The left and right dotted lines for each CSM's model respectively marks $R_{\rm 0}$ and $R_{\rm CSM}$. The compact shell scenario, as well as being the most confined  \citep[$R_{\rm CSM}\lesssim 5\times 10^{14}\,\text{cm}$; see, e.g.,][]{MorzovaPiroValenti2018}, is detached from the stellar surface, since $R_{\rm 0}=7\times 10^{13}\,\text{cm}$ is greater than $R_{\rm \star}$.}\label{fig:density}
\end{figure}

The generation of HE-$\nu$ through pp-collisions is intrinsically linked to the acceleration of protons within the shock region. The dynamic progression of the shock, in turn, relies on the densities of both the progenitor and the CSM at the moment of the explosion. Specifically, the progenitor star's density determines the initial distribution of the ejected mass ($M_{\rm ej}$) during the onset of the explosion, that can be expressed as:
\begin{equation}\label{Eq:prog_density}
\rho_{0}(x)= \frac{M_{\rm ej}\,(n-3)\,(3-\delta)}{4\pi R_{\rm \star,0}^3\,\left[n-\delta-(3-\delta)x_{\rm \star}^{3-n}\right]}\times
\begin{cases}
x^{-\delta}\,& x\le 1\\
x^{-n}\,& 1<x\le x_{\rm \star},
\end{cases}
\end{equation}
where $x= r/R_{\rm \star,0}$ represents the radial coordinate divided by the inner envelope radius ($R_{\rm \star,0}$), while $x_{\rm \star}$ ($= R_{\rm \star}/R_{\rm \star,0}$) is its value for the outer stellar radius ($R_{\rm \star}$). In this way, the ejecta's density profile is divided into two distinct parts (see also Fig. \ref{fig:density}): the internal region ($r \le R_{\rm \star,0}$) and the external one ($R_{\rm \star,0} < r \le R_{\rm \star}$). These regions are characterized by two distinct power-law exponents, $\delta$ ($\simeq 0-1$) and $n$ ($\simeq 8-12$), whose values are determined by the progenitor type \citep[see e.g][]{1999ApJ...510..379M,Moriya2013}.
Whereas, the CSM's density is determined by the mass loss mechanism which has been characterized by the stellar evolution before the explosion. In this case, the matter density profile can be modelled by:
\begin{equation}\label{Eq:CSM_density}
    \rho_{\rm CSM}(r)=\rho_{\rm s,0}\times (r/R_{\rm 0})^{-s}\quad\text{for } r\in [R_{\rm 0}, R_{\rm CSM}],
\end{equation}
in which the internal CSM density
\begin{equation}
\rho_{\rm s,0}=\frac{(3-s)\,M_{\rm CSM}}{4\pi f_{\rm \Omega}\,R_{\rm 0}^3}\times\left[(R_{\rm CSM}/R_{\rm 0})^{3-s}-1\right]^{-1}
\end{equation}
depends on the entire CSM mass ($M_{\rm CSM}$), which is confined between $R_{\rm 0}(\ge R_{\rm \star})$ and $R_{\rm CSM}$ in a solid angle of $4\pi f_{\rm \Omega}$. Like in the ejecta, $\rho_{\rm CSM}$ is radially symmetric\footnote{Although there is strong evidence about the anisotropy and the irregularity of CSM surrounding massive stars \citep[see, e.g.,][and ref. therein]{2017hsn..book..875C}, the radial symmetry approximation into a solid angle $4\pi f_{\rm \Omega}$ is sufficient to estimate the order-of-magnitude for neutrino flux \citep[][]{Fang2020}{}{}.} and its value falls according to the exponent $s<3$. The latter permits the discrimination of three main scenarios: the uniform-dense shell ($s=0$), where the bulk of CSM is produced in a short time before the explosion epoch; the steady-state wind ($s=2$), when the mass-loss rate $\dot{M}$ and the stellar wind velocity $v_w$ are constant for a long time up to the explosion, i.e. $\rho_{\rm CSM}\propto r^{-2}\times \dot{M}/v_{\rm w}$; and the accelerated stellar wind ($s>2$), in case of the loss matter rate increases during the last stages of the progenitor evolution (see Fig. \ref{fig:density}).
 Moreover, the CSM inner boundary $R_{\rm 0}$, when it is detached from $R_{\rm \star}$, describes the scenario where the explosion is much later than the high mass loss phase of the progenitor \citep[see, e.g.,][]{2012ApJ...756L..22M,2023A&A...677A.105D}{}{}.\par
Differently from SLSNe and peculiar massive SNe IIn, the majority of H-rich SNe presents $M_{\rm CSM}\lesssim M_{\rm ej}$ \citep[see, e.g.,][]{MorzovaPiroValenti2018,2023MNRAS.524.3366P}{}{}. 
In this scenario, the ejected material homologously expands within a low-mass CSM, keeping its kinetic energy $E_{\rm k}$ almost unchanged \citep[see, e.g.,][]{Moriya2013}. Consequently, the density profile of the ejecta evolves as follows:
\begin{equation}\label{Eq:ej_density}
\rho_{\rm ej}(x,t)=\frac{\rho_0(x)\, R_{\rm \star,0}^3}{\left(v_{\rm sc}\>t+R_{\rm \star,0}\right)^{3}}\simeq \rho_0(x)\times \left(\frac{t}{t_{\rm e}}\right)^{-3}\> \text{for }t>>t_{\rm e},
\end{equation}
in which $t_{\rm e}=R_{\rm \star,0}/v_{\rm sc}$ represents the SN expansion time and $v_{\rm sc}$ denotes the ejecta's scale velocity for the speed profile\footnote{According to the hypothesis of homologous expansion, the speed profile can be expressed as $v_{\rm ej}(x) = v_{\rm sc} \, x$, where $v_{\rm sc}$ satisfies the following relationship:
$$E_{\rm k}=\frac{1}{2}v_{\rm sc}^2\,R_{\rm \star,0}^3\times \int_0^{x_{\rm \star}}\rho_0(x)x^4dx,$$ from which the equation (\ref{Eq:ejecta_velocity}) is derived.\label{Note:v_ej}}:
\begin{equation}\label{Eq:ejecta_velocity}
    v_{\rm sc}=\sqrt{\frac{2(n-5)(5-\delta)}{(n-3)(3-\delta)}\times\frac{n-\delta-(3-\delta)x_{\rm \star}^{3-n}}{n-\delta-(5-\delta)x_{\rm \star}^{5-n}}\times\frac{E_{\rm k}}{M_{\rm ej}}}.
\end{equation}
\par 
Under these assumptions, the interaction between the ejecta and CSM starts at $t_{\rm 0}\simeq R_{\rm 0}/v_{\rm max}$, where $v_{\rm max}=v_{\rm sc}\, x_{\rm \star}$ represents the maximum velocity of the outest ejecta. This marks the onset of the collision between the two media, leading to the formation of two distinct shock fronts: the backward shock, which recedes inward the ejecta center, and the forward shock progressing within the CSM. The forward shock exhibits greater efficiency in radiative cooling compared to the backward shock, augmenting its efficiency in particle ignition \citep[see, e.g.,][and references therein]{Suzuki_2020,Fang2020}{}. As a result, our exclusive focus is directed towards the forward shock, where the efficient cooling reduces the shock front in a thin shell of thickness equal to $\Delta R_{\rm sh}$, significantly smaller than its radial coordinate $R_{\rm sh}$. By invoking the conservation of momentum in the regime where $t$ is significantly grater than $t_{\rm 0}$, we can determine the velocity of the thin shell using a self-similar approach  \citep[see, e.g.,][and references therein]{Moriya2013}. In this way, this velocity can be expressed as follows:
\begin{equation}\label{Eq:shock_velocity}
v_{\rm sh}(t)\simeq v_{\rm sh,0}\times
\begin{cases}
\left(t/t_{\rm 0}\right)^{-(3-s)/(n-s)}\,& t<t_{\rm t},\\
(\alpha\, x_{\rm \star})^{\frac{n-4}{4-s}}\, \left(t/t_{\rm 0}\right)^{-(3-s)/(4-s)}\,& t\ge t_{\rm t},
\end{cases} 
\end{equation}
where $v_{\rm sh,0}$ is the initial shock velocity. This velocity can be obtained by the following relation:
\begin{equation}
v_{\rm sh,0} = \alpha\times  v_{\rm max}\times(n-3)/(n-s),
\end{equation}
where $\alpha$ is defined as
\begin{equation}\label{Eq:alpha}
    \alpha = A_{\rm n,\delta}^{s,\Omega}\times \left[\frac{M_{\rm ej}}{M_{\rm CSM}}\times \left(\frac{R_{\rm CSM}^{3-s}}{R_{\rm 0}^{3-s}}-1\right)\right]^{1/(n-s)},
\end{equation}
with
\begin{equation}\label{Eq:shock_coeff}
    A_{\rm n,\delta}^{s,\Omega}=\left[\frac{(4-s)(3-\delta)/(n-4)}{(n-\delta)x_{\rm \star}^{n-3}-(3-\delta)}\times f_{\rm \Omega}\right]^{1/(n-s)}.
\end{equation}
Note that $\alpha$ and $A_{\rm n,\delta}^{s,\Omega}$ are constant in time and solely depend on the CSM-ejecta matter distribution profiles. However, the global behaviour of $v_{\rm sh}$ in equation (\ref{Eq:shock_velocity}) changes when the shock moves from the external to inner ejecta\footnote{While this effect has been considered in several previous studies \citep[either numerically or in specific regimes; see e.g.][]{Murase11,Petropoulou17,Murase24}, it is often not discussed explicitly in analytical treatments of HE-$\nu$ emission from Type II SNe, particularly in cases where $M_{\rm CSM}<<M_{\rm ej}$ \citep[e.g.,][]{Murase18}. However, this effect can become significant when describing the LC of low-mass CSM interacting SNe in cases where the shock break-out occurs well before the forward shock reaches the edge of the CSM \citep[e.g.,][]{2011ApJ...729L...6C,Tsuna_2019}, such as in SN 2022qml \citep{Salmaso24}.
Consequently, here we have reformulated this transition analytically to consistently connect the shock evolution with both neutrino and electromagnetic emission over a wide CSM-ejecta parameter space (see also Section \ref{Subsec:par_mod}).}. The transition time ($t_{\rm t}$) corresponding to this change, is given by:
\begin{equation}\label{Eq:transition_time}
R_{\rm sh}(t_{\rm t})\simeq v_{\rm sc}\,t_{\rm t}\longrightarrow t_{\rm t}= t_{\rm 0}\times (\alpha x_{\rm \star})^{(n-s)/(3-s)}.
\end{equation}
By assuming continuity in the shock velocity and position at $t_t$, the shock radius can be determined through time integration of equation (\ref{Eq:shock_velocity}), obtaining the following relation:
\begin{align}\label{Eq:shock_radius}
    R_{\rm sh}(t)=\,&\int_{\rm t_{\rm 0}}^tv_{\rm sh}(t')dt' +R_{\rm 0}\nonumber\\
    \simeq\,& \alpha R_{\rm 0}\times
\begin{cases}
\left(t/t_{\rm 0}\right)^{(n-3)/(n-s)}\,& t<t_{\rm t}\\
-A_{\rm t}+B_{\rm t}\, (t/t_{\rm 0})^{1/(4-s)}\,& t\ge t_{\rm t},
\end{cases}    
\end{align}
where
\begin{equation}\label{Eq:A_Rh_coef}
A_{\rm t} = (\alpha x_{\rm \star})^{\frac{n-3}{3-s}} \times \left[(4-s)\,(3-s)/(n-s)-1\right]
\end{equation}
and
\begin{equation}\label{Eq:B_Rh_coef}
B_{\rm t} =  (\alpha x_{\rm \star})^{\frac{n-4}{4-s}}\times (4-s)\,(n-3)/(n-s).
\end{equation}
According to the equations (\ref{Eq:shock_radius})-(\ref{Eq:B_Rh_coef}), the shock dynamics within the CSM and the consequent proton injection mechanism finish when the entire CSM is swept by the thin shell, i.e. as long as the relation $R_{\rm sh}(t_{\rm f})\simeq R_{\rm CSM}$ is verified. Therefore, by using equation (\ref{Eq:shock_radius}), $t_{\rm f}$ can be analytically expressed as follows:
\begin{equation}\label{Eq:t_f}
t_{\rm f}\simeq t_{\rm 0}\times
\begin{cases}
\left(\frac{R_{\rm CSM}}{\alpha R_{\rm 0}}\right)^{(n-s)/(n-3)}\,& \alpha>\alpha_{\rm t},\\
\left(\frac{R_{\rm CSM}}{\alpha B_{\rm t} R_{\rm 0}}+\frac{A_{\rm t}}{B_{\rm t}}\right)^{4-s}\,&\alpha\le\alpha_{\rm t},
\end{cases}
\end{equation}
in which
\begin{equation}\label{Eq:alpha_t}
\alpha_{\rm t} = (R_{\rm CSM}/R_{\rm 0})^{(3-s)/(n-s)}\times x_{\rm \star}^{-(n-3)/(n-s)}
\end{equation}
represents the value of $\alpha$ when $t_{\rm f}$ equals $t_{\rm t}$.
Note that the relations in equations (\ref{Eq:shock_velocity})-(\ref{Eq:alpha_t}) are valid for a general CSM density profile with $s<3$. As a consequence, our formulation extends the asymptotic solution for the shock velocity to encompass all three main types of CSM described above, including the case of $s=2$ previously derived by \cite{Moriya2013}. This generalization allows for a unified analytical description applicable to low-mass CSM interaction scenarios with interior shock break-out \citep[e.g.,][]{KK_2023}. Specifically, the forward shock shell can penetrate into the inner ejecta if the shock transition occurs before the end of the interaction, i.e. for $t_t<t_f$. Assuming a thin outer ejecta layer ($x_\star\simeq 1$) and an extended CSM ($R_{\rm CSM}>>R_0$), the occurrences of this penetration is directly linked to the mass ratio between the CSM and the ejecta, because $t_t<t_f$ implies:
\begin{equation}\label{Eq:trans_condition}
\frac{M_{\rm CSM}}{M_{\rm ej}}\gtrsim \frac{f_\Omega (4-s)(3-\delta)}{(n-4)(n-3)}.
\end{equation}
In the case of RSG-like ejecta of $10\,$M$_\odot$ and a spherical CSM, the relation (\ref{Eq:trans_condition}) translates to $M_{\rm CSM}\gtrsim 0.5^{+.5}_{-.2}\,$M$_\odot$ for $s\in[0,3[$. These $M_{\rm CSM}$ values are consistent with the assumption of a low-mass CSM, in which the total dissipation energy is less than a tenth of $E_k$ and the deceleration radius exceeds $R_{\rm CSM}$ \citep[see also, e.g.,][]{Murase14,Murase18}.

\par
Once the forward shock's dynamic evolution has been determined, it is possible to assess whether and when the acceleration process of protons begins in the shocked shell.
Indeed, the onset of the proton injection phase depends on the radiation-matter energy exchanges in the shocked shell, as well as on the shock dynamic evolution. Generally, at the beginning of the CSM-ejecta interaction, the shock is radiation dominated. In this regime, the energy dissipation primarily occurs through free-free radiation processes, trapping the energy within an optically thick region in the form of thermal energy. In other words, the high optical depth of the CSM, resulting from Thomson electron scattering, satisfies the radiation-mediated shock condition \citep[$\tau_{\rm s} \gg \tau_{\rm bo}$; see ][]{2023MNRAS.524.3366P}, expressed by:
\begin{equation}\label{Eq:tau}
\tau_{\rm s}\equiv k_{\rm T}\times\int_{\rm R_{\rm sh}}^{R_{\rm CSM}}\rho_{\rm CSM}(r)\,dr \gg \tau_{\rm bo}\equiv c/v_{\rm sh},
\end{equation}
in which the Thomson opacity $k_{\rm T}$ is about $0.34 \, \text{cm}^2 \, \text{g}^{-1}$ for H-rich CSM with solar-like abundances \citep[see, e.g.,][]{2013MNRAS.433..838P}. The shell remains in this state until shock break-out ($t_{\rm bo}$), i.e. when the following relation is valid:
\begin{equation}\label{Eq: t_bo}
\tau_{\rm s}(t_{\rm bo}) =\tau_{\rm bo}(t_{\rm bo}).
\end{equation}
This relation implies that the $R_{\rm sh}(t_{\rm bo})$ coincides with the break-out radius ($R_{\rm bo}$), i.e. $R_{\rm sh}(t_{\rm bo})=R_{\rm bo}(t_{\rm bo})$, where:
\begin{equation}\label{Eq:brakout_time}
R_{\rm bo}(t)=R_{\rm CSM}\times
    \begin{cases}
    \exp{\left[-\beta\, \tau_{\rm bo}(t)\right]}\,& \text{for }s=1,\\
    \\
 \sqrt[1-s]{1-\beta(1-s)\, \tau_{\rm bo}(t)}\,& \text{for }s\not= 1;
\end{cases}
\end{equation}
with
\begin{equation}\label{Eq:beta}
 \beta =\left(k_{\rm T}\,\rho_{\rm s,0}\,R_{\rm CSM}\right)^{-1}\times (R_{\rm CSM}/R_{\rm 0})^{s}.
\end{equation}
\par
After $t_{\rm bo}$, the electromagnetic radiation can easily escape from the shock, and the thermodynamic state of the shocked gas is predominantly influenced by the collisional interactions among ions and electrons.
Here, indeed, the shock transitions into a collisional phase, where the plasma may exhibit thermal anisotropy, giving rise to the development of electromagnetic instabilities that contribute to the generation of a magnetic field \citep[e.g.,][]{PhysRevLett.87.071101,2021ApJ...922....7I}. 
This magnetic field injects relativistic protons into the shock, initiating their acceleration over a timescale $t_{\rm acc}$. Simultaneously, radiation and collisional processes persist, gradually diminishing the energy of these protons throughout a cooling timescale $t_{\rm p,cool}$ \citep[e.g.,][]{Petropoulou17,Pitik2022}. Therefore, the particle acceleration efficiency grows up when the shock becomes collisionless ($t_{\rm acc} < t_{\rm p,cool}$), facilitating non-thermal particle acceleration \citep[][]{1976ApJS...32..233W}, in analogy with the supernova remnants shock \citep[e.g.,][]{1988ApJ...329L..29C,2007A&A...465..695Z,2010A&A...513A..17O}.
In particular, the acceleration timescale of protons,  given by the first-order Fermi mechanism in the Bohm limit \citep[$\eta \simeq 1$; see, e.g.,][]{1983RPPh...46..973D,2004PASA...21....1P}, depends on the proton gyro-radius ($r_{\rm g} \equiv \gamma\, m_{\rm p}\,c^2 / eB_{\rm sh}$) and is defined as follows \citep[e.g.,][]{Murase14}:
\begin{equation}\label{Eq:t_acc}
t_{\rm acc} \equiv \eta \times \frac{20\,r_{\rm g} c}{3\,v_{\rm sh}^2} \simeq \frac{20\,\gamma \, m_{\rm p}/e}{3\,B_{\rm sh}(t)/c} \times \left[\frac{c}{v_{\rm sh}(t)}\right]^2,
\end{equation}
where $\gamma$ denotes the Lorentz factor for protons, $m_{\rm p}/e$ is the proton mass-charge ratio, and $B_{\rm sh}\equiv \sqrt{9\pi\epsilon_{\rm B}\rho_{\rm CSM}v_{\rm sh}^2}$ represents the equipartitioned magnetic field strength near the shock. This magnetic field strength is proportional to a fraction $\epsilon_{\rm B}\sim 10^{-2}-10^{-1}$ of the postshock thermal energy density, $U_{\rm th}\equiv(9/8)\rho_{\rm CSM}v_{\rm sh}^2$, as derived using the Rankine-Hugoniot (RH) thermal condition for a monoatomic gas \citep[see, e.g.,][]{Chevalier2006,2016MNRAS.460...44P,Tsuna_2019}. The primary proton-cooling mechanisms, instead, operate on the following timescale:
\begin{equation}\label{Eq:t_pcool}
    t_{\rm p,cool}\equiv(t_{\rm pp}^{-1}+t_{\rm ad}^{-1})^{-1},
\end{equation}
where $t_{\rm pp}$ and $t_{\rm ad}$ account for the inelastic pp-collisions effects and the plasma cooling due to the adiabatic expansion of the shocked shell, respectively. The latter effects consistently impose limits on the maximum duration of the acceleration process ($t_{\rm acc} \le t_{\rm p,cool}$).
According to this, the protons with the maximum Lorentz factor $\gamma_{\rm max}$, i.e. having a relativistic energy
\begin{equation}\label{Eq:Ep_max}
E_{\rm p}^{\rm max} \equiv \gamma_{\rm max}\, m_{\rm p}c^2,
\end{equation}
have to satisfy the collisionless condition limit given by $t_{\rm acc} = t_{\rm p,cool}$. In this way, equating equations (\ref{Eq:t_acc}) and (\ref{Eq:t_pcool}), $\gamma_{\rm max}(t)$ can be derived by the following implicit equation dependent on $t$:
\begin{equation}\label{Eq:gamma_max_cond}
 \gamma_{\rm max}= \frac{\Gamma (t)/ \sigma_{\rm pp}(\gamma_{\rm max})}{1+t_{\rm pp}(\gamma_{\rm max},t)/t_{\rm ad}(t)}\quad \text{with}\quad\gamma_{\rm max}\ge 1,
\end{equation}
where 
\begin{equation}\label{Eq:Gamma}
    \Gamma(t)=\frac{9\,e\mu\sqrt{\pi\epsilon_{\rm B}}}{80\,ck_{\rm pp}}\times\left[\frac{v_{\rm sh}(t)}{c}\right]^3\times \rho_{\rm CSM}^{-1/2}\left[R_{\rm sh}(t)\right]
\end{equation}
is an efficiency area for the proton acceleration mechanism, proportional to the ratio between $t_{\rm pp}/t_{\rm acc}$. Indeed, $\Gamma$ operates inversely to the pp-inelastic cross-section ($\sigma_{\rm pp}$), which, within the range of energies considered here, becomes \citep[][]{Kelner2006}{}{}:
\begin{equation}\label{Eq:sigma_pp}
    \sigma_{\rm pp}(\gamma)\simeq(33.4-1.61\,\log\gamma+0.25\,\log^2\gamma)\times 10^{-27}\,\text{cm}^2.
\end{equation}
\par
Note that, in equation (\ref{Eq:gamma_max_cond}), the presence of $\sigma_{\rm pp}$ derives from the pp-interaction timescale of the shell expressed as
\begin{equation}\label{Eq:t_pp}
    t_{\rm pp}(\gamma,t)=\left\{c\,k_{\rm pp}\,\sigma_{\rm pp}(\gamma)\times n_{\rm sh}[R_{\rm sh}(t)]\right\}^{-1},
\end{equation}
where $k_{\rm pp}=0.5$ is the constant inelasticity and $n_{\rm sh}$ is the particle numerical density inside the shell \citep[e.g.,][]{Murase14,Sarmah_2022}. Specifically, $n_{\rm sh}$ has been directly inserted in equation (\ref{Eq:Gamma}) by using the RH conditions for a mono-atomic gas, i.e. $n_{\rm sh}\equiv 4\rho_{\rm CSM}/\mu m_{\rm p}$. Moreover, considering a H-rich CSM with solar abundance, the mean molecular weight $\mu$ for a neutral gas is approximately $1.3$ \citep[][]{2019arXiv191200844L}{}{}.\par
Additionally, in equations (\ref{Eq:t_pcool}) and (\ref{Eq:gamma_max_cond}), the adiabatic cooling timescale is the minimum value between the dynamical expansion time and the cooling time of the gas behind the shock \citep[][]{Fang2020}, so it can be expressed as
\begin{equation}\label{Eq:ad_time} 
  t_{\rm ad}\equiv min[t_{\rm dyn},t_{\rm cool}],
\end{equation}
where $t_{\rm dyn}$ and $t_{\rm cool}$ are respectively defined by the equations
\begin{equation}\label{Eq:dyn_time}
t_{\rm dyn}\equiv \frac{R_{\rm sh}}{v_{\rm sh}}\simeq t\times\frac{n-s}{n-3}\times
\begin{cases}
1\,& t<t_{\rm t}\\
\frac{B_{\rm t}-A_{\rm t}\, (t/t_{\rm 0})^{-1/(4-s)}}{(\alpha x_{\rm \star})^{(n-4)/(4-s)}}\,& t\ge t_{\rm t}
\end{cases}  
\end{equation}
and
\begin{equation}\label{Eq:cool_time}
    t_{\rm cool}\equiv\frac{\bar{E}^k_{\rm e,i}/n_{\rm sh}}{\Lambda(T_{\rm sh})}=\frac{3\,k_{\rm B}\,\mu\, m_{\rm p}/8}{\rho_{\rm CSM}\left[R_{\rm sh}(t)\right]}\times \frac{T_{\rm sh}(t)}{\Lambda[T_{\rm sh}(t)]},
\end{equation}
being $\bar{E}^k_{\rm e,i}$ the average electron-ion kinetic energy and $\Lambda(T)$ the radiative cooling function \citep[see, e.g.,][]{2011piim.book.....D,Margalit_2022}{}{}. In particular, $\bar{E}^k_{\rm e,i}$ depends on the temperature inside the shell given by the RH conditions:
\begin{equation}\label{Eq:RH_condition}
\bar{E}^k_{\rm e,i}\equiv\frac{\Tilde{\mu}\,U_{\rm th}}{\mu\,n_{\rm sh}} \equiv \frac{3}{2}k_{\rm B}\,T_{\rm sh}\quad\longrightarrow\quad    T_{\rm sh}(t)=\frac{3\,\Tilde{\mu}\,m_{\rm p}}{16\,k_{\rm B}}\, v_{\rm sh}^2(t),
\end{equation}
where $k_{\rm B}$ is the Boltzmann constant, and $\Tilde{\mu}\simeq 0.6$ is the molecular weight for fully ionized H-rich CSM with solar composition \citep[][]{2023MNRAS.524.3366P}{}{}. Moreover, for $T>10^5\,\text{K}$, $\Lambda(T)$ can be expressed as
\begin{equation}\label{Eq:Lambda}
\Lambda(T)\simeq 1.6\times 10^{-23}\times
\begin{cases}
(T/T_{\rm *})^{-0.6}\,& T\lesssim T_{\rm *}\\
(T/T_{\rm *})^{0.5}\,& T>T_{\rm *}
\end{cases}
\quad\left[\frac{\text{erg}\, \text{cm}^3}{\text{s}}\right],
\end{equation}
where $T_{\rm *} = 4.7 \times 10^7\,\text{K}$ is the transition temperature from the emission regime dominated by free-free processes ($T \gtrsim T_{\rm *}$) to the one where atomic-line emission becomes relevant ($T \lesssim T_{\rm *}$), as noticed in \cite{1994ApJ...420..268C}.

\subsection{Production rate of neutrinos from pp-collisions}\label{Subsec:nu_flux}
The rate of pp-collisions is primarily influenced by the proton energy distribution within the shell throughout the entire shock evolution. Consequently, we need to compute the number of protons ($N_{\rm p}$) at time $t$, with proton Lorentz factors ranging between $\gamma$ and $\gamma+d\gamma$, by integrating the following continuity equation \citep[e.g.,][]{RevModPhys.42.237,Petropoulou17}:
\begin{equation}\label{Eq:continuity}
    \frac{\partial N_{\rm p}(\gamma,t)}{\partial t}+\frac{N_p(\gamma,t)}{t_{\rm esc}(\gamma,t)}+ \frac{\partial}{\partial\gamma}\left[\dot{\gamma}(t)\, N_{\rm p}(\gamma,t)\right]=Q_{\rm p}(\gamma,t),
\end{equation}
where $Q_{\rm p}$, $\dot{\gamma}$, and $t_{\rm esc}^{-1}$ are the rates of proton injection into the shock, adiabatic energy loss, and proton escape, respectively.\par 
In particular, $Q_{\rm p}$ can be expressed in units of time and $\gamma$ by the following relation \citep[][]{2012ApJ...751...65F}:
\begin{equation}\label{Eq:Q_p}
    Q_{\rm p}(\gamma,t)= 
    \begin{cases}
        \frac{q(t)\times\gamma^{-2}}{\log{[\gamma_{\rm max}(t)]}}\,&\gamma\le \gamma_{\rm max}(t)\\
        0\,& \gamma> \gamma_{\rm max}(t).
    \end{cases}
\end{equation}
In this way, $Q_{\rm p}$ dynamically enhances the proton population up to the maximum energy limit [$\propto\gamma_{\rm max}$, cf. equation (\ref{Eq:Ep_max})], achieved through the implicit solution of equation (\ref{Eq:gamma_max_cond}). As in \citet{Pitik2022}, when $\gamma\le \gamma_{\rm max}$, $Q_{\rm p}$ follows a power-law behaviour with a proton spectral index of 2 ($\propto \gamma^{-2}$) and a minimum proton energy level set at $m_{\rm p}c^2$ ($\gamma_{\rm min}=1$). Moreover, the expression for $Q_{\rm p}$ involves the shell-dependent quantity:
\begin{equation}\label{Eq:q_fact}
    q(t)=9\pi f_{\rm \Omega}\, \epsilon_{\rm p}\, R_{\rm sh}^2(t)\, v_{\rm sh}^3(t)\,\rho_{\rm CSM}\left[R_{\rm sh}(t)\right]/(8 m_{\rm p} c^2),
\end{equation}
in which $\epsilon_{\rm p}$ represents the fraction of kinetic energy used to accelerate protons \citep[$\simeq 0.1$ for a parallel or quasi-parallel shock; see, e.g.,][]{2016MNRAS.460...44P}.\par
As for $\dot{\gamma}$, it is determined by the adiabatic expansion of the accelerating shock region\footnote{\label{Note:cooling}Although other energy loss channels, such as radiative emission, are often neglected \citep[see, e.g.,][]{Murase11,Fang2020}, the adiabatic energy loss resulting from shock-shell expansion can become relevant \citep[see, e.g.,][]{2023MNRAS.524.3366P}{}{}. When the cooling rate becomes greater than the expansion one, indeed, the radial width $\Delta R_{\rm \text{sh}} \equiv R_{\rm \text{sh}}\times (t_{\rm \text{ad}}/t_{\rm \text{dyn}})$ may become smaller than $R_{\rm \text{sh}}$, thereby diminishing $|\dot{\gamma}|$ [cf. equation (\ref{Eq:gamma_dot})].}. Therefore, the volume of the radiative shock-shell ($V\propto R_{\rm sh}^2\Delta R_{\rm sh}$), linked to its radial width $\Delta R_{\rm sh}\simeq v_{\rm sh}\, t_{\rm ad}$, defines the adiabatic energy loss-rate as \citep[][]{1975ApJ...196..689G}{}{}:
\begin{equation}\label{Eq:gamma_dot}
\Dot{\gamma}\equiv-\frac{\gamma}{3}\times \frac{d\ln{V}}{dt}=-\frac{\gamma}{t_{\rm dyn}}-\frac{\gamma}{3}\times \frac{\partial\ln{\left(t_{\rm ad}/t_{\rm dyn}\right)}}{\partial t}.
\end{equation}
\par
\begin{figure*}
\centering
	\includegraphics[scale=0.5]{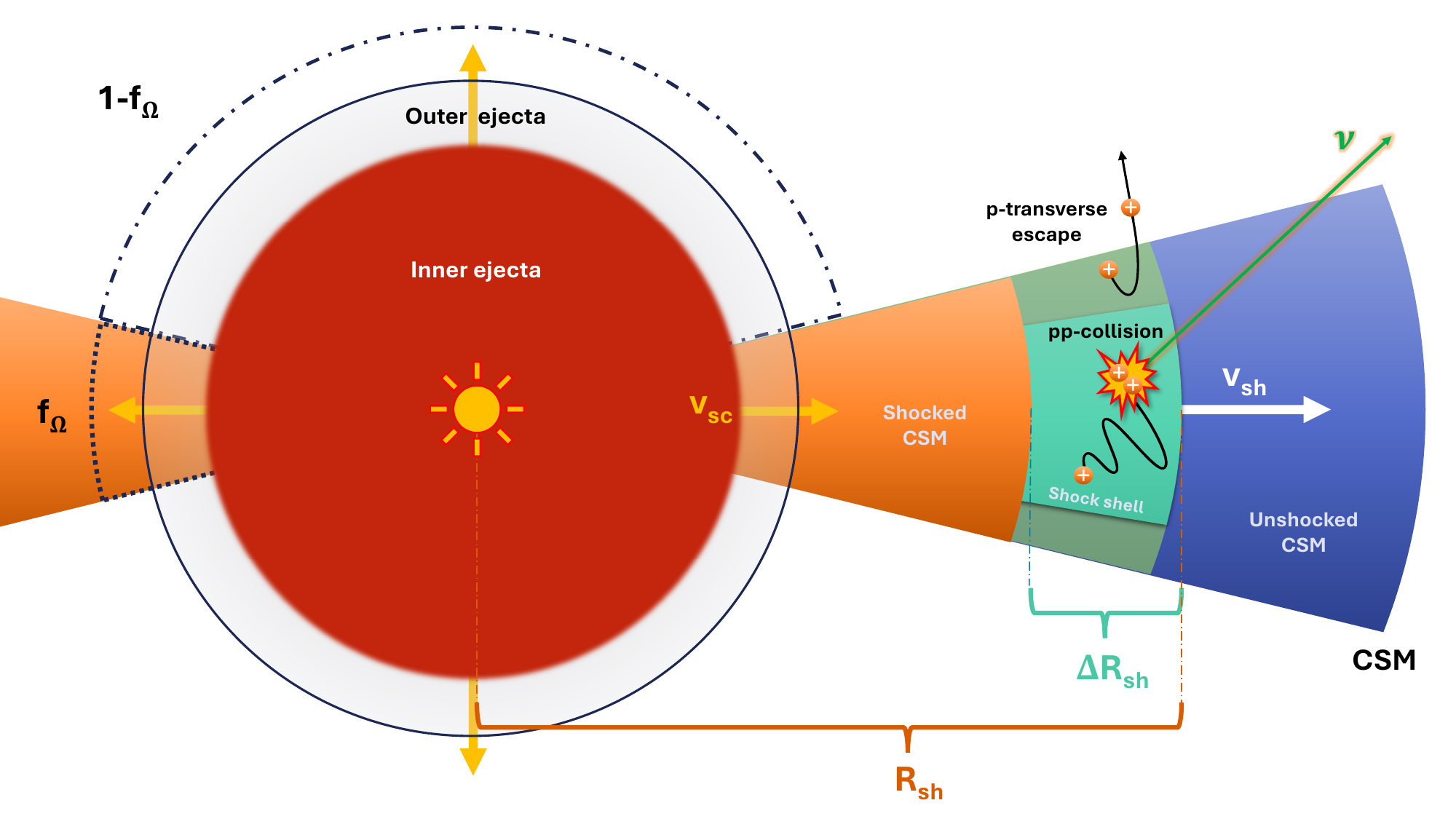}
    \caption{Schematic view of a SN explosion interacting with a discoidal CSM with a solid angle fraction $f_\Omega$. The SN ejecta, following a homologous expansion (cf. footnote \ref{Note:v_ej}), is divided into an inner dense envelope and an outer region with a steeper density profile. The forward shock front, located at radius $R_{\rm sh}$ propagates through the CSM with velocity $v_{\rm sh}$, generating a hot shocked shell of thickness $\Delta R_{\rm sh}$. In this region, particle acceleration can occur, and protons can be destroyed via inelastic pp-collisions, producing neutrinos that escape unimpeded through unshocked CSM. However, protons near the lateral edge of the shocked shell can also escape from the discoidal CSM due to magnetic instabilities and its finite angular extent. This transverse escape introduces an additional loss channel into the proton balance of the shock-shell, which depends on the open solid angle fraction $1-f_\Omega$.
     \label{fig:cartoon}}
\end{figure*}
As far as the term proportional to $t_{\rm esc}^{-1}$, it describes the reduction of $N_{\rm p}$ due to pp-collisions and, for a nonspherical CSM, also the { transverse} loss (see Fig. \ref{fig:cartoon}):
\begin{equation}
 t_{\rm esc}^{-1}\equiv t_{\rm pp}^{-1}+t_{  \perp}^{-1}.
\end{equation}
Inelastic pp-collisions are considered catastrophic energy loss mechanisms \citep[e.g.,][]{Sturner_1997,Petropoulou17}, in contrast to the energy losses associated with the adiabatic expansion of the shell [cf. $\Dot{\gamma}$ in equation (\ref{Eq:continuity})]. Other processes, such as photomeson cooling, are deemed negligible in this context \citep[see also][]{Murase11}. Additionally, in the spherical case, the proton loss timescale due to escape from the shell is usually longer than other characteristic timescales in the system\footnote{In the case of low massive and spherical CSM, assuming the Bohm limit for the proton diffusion length \citep[$\equiv cr_{\rm g}/3v_{\rm sh}$; e.g.][]{Murase14}, the maximum energy that a proton can acquire before escaping from the acceleration region large $\Delta R_{\rm sh}$ is equal to $3eB_{\rm sh}\,(v_{\rm sh}^2/c)\,t_{\rm ad}$ \citep{Fang2020}. This energy is always greater than $E_{\rm p}^{\rm max}=(3/20)eB_{\rm sh}\,(v_{\rm sh}^2/c)\,t_{\rm p,cool}$, which is limited by proton cooling processes [cf. equations (\ref{Eq:t_acc})-(\ref{Eq:t_pcool})], because $t_{\rm p,cool}\le t_{\rm ad}$. However, in scenarios where $M_{\rm CSM}>>M_{\rm ej}$ that are outside the low-mass CSM regime considered in this work, the shock rapidly reaches the deceleration radius, so the proton escape timescale can become shorter than the cooling one, as discussed in \citet{Murase11}.
} \citep[e.g.][]{Petropoulou17}. 
On the other hand, the nonspherical geometry can lead to transverse proton leakage from the acceleration region \citep[see, e.g.,][]{Fang2020}. For a discoidal CSM  where $f_\Omega<{(}3{/20)}v_{\rm sh}/c$, this effect {can become particularly} relevant. Indeed, the transverse loss rate ($t_{  \perp}^{-1}$) is proportional to the fraction of protons whose gyroradius is comparable to the transverse thickness of the shock shell—approximately $R_{\rm sh}\,f_{\rm \Omega}/r_g$—multiplied by the loss rate of accelerated particles, which corresponds to $t_{\rm acc}^{-1}$ for strong shocks \citep[][]{2004PASA...21....1P}.
As a novel contribution, we explicitly account for this effect by defining
\begin{equation}\label{Eq:t_loss}
    t_{ \perp}\equiv \frac{R_{\rm sh}\,f_{\rm \Omega}}{r_{\rm g}}\times \frac{t_{\rm acc}}{\phi}\simeq  \frac{f_{\rm \Omega}}{1-f_{\rm \Omega}}\times \frac{c}{v_{\rm sh}(t)}\times \frac{t_{\rm dyn}(t)}{3{/20}},
\end{equation}
where $\phi\simeq 1-f_{\rm \Omega}$ approximates the geometrical escape probability for protons\footnote{In the spherical CSM case ($f_{\rm \Omega}=1$), as $\phi\rightarrow 0$, the timescale $t_{  \perp}$ tends towards infinity, corresponding to the absence of  transverse losses.} (see Fig. \ref{fig:cartoon}). 
 Unlike the escape-limited scenario \citep{2010A&A...513A..17O,Murase24}, this purely geometric effect, being independent of proton energetics, facilitates cosmic ray escape, thereby reducing the neutrino flux (see also Section \ref{Subsec:par_mod}).

\par
To solve equation (\ref{Eq:continuity}), we require both energy and time boundary conditions. Consistently with the former hypotheses, we assume that the number of accelerated protons at $t_{\rm bo}$ is zero across all energy states, so the following relation is valid:
\begin{equation}\label{Eq:Bound1}
    N_{\rm p}(\gamma,t_{\rm bo})=0 \qquad\,\forall\, \gamma\ge1.
\end{equation}
Additionally, we consider that the maximum energy state remains unpopulated by any protons throughout the entire interaction period (i.e. from $t_{\rm bo}$ to $t_{\rm f}$) and, consequently, the relation
\begin{equation}\label{Eq:Bound2}
    N_{\rm p}\left(\gamma^{MAX},t\right)=0 \qquad \,\forall\, t\in[t_{\rm bo},t_{\rm f}];
\end{equation}
is valid, where $\gamma^{MAX}$ is the maximum protons' Lorentz factor, defined as
\begin{equation}
 \gamma^{MAX}= \underset{t\in[t_{\rm bo},t_{\rm f}]}{MAX}\left[\gamma_{\rm max}(t)\right].   
\end{equation}
\par
After numerically obtaining the distribution of protons measured in the shock\footnote{Given the advection equation (\ref{Eq:continuity}) with $\dot{\gamma}<0$ and its boundary equations (\ref{Eq:Bound1})-(\ref{Eq:Bound2}), the down wind integration method \citep{Courant52} is among the most suitable and stable to calculate the relativistic proton energy distribution in time, i.e. $N_{\rm p}(\gamma,t)$.}, the production rate of neutrinos and antineutrinos [(anti-)neutrinos] in $\text{GeV}^{-1}s^{-1}$ units with energy $E_{\rm \nu}$ at $t$ can be computed for muon and electron flavours $i=[\mu, e]$ using the relation: 
\begin{equation}
    Q_{\rm \nu_i+\Bar{\nu}_i}(E_{\rm \nu},t)= \frac{c\, n_{\rm sh}\left[R_{\rm sh}(t)\right]}{0.938\>\text{GeV}}\times \bar{Q}_{\rm \nu_i+\Bar{\nu}_i}(E_{\rm \nu},t),
\end{equation}
where $\bar{Q}_{\rm \nu_i+\Bar{\nu}_i}$ are respectively given by \citep[see, e.g.,][]{Kelner2006}{}{}:
\begin{align}\label{Eq:Q_mu}
    \bar{Q}_{\rm \nu_\mu+\Bar{\nu}_\mu}(E_{\rm \nu},t)=\int_0^1\,& \sigma_{\rm pp}\left(\frac{E_{\rm \nu}}{xm_{\rm p}c^2}\right)\times N_{\rm p}\left(\frac{E_{\rm \nu}}{xm_{\rm p}c^2} ,t\right)\times \\ 
    \,&\left[F_{\rm \nu_\mu}^{(1)}\left(x,E_{\rm \nu}/x\right)+F_{\rm \nu_\mu}^{(2)}\left(x,E_{\rm \nu}/x\right)\right]\, d(\ln{x})\nonumber
\end{align}
and
\begin{align}\label{Eq:Q_e}
    \bar{Q}_{\rm \nu_e+\Bar{\nu}_e}(E_{\rm \nu},t)=\int_0^1\,& \sigma_{\rm pp}\left(\frac{E_{\rm \nu}}{xm_{\rm p}c^2}\right)\times N_{\rm p}\left(\frac{E_{\rm \nu}}{xm_{\rm p}c^2} ,t\right)\times \\ 
    \,&F_{\rm \nu_e}^{(1)}\left(x,E_{\rm \nu}/x\right)\, d(\ln{x}),\nonumber
\end{align}
in which the functions $F_{\rm \nu_\mu}^{(1)}$, $F_{\rm \nu_\mu}^{(2)}$, and $F_{\rm \nu_e}^{(1)}$ describe the decay channels of pions and leptons resulting from the pp-collision, ultimately leading to neutrino production [cf. equations (62) and (66) of \citet{Kelner2006} for the complete expressions of these functions]. Note that the applicability of equations (\ref{Eq:Q_mu})-(\ref{Eq:Q_e}) is limited by the approximations on $\sigma_{\rm pp}$ of equation (\ref{Eq:sigma_pp}) valid for $E_{\rm \nu}\ge 0.1\,\text{TeV}$, which yet serves the aims of this paper.

\begin{table}
	\centering
	\caption{Summary of all modelling parameters with their typical ranges for the ejecta and CSM of common H-rich SNe \citep[see, e.g.,][]{MorzovaPiroValenti2018,Piro_2021}{}{}. The adopted physical constants and fixed parameter values have been taken considering a H-rich CSM of solar-like composition, according to other cited works (see where they have been introduced in the text). $M_{\rm Ni}$ and $\epsilon_{\rm rad}$  parameters are included here because they are involved in the SN electromagnetic emission modelling presented in Section \ref{sec:EM_emi}.}
	\label{tab:parameters}
	\begin{tabular}{lccc} 
		\hline
		Symbol & Name & Values ranges & Units\\
		\hline
        \multicolumn{4}{c}{Supernova Ejecta}\\
        \hline
		$E_{\rm k}$ & Kinetic Energy & $[0.5-5]$ & $\text{foe}$\\
		$M_{\rm ej}$ & Ejected Mass & $[5-25]$ & $\text{M}_{\rm \odot}$\\
		$R_{\rm \star}$ & Progenitor radius & $[0.1-10]$ & $10^{13}\,\text{cm}$\\
		$M_{\rm Ni}$ & Ejected $^{56}$Ni Mass & $[0.01-0.1]$ & $\text{M}_\odot$\\
		$\delta$ & Internal density profile exp. & $[0-1]$ & {}\\
        $n$ & External density profile exp. & $[8-12]$ & {}\\                		
        $x_{\rm \star}$ & External normalized boundary & $[1-1.2]$ & {}\\
		\hline
          \multicolumn{4}{c}{Circum-Stellar Medium}\\
        \hline
		$M_{\rm CSM}$ & Ejected Mass & $[0.05-0.8]$ & $\text{M}_{\rm \odot}$\\
		$R_{\rm 0}$ & CSM internal radius &$[0.1-10]$ & $10^{13}\,\text{cm}$\\
  		$R_{\rm CSM}$ & CSM external radius & $[1-10]$ & $10^{15}\,\text{cm}$\\
		 $s$ & CSM density profile exp. & $[0-3[$ & {}\\
      	$f_{\rm \Omega}$ & CSM angular distribution & $[0.0{01}-1]$ & {}\\
		\hline
                \multicolumn{4}{c}{Fixed CSM-ejecta interaction parameters}\\
        \hline
        $k_{\rm T}$ & Thomson opacity for $e^-$ & $0.34$ & $\text{cm}^2\text{g}^{-1}$\\
        $T_{\rm *}$ & Cooling transition temperature & $4.7\times 10^7$ & $\text{K}$\\
        $\epsilon_{\rm B}$ & Magnetic energy fraction & $3\times 10^{-2}$ & {}\\
        $\epsilon_{\rm p}$ & $p$-accelerating energy fraction & $0.1$ & {}\\
        $k_{\rm pp}$ & Inelasticity of pp-interaction & $0.5$ & {}\\
        $\mu$ & Mol. weight for neutral gas & $1.3$ & {}\\
        $\Tilde{\mu}$ & Mol. weight for fully ionized gas & $0.6$ & {}\\
        $\epsilon_{\rm rad}$ & Radiation energy fraction & $0.44$ & {}\\      
        \hline
	\end{tabular}
\end{table}
\subsection{SN modelling parameters \& Neutrino emission features}\label{Subsec:par_mod}
The semi-analytical model presented in Sections \ref{Subsec:p_acc} and \ref{Subsec:nu_flux} permits to derive the HE-$\nu$ spectra using twelve SN modelling parameters, which describe the physical configuration of the ejecta and CSM at the explosion (see Tab. \ref{tab:parameters}).
\begin{figure*}
       \centering
        \begin{multicols}{2}
           \subcaptionbox{Ejecta explosion parameters; \label{fig:Nnu_Ek_Mej}}{\includegraphics[width=\linewidth]{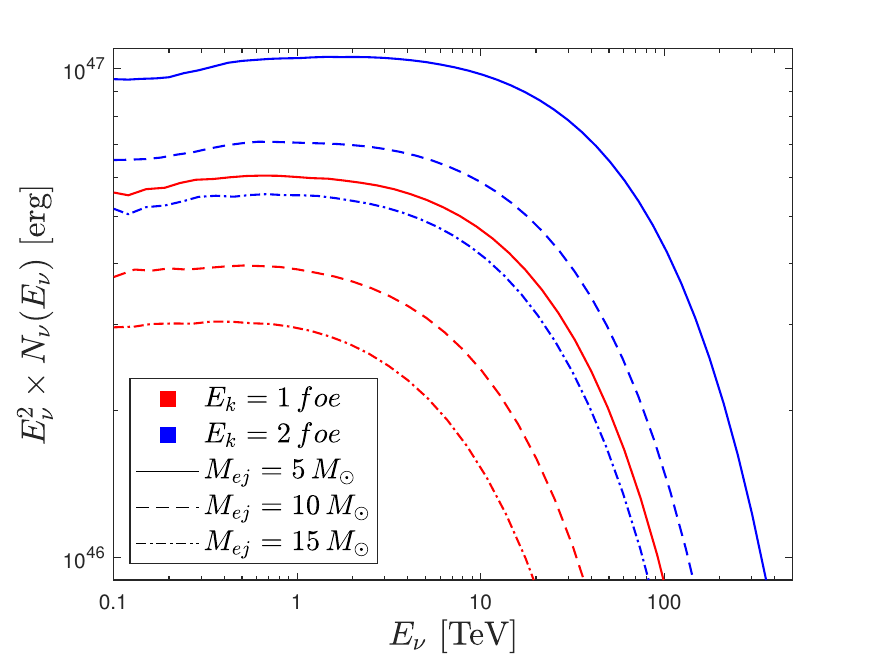}}\par 

            \subcaptionbox{CSM's matter density distribution; \label{fig:Nnu_Mcsm_s}}{\includegraphics[width=\linewidth]{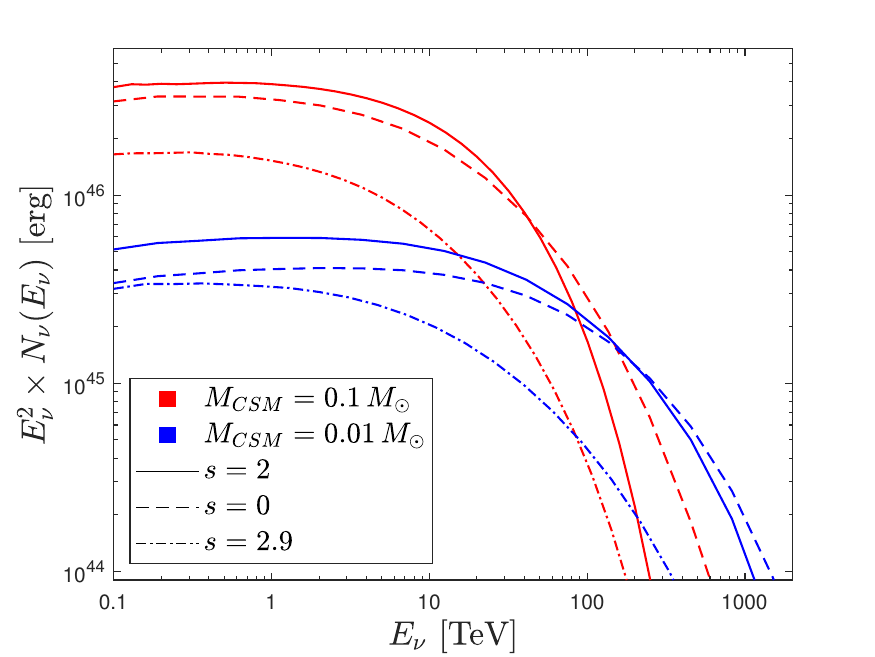}}\par 
         \end{multicols}

        \begin{multicols}{2}
            \subcaptionbox{Outer ejecta density profile; \label{fig:Nnu_xs_n}}{\includegraphics[width=\linewidth]{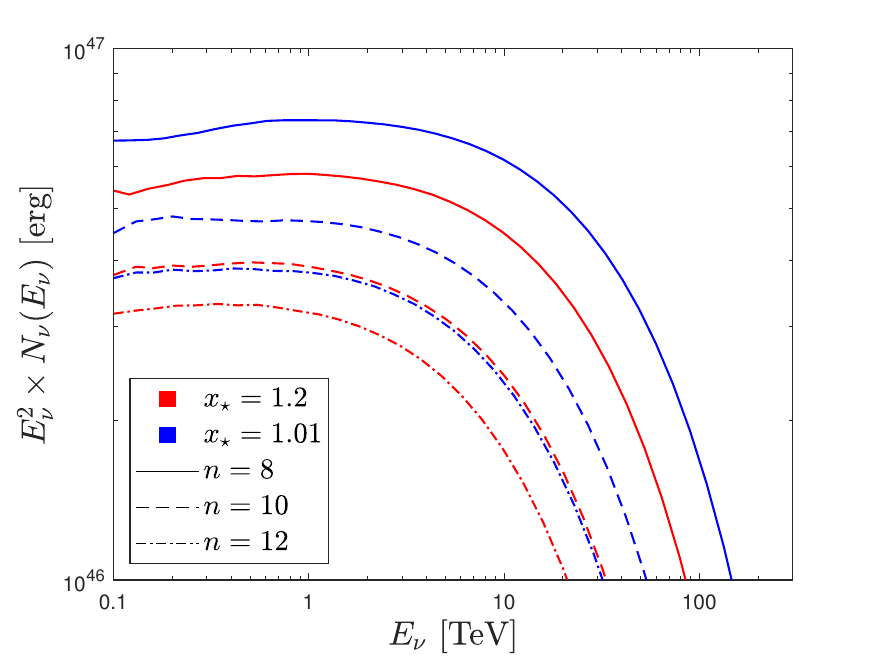}}\par 

            \subcaptionbox{CSM's radial and angular extension.
 \label{fig:Nnu_Rcsm_fOmega}}{\includegraphics[width=\linewidth]{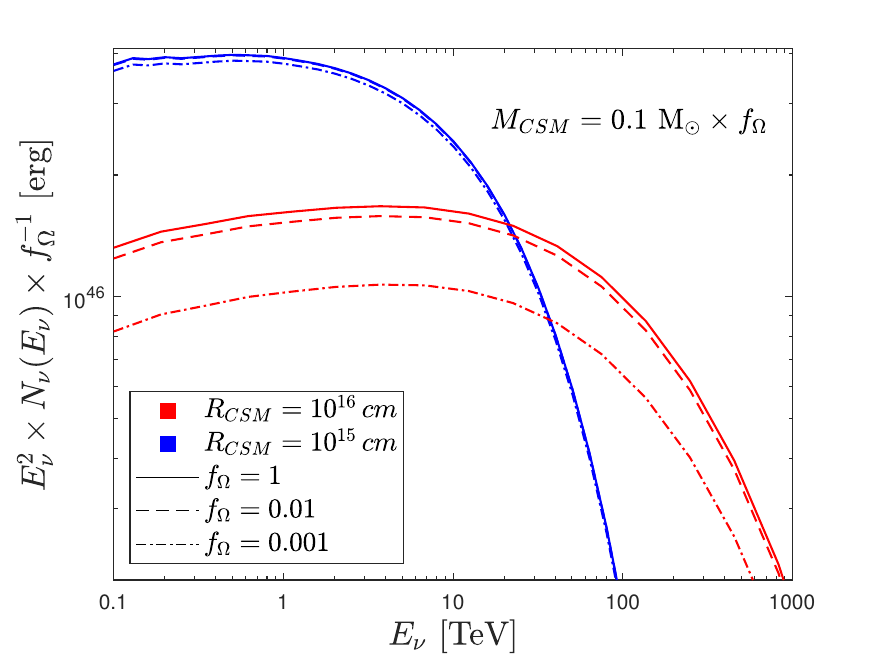}}\par 
 \end{multicols}
\caption{Simulated neutrinos' energy spectra from ejecta-CSM interaction with different values of the main modelling parameters. In all figures, not explicitly specified quantities are set as following: $E_{\rm k}=1\,$foe, $M_{\rm ej}=10\,\text{M}_{\rm \odot}$, $R_{\rm \star}=R_{\rm 0}=10^{13}\,\text{cm}$, $x_{\rm \star}=1.2$, $n=10$, $\delta=1$, $M_{\rm CSM}=0.1\text{M}_{\rm \odot}$, $R_{\rm CSM}=10^{15}\,\text{cm}$, $s=2$ and $f_{\rm \Omega}= 1$.  Each figure shows neutrinic spectra obtained from varying only two paramenters at a time. Only in the case of Fig. \ref{fig:Nnu_Rcsm_fOmega}, the mass of the CSM and the neutrino number in ordinate proportionally change with the angular size (see the text for more details). \label{fig:Nnu}}
\end{figure*}
\begin{figure}
	\includegraphics[width=\columnwidth]{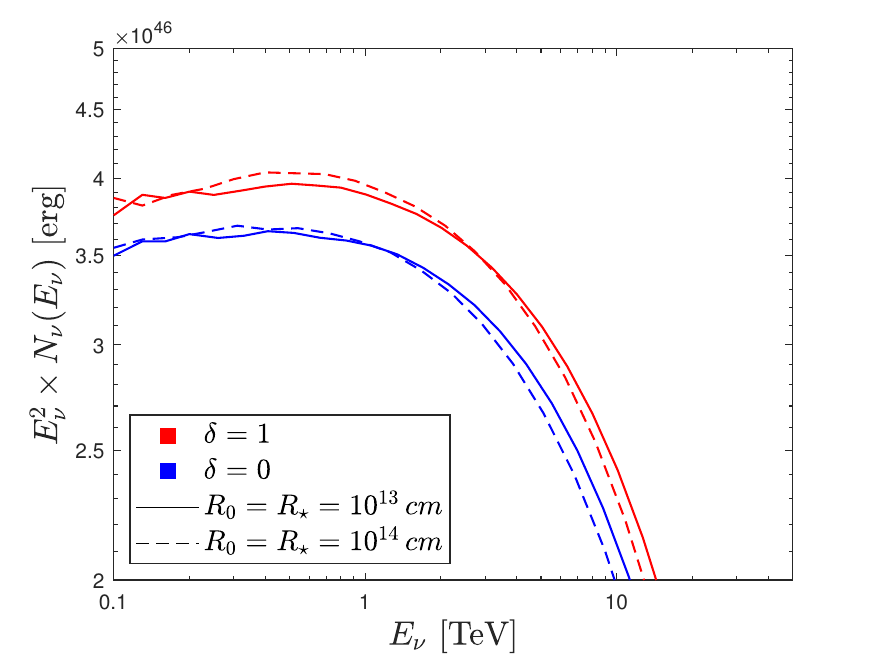}
    \caption{Similar to Fig. \ref{fig:Nnu}, but for different values of $\delta$ and $R_{\rm 0}$.}
    \label{fig:Nnu_delta_R0}
\end{figure}

The H-rich CSM composition having solar-like abundances allows for fixing some model constants, such as the molecular weights, the optical opacity and the cooling parameters (e.g. $T_{\rm *}$). Moreover, in order to analyse the neutrino energy spectrum's behaviour based only on the CSM-ejecta configuration, we specifically selected and set the coefficients governing proton acceleration efficiency \citep[see, e.g.,][and references therein]{Pitik2022}.\par

By integrating in time the any-flavour neutrino rates between the epochs $t_{\rm bo}$ and $t_{\rm f}$, we derive the over-all emitted (anti-)neutrino energy distribution
\begin{equation}
    N_{\rm \nu}(E_{\rm \nu})=\int_{ t_{\rm bo}}^{t_{\rm f}} dt \sum_iQ_{\rm \nu_i+\Bar{\nu}_i}(E_{\rm \nu},t),
\end{equation}
 which, under the previous assumptions, can be used to evaluate the dependency of the neutrinos emission on key ejecta-CSM parameters, such as $E_{\rm k}$, $M_{\rm ej}$, $x_{\rm \star}$, $n$, $M_{\rm CSM}$, $R_{\rm CSM}$, $s$ and $f_{\rm \Omega}$. Indeed, these dependencies are evident when comparing the $N_{\rm \nu}$ energy spectra resulting from different SN configurations, as shown in Fig. \ref{fig:Nnu}.
Among the most pronounced effects, Fig. \ref{fig:Nnu_Ek_Mej} shows that $N_{\rm \nu}$ grows up in both intensity and energy as the explosion energy increases and the ejecta's mass decreases. This is expected since the neutrino emission, sustained by the proton injection rate, is strongly related to the shock speed [$Q_{\rm p}\propto v_{\rm sh}^3$, cf. equations (\ref{Eq:Q_p}) and (\ref{Eq:q_fact})], which in turn is proportional to the ejecta expansion velocity [$v_{\rm sc}\propto \sqrt{E_{\rm k}/M_{\rm ej}}$, cf. equation (\ref{Eq:ejecta_velocity})]. 
Moreover, concerning the $M_{\rm ej}$ dependency, models with the same $E_{\rm k}/M_{\rm ej}$ ratio but different ejecta mass (see the red solid line and the blue dashed line in Fig. \ref{fig:Nnu_Ek_Mej}) may show slight differences attributable to the $M_{\rm ej}/M_{\rm CSM}$ ratio affecting the shock velocity [see equation (\ref{Eq:alpha})].
On the other hand, the CSM mass directly influences the CSM density, thereby modifying the number of protons that can be accelerated by the shock. Although the neutrinos' intensity increases with the CSM's mass (see red and blue curves in Fig. \ref{fig:Nnu_Mcsm_s}), the maximum achievable proton energy [also corresponding to the maximum energy for HE-$\nu$, cf. equations (\ref{Eq:Q_mu}) and (\ref{Eq:Q_e})]
\begin{equation}
E_{\rm p}^{M}= \underset{t\in[t_{\rm bo},t_{\rm f}]}{MAX}\left[E_{\rm p}^{max}(t)\right] =\gamma^{MAX}\, m_{\rm p}c^2
\end{equation}
decreases due to the pp-collisional processes, which limit proton acceleration, especially in denser CSM environments. This behaviour also depends on the CSM density profile linked to $s$-slope, which meaningfully modifies the position of the neutrino energy spectra's "knee" ($E_{\rm \nu}^{k}$), here defined as the neutrino energy value in which the spectral slope is one unit less than that of the proton injection profile $Q_{\rm p}$ equal to $-2$ [see equation (\ref{Eq:Q_p})], that is:
\begin{equation}
\left.\frac{d \log{N_{\rm \nu}}}{d\log{E_{\rm \nu}}}\right|_{\rm E_{\rm \nu}^{k}}=-3.
\end{equation}
\par
\begin{figure*}
        \begin{center}
            \subcaptionbox{Effects of the initial shock velocity $v_{\rm sh,0}$ on the HE-$\nu$ spectra features; \label{fig:Nnu_grids_vsh}}{\includegraphics[width=5.8in]{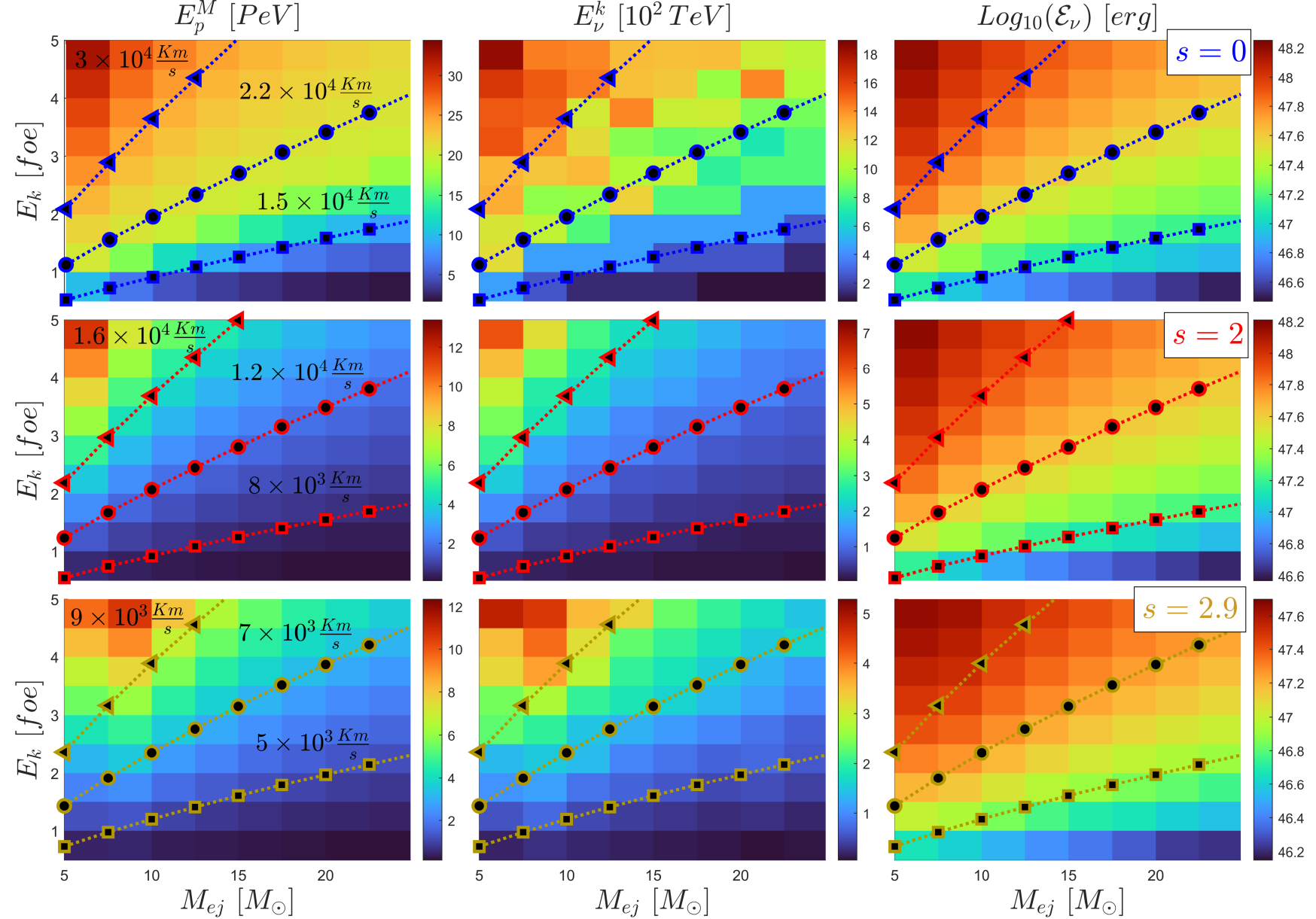}}\par
                  \vspace{0.2in}
			\subcaptionbox{Effects of the internal CSM density $\rho_{\rm s,0}$ on the HE-$\nu$ spectra features; \label{fig:Nnu_grids_rhos0}}{\includegraphics[width=6.in]{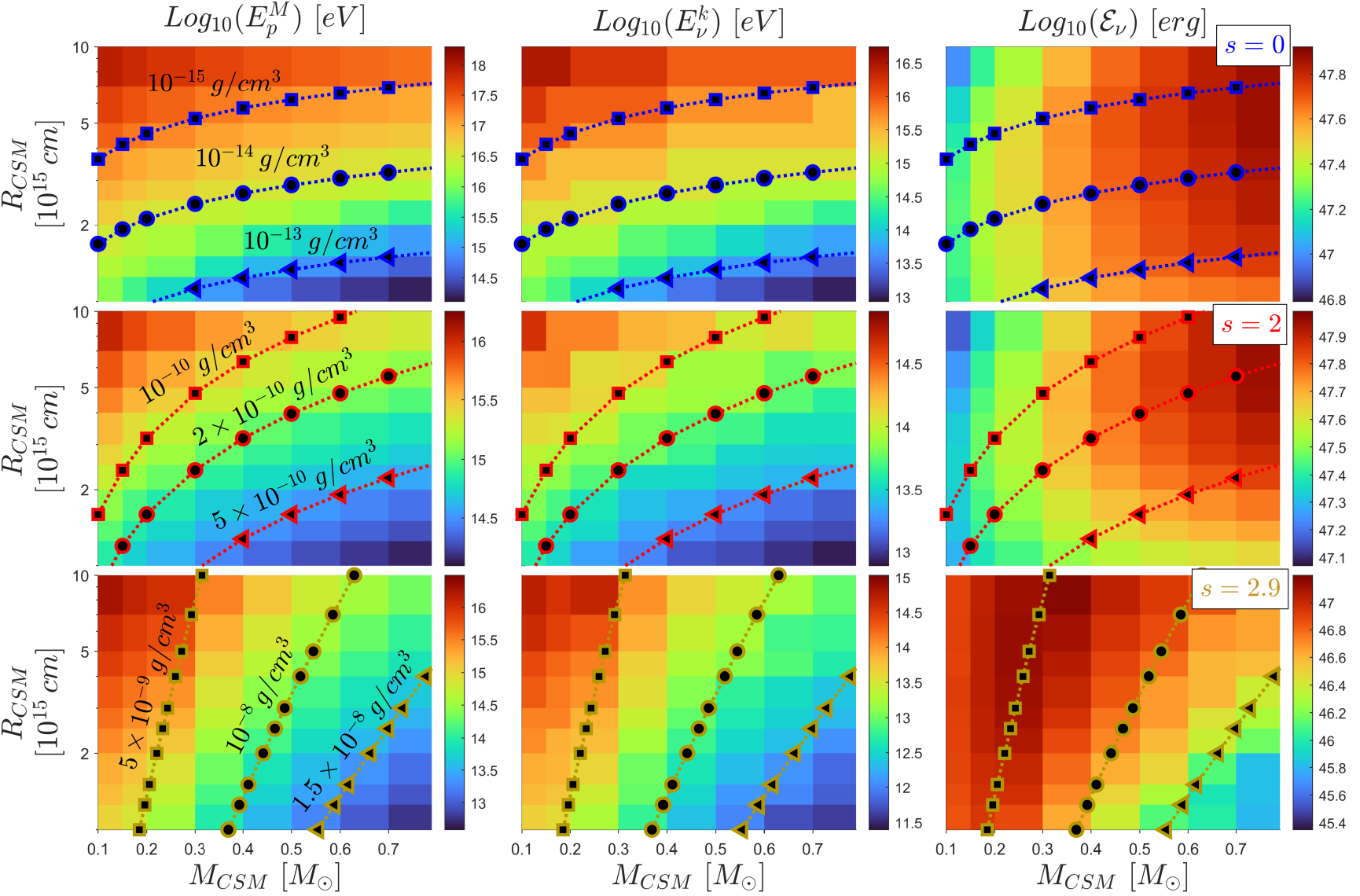}}\par 
            \caption{Color maps illustrating the primary features of neutrino spectra ($E_{\rm p}^{M},E_{\rm \nu}^{k},\mathcal{E}_{\rm \nu}$) across various configurations of SN ejecta-CSM modelling parameters are presented. In Fig. \ref{fig:Nnu_grids_vsh}, dashed lines denote configurations sharing the same initial shock velocity, i.e. $v_{\rm sh,0}$, with each row specifying the velocity value in the first column, while the white boxes in the third column represent the $s$ parameter used for the entire row. Similarly, Fig. \ref{fig:Nnu_grids_rhos0} features dashed lines representing iso-density curves of the CSM in terms of $\rho_{\rm s,0}$. Unaltered modelling parameters remain consistent with those referenced in Fig. \ref{fig:Nnu}.\label{fig:Nnu_grids}}        
        \end{center}
\end{figure*}

Even though the impact of the $s$ exponent on neutrino emission intensity is generally less pronounced compared to that of $M_{\rm CSM}$ (approximately three times less, as shown in Fig. \ref{fig:Nnu_Mcsm_s}), its variation can modify the total energy emitted by neutrinos with $E_{\rm \nu}\ge 0.1\,\text{TeV}$:
\begin{equation}\label{Eq: energy_nu}
    \mathcal{E}_{\rm \nu}= \int_{\rm 0.1TeV}^{E_{\rm p}^{M}} E_{\rm \nu}^2\times N_{\rm \nu}(E_{\rm \nu})\, d(\ln{E_{\rm \nu}}).
\end{equation}
This effect is particularly notable when transitioning from the cases of uniform ($s=0$) or steady wind ($s=2$) CSMs, for which $\mathcal{E}_{\rm \nu}$ is substantially unaffected, to internally denser profiles like the accelerated wind ($s=2.9$, cf. Fig. \ref{fig:density}), in which the total emitted energy is approximately 2-3 times lower (for further details see below and Fig. \ref{fig:Nnu_grids_rhos0}).
Since the number of emitted neutrinos depends on the amount of protons that are hit by the shock after the break-out epoch and, in particular, once the shock is expected to be not radiation-mediated\footnote{ 
The proton acceleration can be suppressed by radiative shock mediation in dense CSM environments. Although our model neglects radiative losses as a direct proton energy loss channel (c.f. footnote \ref{Note:cooling}), this effect is partially included by imposing the physical condition $t_{\rm acc} < t_{\rm p,cool}$, which limits the maximum proton energy considering both hadronic and radiative losses [see equations (\ref{Eq:t_pcool}-\ref{Eq:Lambda})]. As such, our estimates for neutrino production should be regarded as an upper limits in highly radiative regimes.} and collisionless, the reduction of total emitted energy through neutrinos can be explained by the differences in the distribution of CSM density (see the scenarios with the same CSM and ejecta masses in Fig. \ref{fig:Nnu_Mcsm_s}).
Indeed, $t_{\rm bo}$ is generally later for CSMs with a denser inner region [cf. equations (\ref{Eq:tau}) and (\ref{Eq: t_bo})], as seen in accelerated wind cases ($s>2$), where the subshock is weakened by radiative acceleration \citep[e.g.,][]{2012IAUS..279..274K,Murase_2019,Tsuna_2023}. Consequently, in these configurations, the number of protons above $R_{\rm bo}$ swept by the shock shell is lower than that of density profiles with $s\le2$. As a result, $N_{\rm \nu}$ decreases, leading to a reduction in $\mathcal{E}_{\rm \nu}$  as shown in Fig. \ref{fig:Nnu_grids_rhos0}. Additionally, this explains why stellar envelope break-out events are not associated with significant neutrino emission \citep{Murase_2019}.
 Building on this, the steep {CSM} density profile of SN 2023ixf plays a crucial role in shaping the HE-$\nu$ emission during the early phases immediately following the break-out (see Sections \ref{subsec:SN2023ixf}-\ref{Subsec:real} for more details).\par

Further significant effects contributing to the variation in cumulative neutrino spectrum emission arise from the external density profile of the ejecta and the both radial and angular extent of the CSM (refer to Figs. \ref{fig:Nnu_xs_n}-\ref{fig:Nnu_Rcsm_fOmega}). Specifically, an ejecta with a relatively denser outer layer (i.e. model with $x_{\rm \star}=1.01$ and $n=8$) yields a neutrino energy emission approximately three times greater than that of an ejecta with a less dense and extended outer shell (i.e. $x_{\rm \star}=1.2$ and $n=12$). However, models with different values of $x_{\rm \star}$ and $n$ can produce very similar emissions (see model having $x_{\rm \star}=1.2$ and $n=10$ with that having $x_{\rm \star}=1.01$ and $n=12$ in Fig. \ref{fig:Nnu_xs_n}), underscoring a degeneracy issue in the parameters describing the external structure of the ejecta.
Concerning of the radial extension of the CSM, much like its mass, $R_{\rm CSM}$ alters the density of the CSM, thereby impacting the neutrino spectrum in terms of both intensity and energy, while keeping $\mathcal{E}_{\rm \nu}$ approximately constant (see Fig. \ref{fig:Nnu_Rcsm_fOmega}). In particular, comparing models with CSM radii of $10^{15}\,\text{cm}$ and $10^{16}\,\text{cm}$ (represented by the blue and red continuous curves in Fig. \ref{fig:Nnu_Rcsm_fOmega}, respectively), a decrease in maximum intensity by a factor of about 3 is observed in the latter, while $E^k_\nu$ increases tenfold.\par
On another hand, the CSM angular geometry determined by $f_{\rm \Omega}$ (spherical when $f_{\rm \Omega}=1$, or discoidal if it is less), can affect the total emitted energy due to the protons' loss effects [see equation (\ref{Eq:t_loss})].
However, when we change the value of $f_\Omega$ keeping constant $M_{CSM}$, the first effect is that to increase the CSM density in an inversely proportional way [cf. equation (\ref{Eq:CSM_density})], leading to a $N_{\rm \nu}$ behaviour similar to that observed in Fig. \ref{fig:Nnu_Mcsm_s} with the CSM mass increasing. In this way, the effect of CSM density variations due to the volume reduction covers up the neutrino's number decreasing due to transverse proton leaking effect. Therefore in Fig. \ref{fig:Nnu_Rcsm_fOmega}, we choose to represent models with different $f_{\rm \Omega}$ but having the same $\rho_{\rm s,0}$ by proportionally changing the CSM mass ($\rho_{\rm CSM}\sim$ const. $\rightarrow\, M_{\rm CSM}\propto f_{\rm \Omega}$). Moreover, having reduced the CSM mass, the number of emitted neutrinos also proportionally decreases, so instead of studying the number of emitted neutrinos, we considered its value multiplied by the mass correction, i.e. $f^{-1}_{ \rm \Omega}$ (see Fig. \ref{fig:Nnu_Rcsm_fOmega}).
As expected, the geometry effect is less remarkable than the previous ones, and it can be observed only for great CSM radii and without modifying its density.\par 
Even less pronounced effects on the neutrino emission are that due to the variation of $R_{\rm 0}$, $R_{\rm \star}$ and $\delta$ (see Fig. \ref{fig:Nnu_delta_R0}).
In our case, indeed, the difference between $R_{\rm \star}$ and $R_{\rm 0}$ does not exert any discernible effect on the time-integrated neutrino energy distribution\footnote{In the case where $R_0 \gg R_{\star}$, at the time of collision $t_0$, the ejecta may have expanded sufficiently to reduce its density to be roughly equal or less to that of the CSM, i.e. $\rho_{\rm ej}\lesssim \rho_{\rm s,0}$. Consequently, the ejecta may begin to decelerate \citep[see, e.g.,][]{Pitik2022}, causing the shock velocity to deviate from the behaviour described by the equation (\ref{Eq:shock_velocity}). However, this scenario typically applies to SNe exhibiting interaction effects on timescales greater than ten days, which are not considered in this paper because beyond our scope.}; its sole impact lies in delaying the onset of interaction relative to the time of SN explosion. For this reason, throughout this paper, unless explicitly stated otherwise, $R_{\rm 0}$ is equated with $R_{\rm \star}$.
Furthermore, variations in $R_{\rm 0}$ and $\delta$ result in intensity modifications of the neutrino spectra by less than $5\%$, significantly lower than those induced by parameters such as $x_{\rm \star}$, $n$, and $f_{\rm \Omega}$, which exhibit variations around $30\%$.\par

In general Figs \ref{fig:Nnu}-\ref{fig:Nnu_delta_R0} reveal that the neutrino spectrum displays the highest sensitivity (with percentage variations in energy or intensity exceeding $50\%$) to alterations in just five modelling parameters: two pertaining to the SN ejecta ($E_{\rm k}$, $M_{\rm ej}$) and three ones related to the CSM configuration ($R_{\rm CSM}$, $M_{\rm CSM}$, $s$). 
To study the effects of these parameters on the spectra features (i.e. $E_{\rm p}^{M}$, $E_{\rm \nu}^{k}$, and $\mathcal{E}_{\rm \nu}$) within the value ranges outlined in Tab. \ref{tab:parameters},  we present a comprehensive analysis of models divided into two groups of grids in Fig. \ref{fig:Nnu_grids}.

By comparing the grids of Fig. \ref{fig:Nnu_grids_vsh}, we note that both the maximum energy of protons and the intensity of the energy released by HE-$\nu$ emission rise with the speed of the shock, i.e. $v_{\rm sh,0}\propto (E_{\rm k}/M_{\rm ej})^{1/2}$. This trend holds true for all three types of CSM configurations analysed here ($s=0,\,2,\,2.9$), albeit with the shock velocity and spectral characteristic values diminishing as the density slope parameter $s$ increases. As noted above, the density distribution of the CSM affects the energy and intensity of HE-$\nu$ emission. Specifically, as depicted in the first two columns of Fig. \ref{fig:Nnu_grids_rhos0}, it is observed that decreasing internal CSM density, i.e. $\rho_{s,0}$, results in a logarithmic increase in both $E_{\rm \nu}^k$ and $E_{\rm p}^M$. This general behaviour is independent of the type of CSM distribution, indeed, as lower CSM density results in a longer cooling time, enabling the proton acceleration process to be extended.
In terms of the neutrino energy emission, as shown in the right column of Fig. \ref{fig:Nnu_grids_rhos0}, $\mathcal{E}_{\rm \nu}$ shows a transitioning behaviour between the several CSM density configurations identified by the slope parameter $s$. In particular, for both uniform shell and steady wind cases ($s=0,2$), $\mathcal{E}_{\rm \nu}$ generally increases with the CSM mass, whereas in the accelerated wind case it is primarily related to the density $\rho_{sh,0}$. This transition seems to occur when the CSM density comes about sufficiently low ($\lesssim 10^{-9}\,$g cm$^{-3}$, cf. Fig. \ref{fig:Nnu_grids}).
In these cases, the shock has quickly surpassed $R_{\rm bo}$ and, since the amount of energy emitted by neutrinos depends on the number of protons swept by the shock after $t_{\rm bo}$, almost all the protons that constitute the CSM mass contribute to neutrino emission. Differently, for configurations with $s>2$, the protons accelerated above $R_{\rm bo}$ tend to be much fewer compared to the total number of protons in the CSM. Therefore, the number of effectively accelerated protons in this case will depend strictly on the position of $R_{\rm bo}$, which in turn depends on $\rho_{\rm s,0}$, hence the dependency of $\mathcal{E}_{\nu}$ in the case of $s=2.9$ (see Fig. \ref{fig:Nnu_grids_rhos0}). 
Notably, this CSM distribution lowers the minimum $M_{\rm CSM}$ required for an internal shock to form to just $0.3\,$M$_\odot$ [cf. equation (\ref{Eq:trans_condition})]. This threshold synonyms with the iso-density line at $5\times 10^{-9}\,$g cm$^{-3}$, where the behaviour of $\mathcal{E}_{\rm \nu}$ changes (see the third row of Fig. \ref{fig:Nnu_grids_rhos0}).
 Moreover, higher internal densities delay the onset of the collisional phase of the shock, thereby inhibiting the formation of magnetic instabilities that underlie the proton acceleration mechanism and the consequent emission of  HE-$\nu$ \citep[e.g.][and references therein]{Petropoulou17}.
\par
Summarising the results, we can therefore conclude that the neutrinic spectra features mainly depend on only four model characteristics: $v_{\rm sh,0}$, $s$, $\rho_{\rm s,0}$, $M_{\rm CSM}$ (or $R_{\rm CSM}$). These parameters may vary among different SN occurrences. Hence, a comprehensive assessment of neutrino emission from the SN event necessitates a thorough characterization of its physical properties, a task currently achievable solely through electromagnetic data and their modelling.

\section{Information from electromagnetic emission} \label{sec:EM_emi}
The post-explosive electromagnetic emission from a SN ``contains'' a lot of information about the SN physical parameters. This information pertains not only to the configuration of the progenitor system at the time of the explosion \citep[see, e.g.,][]{1993ApJ...414..712P,1996snih.book.....A}, but primarily to the nature of the heating mechanisms that can enhance the SN luminosity \citep[see, e.g.,][]{PZ2011,Khatami_2019,Singh2019ApJ...882...68S,KK_2023}.\par

One of the most important heating mechanisms in H-rich SNe involves the radioactive decay of $^{56}$Ni and $^{56}$Co nuclei, which are synthesized during the explosion. This contribution is usually evident in the latter post explosive phases, resulting in an increase of the SN luminosity and extending the hydrogen recombination stage \citep[see, e.g.,][hereafter referred to as \citetalias{PC2025}]{2013MNRAS.434.3445P,PC2025}. On the other hand, the electromagnetic features linked to the interaction between ejecta and CSM are far less common and difficult to observe, especially when the CSM is low massive. In fact, the presence of CSM is typically detectable through optical emissions only during the initial post-explosive stages, when the shock traverses the densest CSM regions \citep[see, e.g.,][]{Moriya_2014}. Consequently, if the SN discovery occurs too later than the explosion, there is a risk of not promptly observing the signs of the ejecta's interaction with the CSM.\par
However, when such observations are available, analysing the SN's LC can provide valuable insights into the structure of the CSM and the dynamics of shock propagation within it. To find these physical information governing both the HE-$\nu$ emission and the LC shape, we have developed a semi-analytic model that remains consistent with all assumptions made by the HE-$\nu$ emission model (see Section \ref{sec:HEnuSNe}). In the next Section \ref{subsec:LC_peak}, we thus introduce our main equations used to describe the LC peak of interacting SNe, while Section \ref{subsec:SN2023ixf} apply the latter in the specific case of SN 2023ixf.

\subsection{Model of Light Curve peak for interacting SNe}\label{subsec:LC_peak}
The LCs of interacting SNe are often characterized by an initial peak of brightness whose intensity and rise time are closely linked to the CSM's configuration and the explosion energy \citep[see, e.g.,][and references therein]{Ofek_2014,KK_2023}.
As seen in Section \ref{Subsec:p_acc}, indeed, the shock interaction leads to the conversion of the ejecta's kinetic energy into thermal one, according to the following relation:
\begin{equation}
\mathcal{E}_{\rm th}\equiv \int_{\rm V} U_{\rm th}\,dV\equiv\frac{9}{2}\pi f_{\rm \Omega}\times \int_{\rm R_{\rm sh}-\Delta R_{\rm sh}}^{R_{\rm sh}}\rho_{\rm CSM}\,v_{\rm sh}^2\,r^2\,dr,
\end{equation}
becoming then an additional heating source. In this way, the shocked shell transports the energy through the CSM, whereas a fraction\footnote{According to the proton acceleration mechanism, the fraction of radiative energy cannot exceed the residual thermal energy ($1-\epsilon_{\rm p}-\epsilon_{\rm B}$). Furthermore, the thermal radiation from the shock is evenly distributed between inward and outward emission. Therefore, throughout this study, the radiative efficiency of the shock is set at $(1-\epsilon_{\rm p}-\epsilon_{\rm B})/2\simeq 0.44$ (see Tab. \ref{tab:parameters}), which aligns with the findings reported for type IIn SNe \cite[see, e.g.,][]{Fransson_2014}.} $\epsilon_{\rm rad}$ of its variation is radiated outwards \citep{Moriya2013}. Then, the radiative emission of the shock can be described by
\begin{equation}\label{Eq:S_sh}
S_{\rm sh}\equiv\epsilon_{\rm rad}\times \left| \frac{d\mathcal{E}_{\rm th}}{dt}\right|\simeq \frac{9}{2}\pi f_{\rm \Omega}\epsilon_{\rm rad}\,\left.\rho_{\rm CSM}\right|_{\rm R_{\rm sh}(t)}\, v_{\rm sh}^3(t) \, R_{\rm sh}^2(t),
\end{equation}
strictly valid for $t_{\rm dyn}<<t_{\rm cool}$.\par
Before that $S_{\rm sh}$ directly contributes to the SN luminosity, this radiation must pass through the CSM layers above the shock-front up to achieve the photospheric radius ($R_{\rm ph}$), from whom it can freely escape. Therefore, the electromagnetic emission even depends on the radiation diffusion timescale which, inside a full-ionised H-rich CSM and until for $R_{\rm sh}\le R_{\rm ph}$, can be expressed by
\begin{align}\label{Eq:t_diff}
t_{\rm d}\equiv\,& \frac{k_{\rm T}}{c}\times\int_{\rm R_{\rm sh}}^{R_{\rm ph}} \rho_{\rm CSM}(r)\,d\left[(r-R_{\rm sh})^2\right]\nonumber\\
\simeq\,& 2\,k_{\rm T}\,\rho_{\rm s,0}\times\int_{\rm R_{\rm sh}(t)}^{R_{\rm ph}}\frac{\left[r-R_{\rm sh}(t)\right]}{c\,(r/R_{\rm 0})^{s}}\,dr.
\end{align}
The photosphere is typically located where the optical depth is $\tau_{\rm ph}=2/3$, for Eddington's approximation \citep{Ginzburg_2012}. By using equation (\ref{Eq:CSM_density}), hence $R_{\rm ph}$ is assumed to be located at
\begin{equation}\label{Eq:ph_radius}
R_{\rm ph}=R_{\rm CSM}\times
    \begin{cases}
    \exp{\left(-\varphi\right)}\,& \text{with }s=1\\
    \\
 \sqrt[1-s]{1-\varphi(1-s)}\,& \text{with }s\not= 1,
\end{cases}
\end{equation}
being $\varphi$ a dimensionless parameter depending on the main CSM's external boundary features:
\begin{equation}
\varphi= \frac{\tau_{\rm ph}\,k_{\rm T}^{-1}\,R_{\rm CSM}^{-1}}{\rho_{\rm CSM}(R_{\rm CSM})}.
\end{equation}
Finally thus, the outgoing luminosity due to the CSM-ejecta interaction let be modelled by following equation:
\begin{equation}\label{Eq:Lum_sh}
L_{\rm sh}(t)=\int_{\rm t_{\rm bo}}^{t}\frac{S_{\rm sh}(t')}{t_{\rm d}(t')}\times e^{-(t-t')/t_{\rm d}(t')}dt',
\end{equation}
where the shock radiative emission is mediated for the escape probability of photons\footnote{The total amount of energy released by the shock from the time $t_{\rm bo}$ up to $t$, i.e. when it escapes from the photosphere, can be expressed as
$\mathcal{E}_{\rm sh}(t)= \int_{\rm t_{\rm bo}}^t S_{\rm sh}(t')\,\{1-\exp{[-(t-t')/t_{\rm d}(t')}]\}\,dt'$, where the term in curly brackets represents the escaping probability of photons, depending on the diffusion time at $t'<t$. So{, since} $L_{\rm sh}\equiv d\mathcal{E}_{\rm sh}/dt$, one obtains the equation (\ref{Eq:Lum_sh}).}. Using the latter equation, we can examine the luminosity of interacting SNe from the break-out phase $t_{\rm bo}$ until the shock shell emerges from the photosphere, i.e. $R_{\rm sh}= R_{\rm ph}$. Subsequently, the continuous interaction phase starts, and the high photon escape efficiency leads to $L_{\rm sh}$ equating $S_{\rm sh}$, while the scattering-diffusion time rapidly decreases towards zero.\par
\begin{figure}
	\includegraphics[width=\columnwidth]{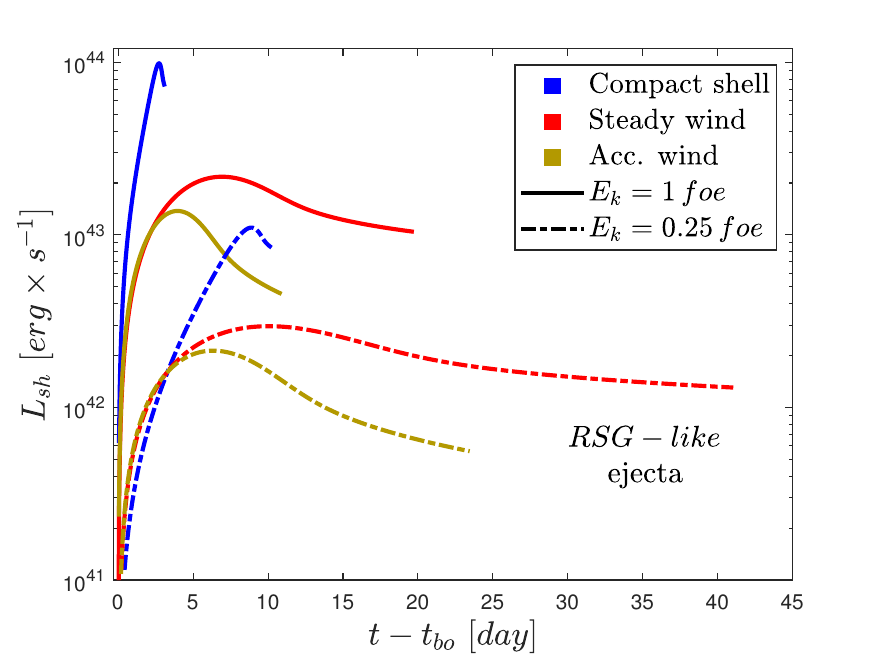}
    \caption{Different types of shock luminosity evolution relative to the break-out epoch are shown. The displayed LCs have been generated using consistent modelling parameters of Fig. \ref{fig:density} for a RSG-like ejecta. Variations in the types of CSM configuration and $E_{\rm k}$ have been applied as detailed in the accompanying box. Here, all LCs have been interrupted to their $t_{\rm f}$.}
\label{fig:Lum_densConf_E}
\end{figure}
\begin{figure}
	\includegraphics[width=\columnwidth]{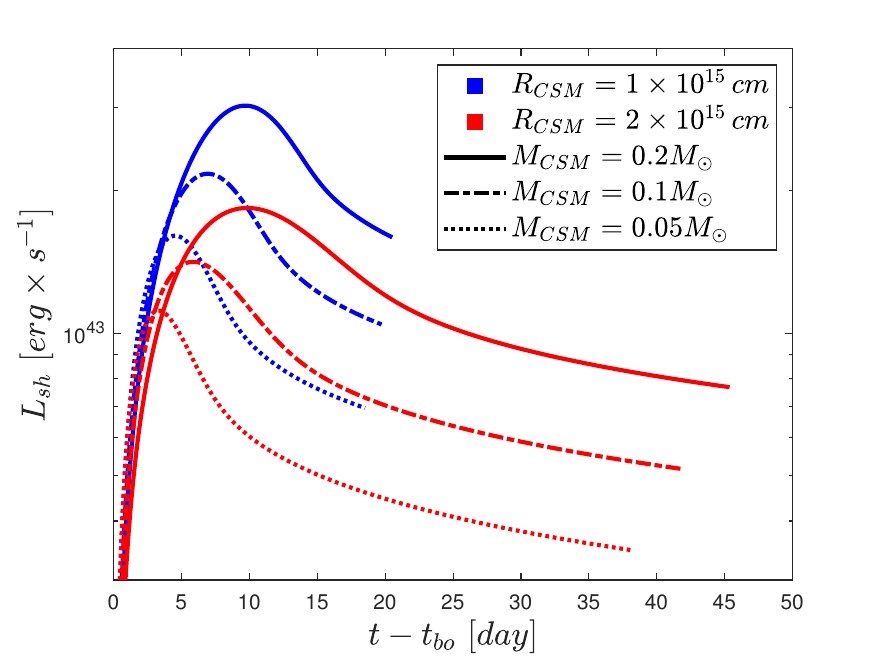}
    \caption{Similarly to Fig. \ref{fig:Lum_densConf_E}, but with variations in $R_{\rm CSM}$ and $M_{\rm CSM}$. Unaltered modelling parameters remain consistent with those used in Fig. \ref{fig:Nnu}.}
 \label{fig:Lum_Rcsm_Mcsm}
\end{figure}
Following the radiative emission model for CSM-ejecta interaction outlined here, equations (\ref{Eq:S_sh}) and (\ref{Eq:t_diff}) show a direct functional dependence of shock luminosity on the quantities $\rho_{\rm s,0}$ and $v_{\rm sh,0}$. Therefore, through the LC analysis, the latter factors can be inferred to provide important constraints on the HE-$\nu$ emission features (see Section \ref{Subsec:par_mod}).
In particular, it is observed that the CSM configuration, set by $s$ and $\rho_{\rm s,0}$ as exemplified in Fig. \ref{fig:density}, holds significant influence over the electromagnetic emission behaviour, particularly discernible between wind-like CSM and uniform shell scenarios (cf. red and blue LC profiles in Fig. \ref{fig:Lum_densConf_E}).
The compact shell configuration, indeed, yields a higher peak luminosity compared to the other wind-like models, generally having a higher average density. Moreover, the differences between these wind scenarios primarily emerge during the declining phase rather than the rising one (cf. red and yellow LCs in Fig. \ref{fig:Lum_densConf_E}).
With regard to the dependence on $v_{\rm sh,0}$, it can be observed in Fig. \ref{fig:Lum_densConf_E} that, the increasing in $E_{\rm k}$ yields a proportional growth in the intensity of the peak, in accordance with the proportion $S_{\rm sh}\propto v_{\rm sh}^3\propto (E_{\rm k}/M_{\rm ej})^{3/2}$. On the other hand, if $v_{\rm sh,0}$ is greater, the evolution of the shock front is faster, resulting in a shorter collisional phase duration, i.e. $t_{\rm f}-t_{\rm bo}$. At low energy, however, the rise time and peak broadening depend mainly on $\rho_{\rm s,0}$.
Indeed, examining the dependence of $\rho_{\rm s,0}$ through its main components $M_{\rm CSM}$ and $R_{\rm CSM}$, it can be seen that the difference in density can change  both the rise time and the peak luminosity. For instance, as depicted in Fig. \ref{fig:Lum_Rcsm_Mcsm} for the steady wind configuration (i.e. $s=2$), both the intensity and the rising time of the luminosity increase with higher CSM density.
It is also noted that differing combinations of mass and radius may result in initial LC phases of similar characteristics, particularly concerning intensity and rising timescales. To try to remove this degeneracy between $M_{\rm CSM}$ and $R_{\rm CSM}$, it is needed a comprehensive analysis of the entire LC, discerning so the overall duration of the interaction and deriving the radial extent of the CSM.\par
Based on these comparisons, we can conclude that to derive all the pertinent parameters for simulating HE-$\nu$ emission, relying solely on the rise time and peak brightness is insufficient. A comprehensive analysis of the SN LC, extending beyond the interaction phase, is necessary. Let us now explore how this can be achieved using the case of SN 2023ixf as an example.

\subsection{Case of SN 2023ixf: Light Curve modelling}\label{subsec:SN2023ixf}
\begin{figure*}
\centering
	\includegraphics[scale=0.65]{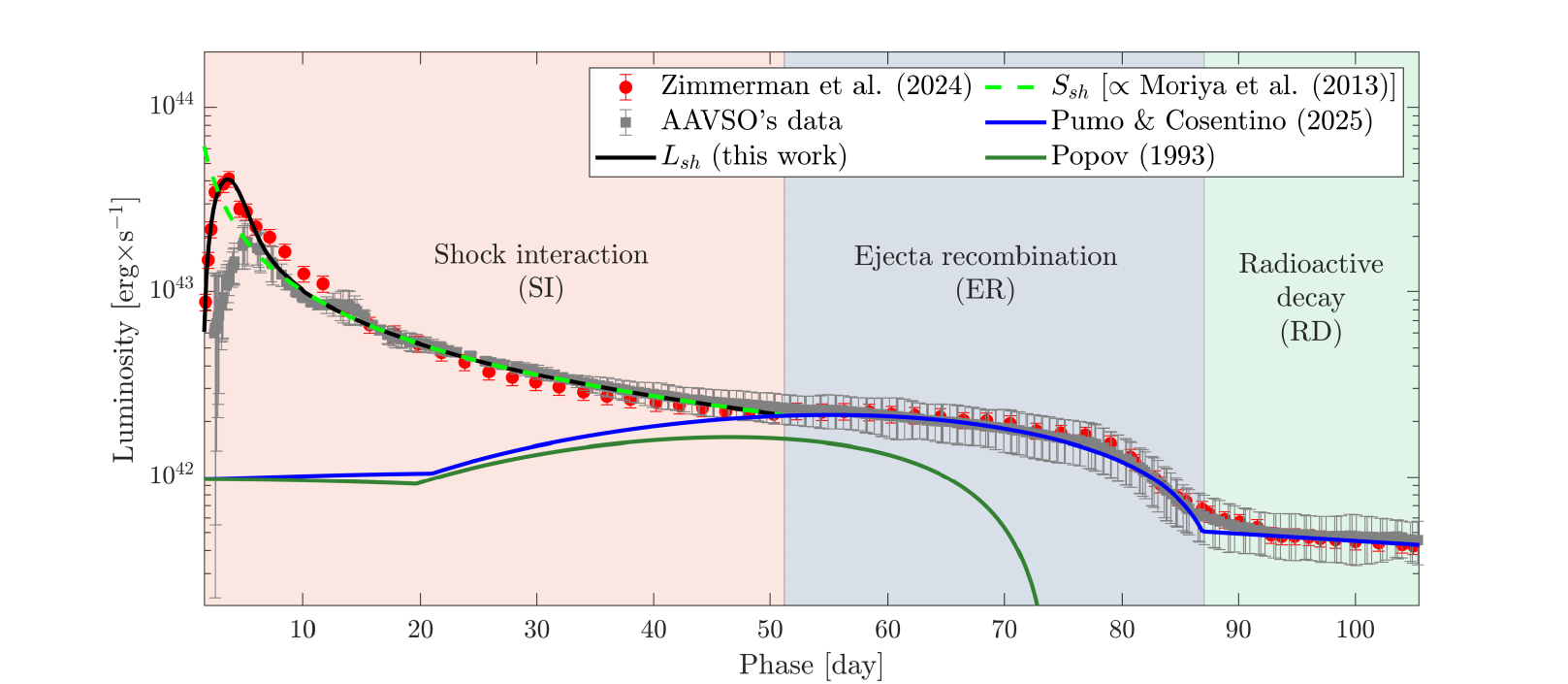}
    \caption{Bolometric luminosity of SN 2023ixf as a function of time since explosion in days (phase). The data in red and grey are two bolometric LCs of this SN respectively obtained from {\citet{zimmerman_complex_2024}} and the data of American Association of Variable Star Observers (AAVSO). For the latter, the bolometric LC has been reconstructed starting by the AAVSO's observations in the BVRI-bands through the procedure of  {\citet{Nicholl_2018}} and considering the distant modulus and the total reddening used by {\citet{Hiramatsu_2023}}, respectively equal to $\mu_{\rm SN}=29.19\,\text{mag}$ and $E(B-V)=0.04\,\text{mag}$ . The plotted curves represent the luminosity for the main LC models discussed in the text and by assuming the modelling parameters reported in Tab. \ref{tab:parameters_23ixf}. The vertical lines split the post-explosive phases based on the type of dominant emission process: shock-interaction (SI), ejecta recombination (ER), and  radioactive decay (RD).      \label{fig:LC_23ixf}}
\end{figure*}

{Our} modelling procedure divides the post-explosive LC of SN 2023ixf into three sub-phases, each represented by a dominant emission mechanism (see Fig. \ref{fig:LC_23ixf}).\par
The first phase is characterized by the shock-interaction (SI), whose beginning corresponds to the shock break-out and ends when the shock achieve the CSM external boundary at $t_{\rm f}$. During this phase we observe the characteristic peak for the interacting SNe more pronounced in \citet{zimmerman_complex_2024}'s data than those reconstructed by AAVSO's observations.
The discrepancy is perfectly justified by the fact that Zimmerman's data also takes into account ultraviolet (UV) bands, whose contribution is particularly important in the first 5-10 days after the explosion \citep[{e.g.,} ][]{Jacobson-Galán_2023}.\par 
As being seen in the Section \ref{subsec:LC_peak}, the equation (\ref{Eq:Lum_sh}) can well describe the Zimmerman's bolometric LC peak, allowing to derive the SN modelling properties of our interest, such as $E_{\rm k}$, $s$, $M_{\rm CSM}$ and $R_{\rm CSM}$. However, by analysing only the LC peak, the latter parameters can not be unambiguously determined, therefore our LC study has been extended even to the later phases to constrain the other ejecta parameters. Subsequently, indeed, the SN luminosity is mainly affected by the ejected-material recombination (ER) rich of the hydrogen, and the LC shape looks like a plateau. The luminosity and the extension of this plateau phase are ruled by the evolution of the wave-front of cooling and recombination, which moves inside the ejecta with a constant temperature of $T_{\rm ion}\simeq 5000\,$K, and whose dynamics is tied to the main ejecta parameters, such as $E_{\rm k}$, $M_{\rm ej}$ and $R_{\rm \star}$ \citep[see also][]{1993ApJ...414..712P}.  In addition to the latter ones, however, the cooling and recombination process may be slowed down by the energy released by the decay of radioactive elements such as $^{56}$Ni \citep[see, e.g., \citetalias{PC2025};][]{2013MNRAS.434.3445P,2011A&A...532A.100U,2009ApJ...703.2205K}, introducing so the $^{56}$Ni ejected mass $M_{\rm Ni}$ between the LC modelling parameters (Tab. \ref{tab:parameters}). Differently from the others, the $M_{\rm Ni}$ parameter can be estimated in a direct way by analysing the exponential decrease of the SN luminosity during the last stage, predominantly driven by its radioactive decay (RD) chain \citep[see, e.g.,][and references therein]{2023MNRAS.521.4801P}. Indeed, when the ejecta is entirely recombined, its nebular emission is solely sustained by the decay of $^{56}$Co, itself produced by the RD of $^{56}$Ni and, with an average lifetime of $111$ days, which is long enough to make a significant contribution during the late SN stages.\par
Such we perform the characterization of the SN 2023ixf's physical parameters starting with the determination of $M_{\rm Ni}\simeq 0.073\,$M$_{\rm \odot}$, through the LC data interpolation during the RD phase using equations (22) and (24) of \citetalias{PC2025}. Our $M_{\rm Ni}$ value is in line with \cite{zimmerman_complex_2024}, indicating a higher $^{56}$Ni ejection compared to that found by \cite{Bersten_AA2024} and \cite{Moriya2024}. This places SN 2023ixf among Type II SNe with significantly higher $^{56}$Ni production at the explosion, exceeding the mean value by a factor of two \citep[e.g.][]{Müller_2017,2021MNRAS.505.1742R}.
Once $M_{\rm Ni}$ has been determined, we use the semi-analytical model\footnote{For SNe with high $M_{\rm Ni}$ values, we demonstrated that the "EXP+IE" sub-model is remarkably effective at inferring SN modelling parameters such as $E_{\rm k}$, $M_{\rm ej}$, and $R_{\rm \star}$ without encountering degeneracy issues (see also Section 4 of \citetalias{PC2025}).} ``EXP+IE'' outlined in \citetalias{PC2025} to characterise the other SN parameters, obtaining the following values: $E_{\rm k}\simeq 1.8\,\text{foe}$, $M_{\rm ej}\simeq 9\,\text{M}_{\rm \odot}$ and $R_{\rm \star}\simeq 1.6\times 10^{13}\,\text{cm}$. The obtained energy and mass values align with the most energetic model investigated by \cite{Bersten_AA2024} and also agree within the error bars with the best model of \cite{Moriya2024} having $E_{\rm k}=2\,$foe. However, our progenitor radius is about three times less than that of \cite{Bersten_AA2024}, although it remains compatible within the error bars with the flash-breaking out observations made by \cite{li_shock_2024}.
\begin{table}
	\centering
\caption{Set of modelling parameters for a spherical and initial attached CSM-ejecta system that best reproduce the entire LC of SN 2023ixf (see Fig. \ref{fig:LC_23ixf} and the text for further details about the LC analysis).}
	\label{tab:parameters_23ixf}
	\begin{tabular}{llll}
		\hline
        \multicolumn{4}{c}{Model parameters of SN 2023ixf from the LC analysis}\\
        \hline
        \vspace{0.1cm}
		$E_{\rm k}$ &$1.8\pm0.2\,\text{foe}$ & $R_{\rm CSM}$ & $(3.6\pm0.5)\times 10^{15}\,\text{cm}$\\
		\vspace{0.1cm}
		$M_{\rm ej}$ & $9\pm0.5\,\text{M}_{\rm \odot}$ & $M_{\rm CSM}$ & $6.5^{+1.5}_{\rm -1}\times 10^{-2}\,\text{M}_{\rm \odot}$\\
		\vspace{0.1cm}		
		$R_{\rm \star}$ & $(1.6\pm0.6)\times 10^{13}\,\text{cm}$ & $x_{\rm \star}$ & $1.01^{+0.04}_{\rm -0.01}$\\
		\vspace{0.1cm}		
$M_{\rm Ni}$ & $7.3^{+1}_{\rm -0.5}\times 10^{-2}\,\text{M}_{\rm \odot}$ &$s$ & $2.90\pm0.03$\\		

	\vspace{0.1cm}		
		$\delta$ & $0$ & $n$ & $8.6\pm0.4$  \\		
		\hline    
	\end{tabular}
\end{table}
By using our ejecta parameters, the outer external boundary $x_{\rm \star}\simeq 1.01$, together with CSM features, are finally determined by modelling LC data in the SI phase with the function seen in equation (\ref{Eq:Lum_sh}).
As far as the outer ejecta-CSM configuration, the post-peak decline of the luminosity in SI phase suggests that the CSM density profile has an accelerated wind distribution (i.e. $s\simeq2.9$).
After 10 days from the explosion, indeed, the bolometric LC decreases according to equation (\ref{Eq:S_sh}), in which $S_{\rm sh}$ depends on $t^{-0.94}$, similarly to the equation (22) of \cite{Moriya2013}.
Moreover, this equation permits us to derive $n\simeq8.6$ as the exponent for the outer density profile of the ejecta. This LC behaviour ends about $51\,$days after the explosion, which can be assumed like the final epoch for the interaction phase $t_{\rm f}$. In this way, the CSM radius can be derived using the equation (\ref{Eq:t_f}), obtaining so $R_{\rm CSM}\simeq 3.6\times 10^{15}\,\text{cm}$. Once all other parameters have been estimated, the CSM mass can be find using the brightness intensity of the initial peak, so we obtain $M_{\rm CSM}\simeq 6.5\times 10^{-2}\,\text{M}_{\rm \odot}$.
Our proposed mass distribution is simpler compared to \cite{zimmerman_complex_2024}'s one. Despite assuming a single-slope CSM density, however, we have successfully reproduced the bolometric LC in the SI phase (Fig. \ref{fig:LC_23ixf}), also verifying that with our proposed accelerated wind profile (i.e. $s=2.9$) the columnar density between phases $4.5\,$day and $11\,$day decreases by about $82\%$, consistent with X-ray observations of \cite{Grefenstette_2023}.
The complete set of derived modelling parameter values with their errors are listed in Tab. \ref{tab:parameters_23ixf}. The error ranges for each free parameter of this modelling procedure have been determined by locally analysing the $\chi^2$ around its minimum identified by the best-fitting. However, refining the analysis for $\delta$ parameter is not feasible due to the general assumptions outlined in \citetalias{PC2025}, which necessitate an uniform density for the ejecta. Nonetheless, characterizing the internal ejecta density falls beyond the scope of this paper, as its effects on neutrino emission can be deemed negligible (cf. Fig. \ref{fig:Nnu_delta_R0}). In addition to the simplified description of an ejecta with uniform internal density, we have introduced assumptions of a spherical and regular CSM distribution in contact with the progenitor's surface (i.e. $R_{\rm 0}=R_{\rm \star}$).
Although some these assumptions contrast with the asymmetrical scenarios proposed by \cite{Vasylyev_2023} and \cite{Soker2023}, their use enable us to simplify both LC modelling and neutrino emission simulation. Nonetheless, this scenario presents parameters that are fully consistent with other hydrodynamical approaches and multiwavelength observations \citep[see, e.g.,][]{Bersten_AA2024,Moriya2024}.\par
After acquiring all the needed physical parameters for simulating HE-$\nu$ emission from SN 2023ixf, it becomes essential to outline some general characteristics of the large-volume neutrino detectors, which enable the observation or determination of limits on such astrophysical neutrinos.

\section{Detection potential for the large-volume neutrino telescopes}\label{sec:detect_pot}
In the last decade projects for large volume neutrino telescopes entered the construction and even data-taking phase. IceCube and KM3NeT have been built aiming at the detection of HE-$\nu$, up to the TeV region and more. While the first is built in the stable environment of the frozen Antarctic region \citep{Aartsen_2017}, the second is an under-sea deployed structure with two separated deployment sites in the Mediterranean sea \citep{2024EPJC...84..885K}. Settled in the opposite hemispheres, their fields of view are almost complementary with an overlap around the zero declination.\par
Their ability to observe neutrino signals from point sources, such as SNe, mostly depends on (anti-)neutrinos arrival energies $E_{\rm \nu}$ and their direction $\Omega_\star$ [$\equiv (\alpha_\star,\delta_\star)$], coincident with the celestial angular coordinates of the source \citep[see, e.g.,][]{Trovato_2017}. Therefore, the measurement of detection efficiency is often described in terms of the Effective Area $A_{\rm eff}(E_{\rm \nu},\Omega_\star)$, which is specific not only to the type of telescope but also to the configuration of its active detectors \citep[see, e.g.,][]{Aartsen_2014,2021arXiv210109836I}.\par
Nowadays the search for HE-$\nu$ originating from young SNe lacks definitive experimental confirmation. In fact, only a handful of events detected by IceCube appear to exhibit statistical coincidence, both in terms of timing and direction of arrival, with electromagnetic transients as SLSNe  \citep[see, e.g.,][]{Pitik2022}. The possibility to detect these types of neutrinos with an arrival energy $E_{\rm \nu}$ at the observer time $t$ from the explosion, can be estimated by the (anti-)neutrino flux at Earth $F_{\rm \nu_\alpha+\bar{\nu}_\alpha}^{(z)}$, whose values in units of $\text{GeV}^{-1}\,\text{s}^{-1}\,\text{cm}^{-2}$ for a generic flavour $\alpha=[e,\mu,\tau]$ are given by the following equation:
\begin{equation}
    F_{\rm \nu_\alpha+\bar{\nu}_\alpha}^{(z)}\equiv\sum_\beta \frac{P_{\rm \nu_\beta\rightarrow\nu_\alpha}}{4\pi D^2}\times    Q_{\rm \nu_\beta+\bar{\nu}_\beta}\left[E_{\rm \nu} (1+z),\frac{t}{1+z}\right],
\end{equation}
where $D$ is the source distance, $z$ is its redshift and, $P_{\rm \nu_\beta\rightarrow\nu_\alpha}$ are the transition probability for the neutrinos flavour changes.
In particular, similarly for the electromagnetic radiation, $F_{\rm \nu_\alpha+\bar{\nu}_\alpha}^{(z)}$ decreases with the square of $D$, which is related in a flat $\Lambda$CDM cosmology to its redshift-$z$ by the following relation:
\begin{equation}
    D(z)=\frac{c}{H_0}\times\int_0^z\frac{dz'}{\sqrt{\Omega_\Lambda+\Omega_M(1+z')^3}},
\end{equation}
where $\Omega_\Lambda=0.685$, $\Omega_M=0.315$ and the Hubble-Lema\i tre constant equals to $H_0=67.4\,\text{km}\,\text{s}^{-1}\,\text{Mpc}^{-1}$ \citep[][]{2020A&A...641A...6P}{}{}. Moreover, during their travel from the SN to Earth, neutrinos undergo flavour changes with the following transitions probabilities:
\begin{align}
    P_{\rm \nu_e\rightarrow\nu_\mu}=\,& P_{\rm \nu_\mu\rightarrow\nu_e}=P_{\rm \nu_e\rightarrow\nu_\tau}=0.25\times\sin^2(2\theta_{\rm 12}),\\
    P_{\rm \nu_\mu\rightarrow\nu_\mu}=\,&P_{\rm \nu_\mu\rightarrow\nu_\tau}=0.5\times\left[1-0.25\times\sin^2(2\theta_{\rm 12})\right],\\
    P_{\rm \nu_e\rightarrow\nu_e}=\,&1-0.5\times\sin^2(2\theta_{\rm 12});
\end{align}
where $\theta_{\rm 12}\simeq 33.5^\circ$ \citep[][]{2020JHEP...09..178E}{}{}, and $P_{\rm \bar{\nu}_\alpha\rightarrow\bar{\nu}_\beta}=P_{\rm \nu_\alpha\rightarrow\nu_\beta}$. Then the initial flavour mix $\nu_e\,:\,\nu_\mu:\,\nu_\tau=1\,:\,2\,:\,0$ changes into the universal flux proportion $1\,:\,1\,:\,1$ \citep[see][]{ANCHORDOQUI20141}{}{}, halving the initial number of muon neutrinos arriving on Earth, which is what we are particularly interested in. Indeed, the muon-induced track events offer improved angular resolution compared to the cascades of other flavours, so our focus lies only on muon (anti-)neutrinos.\par 
The elusivity of the neutrino particles does not allow to directly evaluate their instantaneous flux, so the telescope needs to consider a observation time interval from which it is possible to evaluate the average muon neutrino flow ($\Phi_{\rm \nu_\mu}$) from the source. In the case of SNe, this interval can start from $t_{\rm bo}$ and maybe extended up to $t_{\rm f}$. Along this interval, $\Phi_{\rm \nu_\mu}$ can be express by the following relation:
\begin{equation}
    \Phi_{\rm \nu_\mu}(E_{\rm \nu},t)=(t-t_{\rm bo})^{-1}\times\int_{\rm t_{\rm bo}}^{t}F_{\rm \nu_\mu+\bar{\nu}_\mu}^{(z)}(E_{\rm \nu},t')\,dt',
\end{equation}
hence the average flux extended for the entire collisional phase is defined as $\Phi_{\rm \nu_\mu}^{f}\equiv\Phi_{\rm \nu_\mu}(t_{\rm f})$.\par
It is worth noting that the sensitivity of telescopes for pinpoint sources imposes a lower limit on the observable neutrino flow. This limit is contingent upon various telescope characteristics and can even be influenced by the atmospheric and diffuse astrophysical neutrino background \citep[see, e.g.,][]{Aartsen_2014}.
Among these characteristics, the angular resolution ($\Delta \theta^\circ$) of the detector for the neutrino arrival direction plays a crucial role, as it affects the background flow of muon (anti-)neutrinos:
\begin{equation}	\label{Eq:background}
F^{\rm bk}_{\nu_\mu+\bar{\nu}_\mu}(E_\nu)\simeq \phi_{\nu_{\mu}}^{\rm bk}(E_\nu)\times 2\pi\left[1-\cos{\frac{\Delta \theta^\circ(E_\nu)}{180/\pi}}\right],
\end{equation}
where $\phi_{\nu_{\mu}}^{\rm bk}$ represents the HE-$\nu$ background flux per unit solid angle (sr$^{-1}$), obtained by summing the "conventional" \citep{PhysRevD.75.043006} and "prompt" \citep{PhysRevD.78.043005} atmospheric (atm.) muon neutrino fluxes \citep[e.g.,][]{PhysRevD.83.012001} with the diffuse astrophysical (astro) contribution \citep[e.g.,][]{2014PhRvL.113j1101A,Aartsen_2016}. Moreover, assuming the kinematic limit for the angular resolution \citep[e.g.,][]{Murase16}, we adopt the parametrization $\Delta \theta^\circ(E_\nu)= \theta_1+\theta_2\times(E_\nu/$TeV$)^{-0.5}$, where the angles $\theta_1$ and $\theta_2$ vary for different detectors as listed in Tab. \ref{tab:angular_res}.
\par
Once the muon (anti-)neutrino flow is obtained, it is possible to derive the {expected} detection rate $\Dot{\mathcal{N}}_{\rm \nu_\mu}$  of (anti-)neutrinos by using the following relation, which accounts for the neutrino telescope's effective area $A_{\rm eff}$ {\citep[e.g.,][]{2021arXiv210109836I,Pitik2022}}:
\begin{equation}\label{Eq:rate_N}
    \Dot{\mathcal{N}}_{\rm \nu_\mu}^{{>E_\nu}}(t)=\int_{{E_\nu}}^{\infty}\, A_{\rm eff}(E_{\rm \nu}{'},\Omega_\star)\times F_{\rm \nu_\mu+\bar{\nu}_\mu}(E_{\rm \nu}{'},t)\, dE_{\rm \nu}{'},
\end{equation}
which is integrated over neutrino energy above an energy threshold $E_\nu\ge0.1\,$TeV, due to the poor capability of large-volume telescopes to observe less energetic neutrino events.

In this work, we consider the muon neutrino effective area obtained by Monte Carlo simulations to evaluate the response of detectors such as IceCube86-II \citep[depending on the source's declination $\delta_\star$; see][]{2021arXiv210109836I} and KM3NeT/ARCA21 \citep[see][and references therein]{Muller:2023xO}, as it provides sufficiently accurate results and remains consistent with other studies such as \citet{2023MNRAS.524.3366P}. However, in cases where the background dominates over the signal, a more precise approach would involve calculating the number of through-going muon events using the muon effective area \citep[see, e.g.,][]{Murase18}, which accounts for muon energy loss \citep[see also][for further details]{Murase16}. Although this method would refine the analysis, it goes beyond the scope of this work. 
\begin{table}
{
	\centering
	\caption{{Angular resolution parameters for different large volume neutrino detectors used to constrain the background flow of muon (anti-)neutrinos.} }
	\label{tab:angular_res}
	\begin{tabular}{lccc} 
		\hline
		Detectors & $\theta_1$ & $\theta_2$ & References\\
        \hline
		IceCube86-II & $0.35^\circ$ & $0.7^\circ$ & (a)\\
		IceCube-Gen2 & $0.1^\circ$ & $0.5^\circ$ & (b)\\
		KM3NeT/ARCA21 & $0.25^\circ$ & $0.9^\circ$ & (c)\\
		KM3NeT/ARCA230 & $0.06^\circ$ & $0.75^\circ$ & (d)\\      
        \hline
\multicolumn{4}{l}{(a)~\citet{Aartsen_2014};
(b) \citet{2014arXiv1412.5106I};}\\
 \multicolumn{4}{l}{(c) \citet{Muller:2023xO}; (d) \citet{2024EPJC...84..885K}.}	
	\end{tabular}
	}
	
\end{table}
\par
To assess the statistical significance of the neutrino signal from the interacting SN source under investigation, we rely on two statistical parameters. The first is the ratio $\mathcal{N}_{\rm \nu_\mu}^{\rm  SN}/(\mathcal{N}_{\rm \nu_\mu}^{\rm  SN}+\mathcal{N}_{\rm \nu_\mu}^{\rm bk})$, which provides an estimate of the expected signalness and offers an indication of the probability that a detected neutrino event originates from an astrophysical source \citep[e.g.,][]{2023MNRAS.524.3366P}. The second is the test statistic $\mathcal{N}_{\rm \nu_\mu}^{\rm  SN}/\sqrt{\mathcal{N}_{\rm \nu_\mu}^{\rm bk}}$, which, in background-dominated data \citep{1998PhRvD..57.3873F}, is proportional to the reciprocal of the Model Rejection Factor \citep[MRF $\sim\lambda_{90}$; see][for further details]{2024APh...16202990A}. In both cases, $\mathcal{N}_{\rm \nu_\mu}^{\rm bk}$ represents the expected number of background events, while $\mathcal{N}_{\rm \nu_\mu}^{\rm  SN}$ denotes the expected number of signal events provided by the input SN {HE-$\nu$} flux. These quantities can be computed by integrating equation (\ref{Eq:rate_N}) over time, and their values depend not only on the choice of the lower energy threshold but also on the initial and final times over which the signal and background are integrated \citep[see, e.g.,][]{Murase18}.\par
In the following subsections, we provide flux sensitivity study for existing and upcoming Large Volume Neutrino telescopes, along with their expected event counts for nearby realistic SN explosions such as SN 2023ixf \citep[with $z = 8.04\pm0.07 \times 10^{-4}$;][]{1991rc3..book.....D}, to highlight potential future prospects. However, for accurate predictions, it is essential accounting for a dedicated likelihood analysis preferably made through the muon effective area, that is a task beyond the scope of this study.

\subsection{High energy neutrinos detection by type II SNe}\label{Subsec:real}
The possibility to detect HE-$\nu$ from young SNe is mainly hindered by the intergalactic distances that separate us from these events. Estimates suggest that the occurrence of a SN event within our galaxy \citep[$\sim 60\,$years; see, e.g.,][]{2021NewA...8301498R} could surpass the operational lifetimes of neutrino observatories, therefore it is necessary to move the search at least to a distance of $6\,\text{Mpc}$, where the type II SN rate is about one every 10 years \citep{Perley_2020}. Moreover, both current and future neutrino observatories appear to have limitations that restrict their ability to detect HE-$\nu$ emitted by the most common H-rich SNe (such as Type IIP) to distances within a few Mpc \citep[e.g.,][]{Sarmah_2022,KheiMura_2023}.\par
\begin{figure}
	\includegraphics[width=\columnwidth]{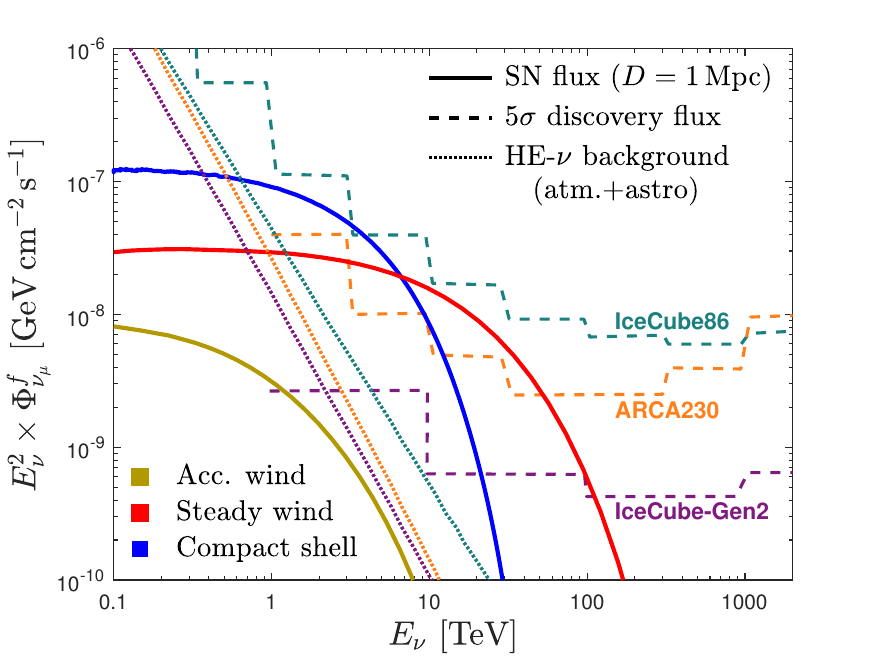}
    \caption{Averaged neutrino energy flow from interacting SNe exploding with an energy of $1\,\text{foe}$ at $1\,\text{Mpc}$ from Earth, having a RSG-like ejecta and a variable CSM configuration according to the scenarios presented in Fig. \ref{fig:density}. The dashed lines show 
the differential sensitivities ($5\sigma$ discovery potential) of IceCube86 \citep{Aartsen_2017} and IceCube-Gen2 \citep{Aartsen_2021} for the detection of point sources at the celestial horizon ($\delta_\star=0^\circ$), while KM3NeT/ARCA230 sensitivity refers to a point source inside its maximum visibility area with $\delta_\star\simeq -72^\circ$ \citep{AIELLO2019100,2024APh...16202990A}. Dotted lines show $F_{\nu_\mu+\bar{\nu}_\mu}^{\rm bk}$ seen by these three HE-$\nu$ detectors [cf. equation (\ref{Eq:background})].    
    }\label{fig:PhiNu_1Mpc_detectors}
\end{figure}
\begin{figure}
	\includegraphics[width=\columnwidth]{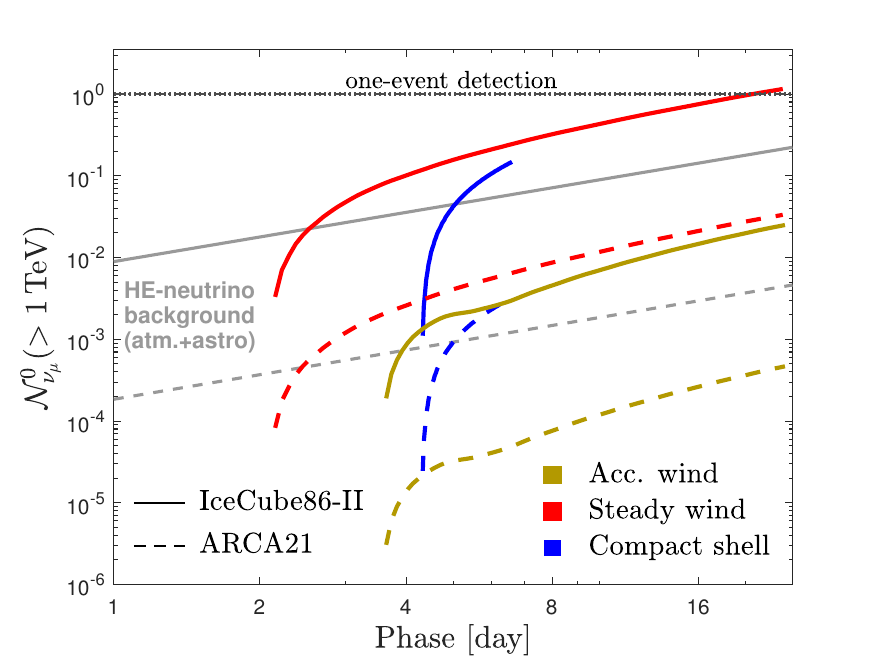}
    \caption{Expected number of detected muon (anti-)neutrinos with energy above $1\,$TeV integrated from the explosion epoch, given by $\mathcal{N}^0_{\rm \nu_\mu}(>1\text{TeV})=\int_0^{t} \Dot{\mathcal{N}}_{\rm \nu_\mu}^{\rm >1\,TeV}dt$, for both the HE-$\nu$ background (atm.+astro) and SN sources shown in Fig. \ref{fig:PhiNu_1Mpc_detectors}. The neutrino detection rate, $\Dot{\mathcal{N}}_{\rm \nu_\mu}^{\rm >1\,TeV}$, has been computed using the effective areas of KM3NeT/ARCA21, for track events of $\nu_\mu+\bar{\nu}_\mu$ CC interactions \citep{Muller:2023xO}, and IceCube86-II \citep[][]{2021arXiv210109836I}, assuming a source declination of $\delta_\star=0^\circ$. Each SN scenario curve begin and finish in correspondence with its own $t_{\rm bo}$ and $t_{\rm f}$, respectively. The dash-dotted horizontal line marks the one-event detection threshold.}\label{fig:ObsNu_number_rates}
\end{figure}

On the other hand, the distance is not the only parameter to take into account when examining Type II SNe as HE-$\nu$ sources \citep[e.g.][]{CosPumChe2024}. In fact, our model has shown how neutrino emission can be strongly influenced by the physical properties characterizing the SN ejecta and CSM, which for type II SNe can substantially change from event to event.
To understand how the SN characteristics affect the possibility of detecting its HE-$\nu$ by large volume neutrino observatories, let us consider the example of three SN scenarios differing only in the configuration of the CSM like depicted in Figs. \ref{fig:PhiNu_1Mpc_detectors}-\ref{fig:ObsNu_number_rates}.

In particular, these SN scenarios refer to H-rich SN events located $1\, \text{Mpc}$ away from Earth, each of them has a RSG-like ejecta of $10\,\text{M}_{\rm \odot}$, which expands with a kinetic energy of $1\, \text{foe}$. The three cases are hence distinguished only for the CSM matter density profile and present the same total CSM mass of $0.1\,\text{M}_{\rm \odot}$.
By considering these information, our model predicts an average neutrino energy {flow} $E_{\rm \nu}^2\times \Phi_{\rm \nu_\mu}^f$ for each scenario within the range of $10^{-8}-10^{-7}\,\text{GeV}\,\text{cm}^{-2}\,\text{s}^{-1}$, which is comparable to the sensitivities of neutrino telescopes inside their maximum visibility area (see also, Fig. \ref{fig:PhiNu_1Mpc_detectors}), in agreement with \cite{Sarmah_2022}.
In more detail, these energy flow remain slightly below the current sensitivity threshold of IceCube \citep{Aartsen_2017}, which is expected to improve by at least an order of magnitude with future telescopes such as KM3NeT \citep{AIELLO2019100} and IceCube-Gen2 \citep{Aartsen_2021}. The CSM configurations such as the compact shell ($s=0$) and the steady wind ($s=2$) will exceed the new sensibility limits even at energies greater than $1-10\,$TeV, where the contamination of atmospheric background is notably reduced. In this energy range, indeed, the signal significance for IceCube86-II, given by $\mathcal{N}_{\rm \nu_\mu}^{\rm  SN}/(\mathcal{N}_{\rm \nu_\mu}^{\rm  SN}+\mathcal{N}_{\rm \nu_\mu}^{\rm bk})$, may reach 84-98\% for the steady wind scenario and 70-88\% for the case with $s=0$, significantly higher than that find for AT2019fdr by \cite{Pitik2022} in correspondence of the observed neutrino event IC200530A.\par
On the other hand, the flux from denser and more compact CSM configurations, such as the accelerated wind case, would still remain below the HE-$\nu$ background level, making it undetectable by these telescopes (see Fig. \ref{fig:PhiNu_1Mpc_detectors}), and reducing the limiting distance to below $100\,$Kpc \citep[{e.g.,}][]{Murase18}. This decrease in the average HE-$\nu$ flow can be attributed to the delayed break-out epoch compared to other less dense models, significantly diminishing the total neutrino energy emission, as explained in Section \ref{Subsec:par_mod}. The cumulative curves of the detected neutrino number reveal indeed a break-out time shift of more than one day between the models of steady and accelerated wind (see Fig. \ref{fig:ObsNu_number_rates}).
These curves, computed for different neutrino detectors and an energy threshold above $1\,$TeV—where background contamination is lower—assess the {HE-$\nu$} detectability of type II interacting SNe, similarly to \citet{Murase18} [cf. his Fig. 3], but at $1\,$Mpc instead of $10\,$kpc. Rescaling the expected neutrino number with this distance ($\mathcal{N}_{\rm \nu\mu}^{\rm SN} \propto D^{-2}$), we find that steady wind scenarios should yield $\sim10^3$ neutrino events in a 7-day time window post-explosion and $\sim10^4$ over the entire interaction phase. As in \citet{Murase18}, cases with $s>2$ lead to about one order of magnitude fewer neutrino numbers than the $s=2$ steady wind, although our model is more sensitive to density variations, as reflected in the shift of the break-out epoch.
Furthermore, the latter curves, which are in any case dependent on the effective area of the specific detector, can be used to understand which phases have the greatest neutrinic production and which most increase the probability of detecting a signal. Specifically, the effective area of ARCA21, with only 21 strings compared to the envisaged 230 \citep{Muller:2023xO}, is approximately two orders of magnitude less than IceCube86-II \citep{2021arXiv210109836I}. However, apart from the discrepancy in the neutrino numbers between the two detectors (explained by their different size), the increase in event numbers follows a similar trend, mainly influenced by the type of CSM model rather than the detector.\par
Besides the difference in $t_{\rm bo}$ between the two wind-like models, we note that the compact shell case exhibits the shortest duration for neutrino emission, spanning from 4 to 7 days after the explosion. Therefore, the detection of even a single HE neutrino, arriving after this specific time window, can offer valuable constraints on the duration of the interaction mechanism and may be used to get information about the type of explosion scenario. Moreover, this information can place a lower limit on the outer radius of the CSM, completely independent on the electromagnetic observations. For instance, in the case of a stationary wind, the interaction phase lasts about 23 days and, considering a supernova located at $1\, \text{Mpc}$, the current sensitivity of IceCube would allow to detect at least one neutrino as early as 21 days after the explosion (cf. red curves in Fig. \ref{fig:ObsNu_number_rates}), well beyond the epoch of the SN photometric peak (see also Fig. \ref{fig:Lum_densConf_E}).\par 
Conversely, when the physical parameters of the SN and its CSM are known—derived, for instance, from electromagnetic modelling as seen with SN 2023ixf—our model can also determine the optimal time window to enhance the search for neutrino signals within the detector.
 \subsection{Forecasting HE-$\nu$ signals from SN 2023ixf}
\begin{figure}
	\includegraphics[width=\columnwidth]{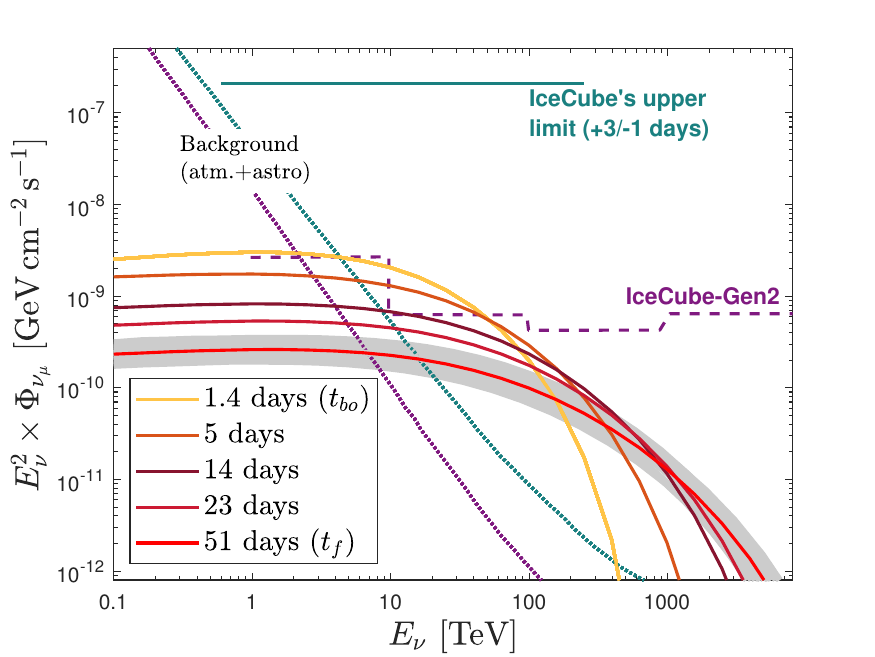}
    \caption{The temporal sequence of the muon neutrino flow from SN 2023ixf since its explosion epoch \citep[$\text{MJD}=60082.74$; ][]{Hiramatsu_2023}. The IceCube's upper limit for the period between 2 days before and after the discovery ($+3/-1$ since the explosion) has been reported as horizontal continuous line \citep{2023ATel16043....1T}, while the IceCube-Gen2' sensitivity and HE-$\nu$ background (atm.+astro) are like showed in Fig. \ref{fig:PhiNu_1Mpc_detectors}. The gray region around the energy spectrum at $t_{\rm f}=51$ day is the error range for $\Phi_{\rm \nu_{\rm \mu}}^f$ due to the deviations on the modelling parameters listed in Tab. \ref{tab:parameters_23ixf}.}\label{fig:Phi_SN2023ixf}
\end{figure}

In Section \ref{subsec:SN2023ixf}, we have studied the real case of SN 2023ixf finding its physical parameters necessary to simulate the expected HE-$\nu$ flux during SI epochs (see Fig. \ref{fig:Phi_SN2023ixf}).
 Specifically, based on the best-fit model parameters of Tab. \ref{tab:parameters_23ixf}, we determine that the break-out epoch occurred $1.4$ days after the explosion, which is less than half a day after SN discovery \citep[$\text{MJD}= 60083.73$, see][]{Hiramatsu_2023}. At this stage, the {energy flow} was most intense, peaking around $1\,$TeV with a value of approximately $3\times10^{-9}\,\text{GeV}\,\text{cm}^{-2}\,\text{s}^{-1}$, which is consistent with that find even by \citet{2024JCAP...04..083S}. This maximum {energy flow} is two orders of magnitude below the upper limit on the muon neutrino flux set by IceCube, calculated within $+/-2$ days since the SN discovery \citep{2023ATel16043....1T}. However, it remains above IceCube-Gen2's sensitivity until $14$ days post-explosion (see Fig. \ref{fig:Phi_SN2023ixf}). {T}he model predictions show a decline in energy flow intensity as the maximum proton energy increases progressively up to $50$ days, marking the end of the interaction phase at $t_{\rm f}$. At this epoch, the average neutrino energy flow decreases to $(2.6\pm 0.8)\times 10^{-10}\,\text{GeV}\,\text{cm}^{-2}\,\text{s}^{-1}$, about ten times less than its initial value.  
This flux corresponds to an all-flavor ($\nu_e+\bar{\nu}_e+\nu_\mu+\bar{\nu}_\mu+\nu_\tau+\bar{\nu}_\tau$) energy fluence at $1\,$TeV of $(3.4\pm1.1)\times 10^{-3}\,\text{GeV}\,\text{cm}^{-2}$, close to that obtained by \citet{Murase24} in his denser CSM scenario \citep[i.e., $D_*=0.1$; see also Appendix in][]{KheiMura_2023}. Meanwhile, both the maximum proton energy and the neutrino spectrum knee rise to the PeV energies, with $E_{\rm p}^{M}=43\pm16\,\text{PeV}$ and $E_{\rm \nu}^k=1.6\pm0.6\,\text{PeV}$, respectively. During the entire interaction phase, hence, the { HE-$\nu$} manage to carry out from the SN 2023ixf an energy of $(2\pm0.4)\times 10^{47}\,$erg, just about one ten-thousandth of the entire SN kinetic energy.\par
\begin{figure}
	\includegraphics[width=\columnwidth]{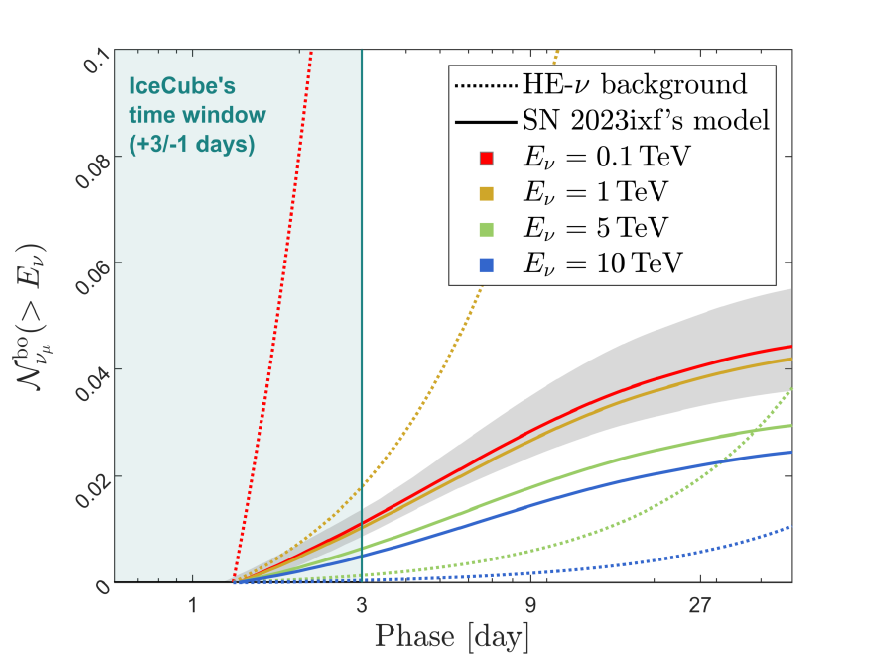}
    \caption{Temporal evolution of expected number of muon (anti-)neutrinos with energy above $E_\nu$ detected by IceCube86-II \citep{2021arXiv210109836I}, with an arrival direction compatible with that of SN 2023ixf ($\delta_\star\simeq 54.3^\circ$, zenith angle $\sim 144.3^\circ$), since its shock break-out, i.e., $\mathcal{N}^{\rm bo}_{\rm \nu_\mu}(>E_\nu)=\int_{t_{\rm bo}}^{t} \Dot{\mathcal{N}}_{\rm \nu_\mu}^{\rm >E_\nu}dt$. Dotted lines represents the expected {HE-$\nu$} background (atm.+astro) $\mathcal{N}_{\rm \nu_\mu}^{\rm  bk}$, while the solid lines refer to the SN contribution $\mathcal{N}_{\rm \nu_\mu}^{\rm  SN}$. The vertical line marks the end of IceCube's time window referred to the SN explosion epoch ($+3/-1$ days). The gray region, such as in Fig. \ref{fig:Phi_SN2023ixf}, displays the errors on the expected (anti-)neutrinos number with energies above $0.1\,$TeV, arising from uncertainties in LC SN modelling parameters.}\label{fig:Nu_SN2023ixf}
\end{figure} 
The flux of {the point source SN 2023ixf, from} a declination of $\delta_\star\simeq 54.3^\circ$ { \citep[e.g.,][]{2023TNSTR1158....1I}}, left no event trace in the IceCube detector during its observation window \citep{2023ATel16043....1T}, although the direction of arrival is within the telescope's field of view.  
Besides, considering SN 2023ixf's distance of $6.9\,\text{Mpc}$ \citep[e.g.,][]{Hiramatsu_2023}, the expected number of muon (anti-)neutrinos with energy above $0.1\,$TeV along the entire interaction phase ($\sim50\,$day) can just achieve about $(4\pm 1)$ $\times 10^{-2}$ (see Fig. \ref{fig:Nu_SN2023ixf}). At this energy threshold, the background neutrino contamination significantly reduces the signalness to about $1\%$. Under these energy and time ranges, the statistical test $\mathcal{N}_{\rm \nu_\mu}^{\rm  SN}/\sqrt{\mathcal{N}_{\rm \nu_\mu}^{\rm bk}}$ reaches its maximum at $t_{\rm ts}^{\rm max}\simeq 7$ days, corresponding to a MRF minimum, and the signalness equates $5\%$. However, as a consequence, the expected number of neutrinos decreases to $\sim(2\pm 0.5)\times 10^{-2}$ (see Tab. \ref{tab:Nu_23ixf} for further details).\par 
It is worth noting that the best time window to search for HE-$\nu$ does not coincide with that \citet{2023ATel16043....1T} chose. Indeed, to maximize the expected neutrino number from this SN the best observation window should start about one day and a half after the SN explosion (about 12 hours after the discovery) and extend up to $t_{\rm ts}^{\rm max}\simeq 7-10\,$ days, where its value settles around two times greater than that obtained at only 3 days after the explosion (see Fig. \ref{fig:Nu_SN2023ixf} and Tab. \ref{tab:Nu_23ixf}). Notably, the choice of the optimal observation window depends on the selected energy threshold. Tab. \ref{tab:Nu_23ixf} also provides the expected neutrino numbers for different energy and temporal observation ranges, along with their respective statistical test values (typically used in background-dominated data) and the corresponding signalness percentage.
\begin{table}
{
	\centering
	\caption{Statistics on the expected number of muon (anti-)neutrinos from SN 2023ixf with energy above the minimum value, integrated from the SN break-out up to the phases $t_{\rm f}$ and $t_{\rm ts}^{\rm max}$ (see also Fig. \ref{fig:Nu_SN2023ixf}).}
	\label{tab:Nu_23ixf}
	\begin{tabular}{c|c||ccc} 
		\hline
		 Min. energy & {Phase} &$\mathcal{N}_{\rm \nu_\mu}^{\rm  SN}$ & \multirow{2}{*}{\centering $\mathcal{N}_{\rm \nu_\mu}^{\rm  SN}/\sqrt{\mathcal{N}_{\rm \nu_\mu}^{\rm bk}}$} & \multirow{2}{*}{\centering$\frac{\mathcal{N}_{\rm \nu_\mu}^{\rm  SN}}{\mathcal{N}_{\rm \nu_\mu}^{\rm  SN}+\mathcal{N}_{\rm \nu_\mu}^{\rm bk}}$} \\
		{[TeV]} &{\centering [day]} & $[\times 10^{-2}]$ &  &  \\		
        \hline
		\hline
		\multirow{2}{*}{\centering 0.1}& $t_{\rm f}=51$ &  $4.4\pm1.0$ & 0.02 & 1\%  \\
		{}& $t_{\rm ts}^{\rm max}=7$ &  $2.4\pm0.5$ & 0.04 & 5\%  \\
		\hline
		\multirow{2}{*}{\centering$1$} & $t_{\rm f}=51$ & $4.0\pm0.9$  & 0.06 & 7\%\\
		{}& $t_{\rm ts}^{\rm max}=7$ & $2.3\pm0.5$  & 0.09 & 27\%\\
		\hline		
		\multirow{2}{*}{\centering$5$} & $t_{\rm f}=51$ & $2.8\pm0.6$ & 0.13 & 39\% \\
		{}& $t_{\rm ts}^{\rm max}=9$ & $1.7\pm0.4$ & 0.21 & 72\% \\
		 		\hline		
		\multirow{2}{*}{\centering$10$} & $t_{\rm f}=51$ & $2.3\pm0.5$ & 0.20 & 64\% \\
		{}& $t_{\rm ts}^{\rm max}=9.4$ & $1.4\pm0.3$ & 0.31 & 87\% \\		
		\hline
	\end{tabular}
	}	
\end{table}
Interestingly, increasing the energy threshold to $1\,$TeV results in only a minor decrease in the expected number of muon (anti-)neutrinos, remaining within the error bars of the values at $0.1\,$TeV (gray region in Fig. \ref{fig:Nu_SN2023ixf}). However, the signal significance increases notably, reaching 27\% (see Tab. \ref{tab:Nu_23ixf}).\par
Although for SN 2023ixf the choice of observation time window does not increase the expected number above the detection limit of one event, for a similar but closer SN explosion set at $1\, \text{Mpc}$ this choice can be crucial. In this case, indeed, the number of neutrino events detected within a time window of 3 days after $t_{\rm bo}$ should be $\mathcal{N}_{\rm \nu_\mu}{^{\rm SN}(<3\,\text{days}) \simeq 0.5}$ less than one. Instead, if we extend the observation time to {$7$} days, i.e. $\mathcal{N}_{\rm \nu_\mu}{^{\rm SN}(<7\,\text{days}) \simeq 1.1}$, we have a high probability of observing one significant astrophysical neutrino with the current IceCube configuration.\par
Summarising, the astrophysical scenario we {inferred} from the analysis of SN 2023ixf{'s electromagnetic LC} is consistent with the non-detection of HE-$\nu$ {by the IceCube detector \citep{2023ATel16043....1T}}.
The neutrino flux limits established by observatories can aid in distinguishing between different SN scenarios, including those involving alternative heating mechanisms such as choked jets  \citep[see, e.g.,][]{Guetta_2023}. Additionally, these limits can impose constraints on the efficiency of the physical mechanisms governing particle acceleration processes around young SN explosions \citep[e.g.,][]{Murase_2019,2024A&A...686A.254M}.

\section{Summary \& further considerations}\label{sec:summary}
In this work, a new astrophysical approach to study HE-$\nu$ emission by low interacting type II SNe is presented and applied to the nearby SN 2023ixf in M101. Our method includes a greater amount of astrophysical information about the nature of the SN progenitor and its CSM. Such an approach, hence, requires a sufficiently accurate analysis of the SN LCs and a detailed modelling procedure to derive the physical characteristics of these events.\par 
In this regard, we have developed a new coherent model capable of describing both the HE-$\nu$ and electromagnetic emissions during the entire SN post-explosive phase. Specifically, we present a generalized analytic description for the hydrodynamical evolution of the forward shock inside the CSM. In line with previous approaches \citep[e.g.,][]{Murase14,Fang2020}, our description of CSM-ejecta interaction includes details about the thermal energy increase and the formation of magnetic instabilities inside the shocked shell, which affect the energy distribution of the CSM protons swept by the shock (see Section \ref{Subsec:p_acc} for details). So we have used this proton energy distribution to evaluate the production rates of HE-$\nu$ obtained by the pp-collisions, considering for the first time even the effects of proton transverse losses due to the CSM asphericity. These neutrino production rates have permitted us to evaluate the overall neutrino energy spectra and study how affect the SN physical parameters considered in our model.\par
By studying the dependence of HE-$\nu$ spectra features on all twelve modelling parameters describing the SN configuration, we find that the neutrino emission is strongly related to four main properties, including the initial shock velocity and the inner CSM density. In addition to these quantities, also be considered in other works \citep[e.g.,][]{Murase18,Sarmah_2022}, we notice it is essential to take into account also the CSM density profile (here characterised by the $s$ parameter) and its radial extension (see Section \ref{Subsec:par_mod} for further details). On the other hand, since four SN properties primarily influence the HE-$\nu$ emission, the two optical conditions derived by measuring the rise time and the peak luminosity are no longer sufficient to derive the expected neutrino spectra  \citep[see also][]{2023MNRAS.524.3366P}.  To break this degeneracy, we have introduced a new approach to SN electromagnetic LC modelling, which is able to give us more information about the SN parameters and improves forecasts of the HE-$\nu$ detection.

\subsection{Supernova Light Curve Modelling in High-Energy Neutrino forecasting: A Novel Approach}
Our SN modelling approach is based on a new semi-analytical description for the SN LCs during the whole interacting phase. Specifically, this LC model is fully consistent with the hypothesis made to simulate the HE-$\nu$ emission and proposes a novel analytical formulation for describing radiation diffusion through the CSM during the entire interaction phase, in line with the general principles adopted in previous works \citep[e.g.,][]{Moriya2013}. In this way, the simulated LCs give an accurate description of the SN luminosity from the break-out to the end of the interaction, reproducing all the photometric characteristics of the interaction peak, including both the rising and the descending phases (see Section \ref{subsec:LC_peak} for more details). However, to fully characterise all necessary parameters, our approach extend the electromagnetic emission analysis also during the  late post-peak phases, introducing a coherent LC modelling procedure which permits us to infer all needed parameters. Then, this procedure combines the new description of the interaction phase with the post-explosive LC model presented in \citetalias{PC2025}. In this way, we can derive all SN parameters for a real SN, such as SN 2023ixf (see also Section \ref{subsec:SN2023ixf}).\par
Applying this procedure to the LC of the nearby SN 2023ixf in M101, our results reconstruct the physical configuration for the progenitor system of this SN event at the time of its explosion, leading to a more comprehensive interpretation of this explosive event (see Tab. \ref{tab:parameters_23ixf}, for all parameter values). In particular, we found that SN 2023ixf has an ejecta mass of approximately $9\,\text{M}_{\rm \odot}$, expanding with a kinetic energy of $1.8\pm0.2\, \text{foe}$. Additionally, the ejected $^{56}$Ni mass synthesized during the explosion is about $0.07\,\text{M}_{\rm \odot}$ and exceeds the median value for the standard type II SNe.
Concerning the CSM mass distribution, our findings align with those of \cite{zimmerman_complex_2024}, indicating a steeper density profile than that produced by a constant stationary wind, i.e. $s\simeq 2.9>2$, which is also consistent with the shock flash break-out analysis by \cite{li_shock_2024}. Moreover, the most dense CSM region, just above the progenitor radius $\sim 1.6\times 10^{13}\,\text{cm}$, extends to an outer boundary around $3.6\times 10^{15}\,\text{cm}$, within which a total mass of about $0.06\,\text{M}_{\rm \odot}$ is contained. 
These CSM features testify to the abrupt increase in the mass loss rate of the progenitor, which, assuming a wind speed of $v_{\rm w}\simeq 55\,\text{Km}\,\text{s}^{-1}$ \citep{ZHANG20232548}, must have started about $20\,$years (i.e. $\simeq R_{\rm CSM}/v_{\rm w}$) before the explosion, in agreement with the results of \cite{xiang_dusty_2023}.
{Moreover, a}dding the mass of the compact remnant to that of the ejected material, we obtain a total stellar mass at explosion of $9.8-11.5\,\text{M}_{\rm \odot}$, to which we can add the mass lost during the entire pre-SN evolution (i.e. since the ZAMS) to obtain
$M_{\rm ZAMS}\simeq 10.4-13.3\, \text{M}_{\rm \odot}$ \citep[for details about the mass loss during the pre-SN evolution see][and references therein]{2017MNRAS.464.3013P}. This result is fully consistent with the estimate of $12^{+2}_{-1}\,\text{M}_{\rm \odot}$ from the direct progenitor detection method of \cite{xiang_dusty_2023} and resolves the missing mass tension highlighted by \cite{zimmerman_complex_2024} without invoking exotic scenarios.
We can conclude that SN 2023ixf is the final explosive event of a RSG with a $M_{\rm ZAMS}\simeq 12\,\text{M}_{\rm \odot}$, as well as one of the ``common'' type II SNe that can occur at a distance of $6-7\,\text{Mpc}$ about every 10 years.\par
In the light of these findings, we have investigated the expected fluxes of HE-$\nu$ from this type of SNe and compared them with the observability limits of current and future neutrino telescopes.
As concerning the HE-$\nu$ emission from SN 2023ixf, we have used above modelling parameters to find an all-flavour energy fluence of $(3.4\pm 1.1)\times 10^{-3}\,\text{GeV}\,\text{cm}^{-2}$ and a maximum muon neutrino energy flux of about $3\times10^{-9}\,\text{GeV}\,\text{cm}^{-2}\,\text{s}^{-1}$, which is comparable with the expected sensitivity of IceCube-Gen2 \citep{Aartsen_2021}. 
Moreover, {while consistent with the estimates of \citet{KheiMura_2023} and \citet{2024JCAP...04..083S}, our results offer greater accuracy due to the tighter parameter constraints imposed by the LC modelling}. In addition, our model allowed us to calculate the maximum energy achieved by protons in the forward shock of SN 2023ixf, i.e. $\sim40\,$PeV, and to evaluate {phase} by {phase} its neutrino flux energy distribution. From this analysis, we have concluded that to search for this kind of HE-$\nu$ signals inside large-volume neutrino telescopes, it is crucial to recognize the optimal time window for the detector. Specifically, as tested for SN 2023ixf, we find that the best time window should begin about 1-2 days after the SN explosion and extended at least to 7-10 days to achieve IceCube's expected detection of $\sim2\times 10^{-2}$ SN neutrino events, increasing to $\sim4\times 10^{-2}$ by the end of the interaction phase.
\subsection{Prospects of multimessanger transient astronomy in the next decade}
Our study highlights that the possibility of observing HE-$\nu$ from interacting SNe is strongly influenced by the distribution of matter within their CSM. Indeed, with all other parameters being fixed, the CSM configuration can modify the expected neutrino flux by up to an order of magnitude (see Section \ref{sec:detect_pot} for details). 
Moreover, for SNe located at a distance of $1\,\text{Mpc}$ from Earth, we have found the average flux expected in the first 20 days is already within the sensitivity range of telescopes such as IceCube. On the other hand, the next-generation telescopes like KM3NeT/ARCA and IceCube-Gen2 will be able to extend the maximum detection distance for HE-$\nu$ from these SNe to about $5-10\,\text{Mpc}$.
 In this way, the origin of HE-$\nu$ from even low-interacting SNe, such as SN 2023ixf, might be confirmed by direct observations.\par

The electromagnetic information derived from the complete analysis of the LC observations and the modelling on the entire post-explosive phase can play a key role in the identification and the characterization of astrophysical source for futures HE-$\nu$ signals. 
Conversely, if such signals were discovered, the information on the energy of neutrinos and their arrival time compared to the time of the explosion would allow to better understand the structure of the CSM, thus obtaining information independent of electromagnetic observations. In particular, this study highlights how the neutrino detection delay respect to the explosion time is intrinsically linked to the radial extension of the CSM, information difficult to deduce from the analysis of electromagnetic emission.\par
In addition, the discovery of this type of neutrino signals would greatly enhance our understanding of the processes affecting particle acceleration and cosmic ray production near young SNe \citep[e.g.,][]{Murase18}. The modelling of these mechanisms currently involves significant uncertainties, primarily due to the lack of direct measure of their efficiency by both neutrino and gamma observations \citep[e.g.,][]{Murase_2019,2024A&A...686A.254M}.
Compared to gamma emission, neutrino one has the advantage of reaching us without being absorbed by the CSM surrounding the shock shell. However, the redshift horizon of gamma telescopes extends well beyond ten Mpc \citep[][]{ACHARYA20133}, and with the advent of the Cerenkov Telescope Array (CTA), even events like SN 2023ixf could be within reach of these telescopes \citep[e.g.,][]{2024JCAP...04..083S,Murase24}. For this reason, we plan to develop models consistent with the one presented here that are capable of simulating even the gamma radiation produced by the interaction between the CSM and the SN ejecta.\par
The Legacy Survey of Space and Time (LSST) at the Vera C. Rubin Observatory is expected to increase the number of SN discoveries by another order of magnitude within a few years \citep{2019ApJ...873..111I}. Then, the capability to connect all the information between the several SNe emission channels will be fundamental for establishing a multi-messenger approach to stellar explosion astrophysics. In this framework, we think that the cooperation between optical (e.g. LSST), gamma (e.g. Fermi, CTA), and neuritic (e.g. KM3NeT, IceCube) communities, with the aim to improve the global interpretation of all their data through models like ours, could significantly increase the discovery potential for each of them, leading so to a better understanding of the physical processes involved during the SN explosions and their post-explosive evolution.\\

\section*{Acknowledgements}
We thank an anonymous referee for his/her critical and constructive feedback. We also sincerely thank Dr. Giovanna Ferrara, researcher of Physics and Astronomy department of Catania University (MUR-PNRR project KM3NeT4RR, IR0000002), for her valuable discussions that contributed to this work. The author thanks Dr. G.E. Cinardi for her assistance in improving the visual design of Fig. 2. We gratefully acknowledge the variable star observations from the AAVSO International Database, contributed by observers worldwide and used in this research.  This paper is supported by the Fondazione ICSC, Spoke 3 Astrophysics and Cosmos Observations. National Recovery and Resilience Plan (Piano Nazionale di Ripresa e Resilienza, PNRR) Project ID CN 00000013 “Italian Research Center on High-Performance Computing, Big Data and Quantum Computing” funded by MUR Missione 4 Componente 2 Investimento 1.4: Potenziamento strutture di ricerca e creazione di “campioni nazionali di R\&S (M4C2-19)” - Next Generation EU (NGEU).


\section*{Data Availability}
The data underlying this article are available in the article.



\bibliographystyle{mnras}
\bibliography{example} 

\begin{thebibliography}{}
\makeatletter
\relax
\def\mn@urlcharsother{\let\do\@makeother \do\$\do\&\do\#\do\^\do\_\do\%\do\~}
\def\mn@doi{\begingroup\mn@urlcharsother \@ifnextchar [ {\mn@doi@}
  {\mn@doi@[]}}
\def\mn@doi@[#1]#2{\def\@tempa{#1}\ifx\@tempa\@empty \href
  {http://dx.doi.org/#2} {doi:#2}\else \href {http://dx.doi.org/#2} {#1}\fi
  \endgroup}
\def\mn@eprint#1#2{\mn@eprint@#1:#2::\@nil}
\def\mn@eprint@arXiv#1{\href {http://arxiv.org/abs/#1} {{\tt arXiv:#1}}}
\def\mn@eprint@dblp#1{\href {http://dblp.uni-trier.de/rec/bibtex/#1.xml}
  {dblp:#1}}
\def\mn@eprint@#1:#2:#3:#4\@nil{\def\@tempa {#1}\def\@tempb {#2}\def\@tempc
  {#3}\ifx \@tempc \@empty \let \@tempc \@tempb \let \@tempb \@tempa \fi \ifx
  \@tempb \@empty \def\@tempb {arXiv}\fi \@ifundefined
  {mn@eprint@\@tempb}{\@tempb:\@tempc}{\expandafter \expandafter \csname
  mn@eprint@\@tempb\endcsname \expandafter{\@tempc}}}

\bibitem[\protect\citeauthoryear{{Aartsen} et~al.,}{{Aartsen}
  et~al.}{2014a}]{2014PhRvL.113j1101A}
{Aartsen} M.~G.,  et~al., 2014a, \mn@doi [\prl]
  {10.1103/PhysRevLett.113.101101}, \href
  {https://ui.adsabs.harvard.edu/abs/2014PhRvL.113j1101A} {113, 101101}

\bibitem[\protect\citeauthoryear{Aartsen et~al.,}{Aartsen
  et~al.}{2014b}]{Aartsen_2014}
Aartsen M.~G.,  et~al., 2014b, \mn@doi [\apj] {10.1088/0004-637X/796/2/109},
  796, 109

\bibitem[\protect\citeauthoryear{Aartsen et~al.,}{Aartsen
  et~al.}{2016}]{Aartsen_2016}
Aartsen M.~G.,  et~al., 2016, \mn@doi [\apj] {10.3847/0004-637X/833/1/3}, 833,
  3

\bibitem[\protect\citeauthoryear{Aartsen et~al.,}{Aartsen
  et~al.}{2017}]{Aartsen_2017}
Aartsen M.~G.,  et~al., 2017, \mn@doi [\apj] {10.3847/1538-4357/835/2/151},
  835, 151

\bibitem[\protect\citeauthoryear{Aartsen et~al.,}{Aartsen
  et~al.}{2021}]{Aartsen_2021}
Aartsen M.~G.,  et~al., 2021, \mn@doi [Journal of Physics G: Nuclear and
  Particle Physics] {10.1088/1361-6471/abbd48}, 48, 060501

\bibitem[\protect\citeauthoryear{Abbasi et~al.,}{Abbasi
  et~al.}{2011}]{PhysRevD.83.012001}
Abbasi R.,  et~al., 2011, \mn@doi [Phys. Rev. D] {10.1103/PhysRevD.83.012001},
  83, 012001

\bibitem[\protect\citeauthoryear{Acharya et~al.,}{Acharya
  et~al.}{2013}]{ACHARYA20133}
Acharya B.,  et~al., 2013, \mn@doi [Astroparticle Physics]
  {https://doi.org/10.1016/j.astropartphys.2013.01.007}, 43, 3

\bibitem[\protect\citeauthoryear{Ahlers \& Halzen}{Ahlers \&
  Halzen}{2018}]{Ahlers2018}
Ahlers M.,  Halzen F.,  2018, \mn@doi [Progress in Particle and Nuclear
  Physics] {https://doi.org/10.1016/j.ppnp.2018.05.001}, 102, 73

\bibitem[\protect\citeauthoryear{Aiello et~al.,}{Aiello
  et~al.}{2019}]{AIELLO2019100}
Aiello S.,  et~al., 2019, \mn@doi [Astroparticle Physics]
  {https://doi.org/10.1016/j.astropartphys.2019.04.002}, 111, 100

\bibitem[\protect\citeauthoryear{{Aiello} et~al.,}{{Aiello}
  et~al.}{2024}]{2024APh...16202990A}
{Aiello} S.,  et~al., 2024, \mn@doi [Astroparticle Physics]
  {10.1016/j.astropartphys.2024.102990}, \href
  {https://ui.adsabs.harvard.edu/abs/2024APh...16202990A} {162, 102990}

\bibitem[\protect\citeauthoryear{Anchordoqui et~al.,}{Anchordoqui
  et~al.}{2014}]{ANCHORDOQUI20141}
Anchordoqui L.~A.,  et~al., 2014, \mn@doi [Journal of High Energy Astrophysics]
  {https://doi.org/10.1016/j.jheap.2014.01.001}, 1-2, 1

\bibitem[\protect\citeauthoryear{{Arnett}}{{Arnett}}{1996}]{1996snih.book.....A}
{Arnett} D.,  1996, {Supernovae and Nucleosynthesis: An Investigation of the
  History of Matter from the Big Bang to the Present}

\bibitem[\protect\citeauthoryear{{Bersten}, {Orellana}, {Folatelli},
  {Martinez}, {Piccirilli}, {Regna}, {Román Aguilar}  \& {Ertini}}{{Bersten}
  et~al.}{2024}]{Bersten_AA2024}
{Bersten} M.~C.,  {Orellana} M.,  {Folatelli} G.,  {Martinez} L.,  {Piccirilli}
  M.~P.,  {Regna} T.,  {Román Aguilar} L.~M.,   {Ertini} K.,  2024, \mn@doi
  [A&A] {10.1051/0004-6361/202348183}, 681, L18

\bibitem[\protect\citeauthoryear{Blumenthal \& Gould}{Blumenthal \&
  Gould}{1970}]{RevModPhys.42.237}
Blumenthal G.~R.,  Gould R.~J.,  1970, \mn@doi [Rev. Mod. Phys.]
  {10.1103/RevModPhys.42.237}, 42, 237

\bibitem[\protect\citeauthoryear{Bostroem et~al.,}{Bostroem
  et~al.}{2023}]{Bostroem_2023}
Bostroem K.~A.,  et~al., 2023, \mn@doi [\apjl L] {10.3847/2041-8213/acf9a4},
  956, L5

\bibitem[\protect\citeauthoryear{Burrows}{Burrows}{1990}]{AnnRev87A}
Burrows A.,  1990, \mn@doi [Annual Review of Nuclear and Particle Science]
  {https://doi.org/10.1146/annurev.ns.40.120190.001145}, 40, 181

\bibitem[\protect\citeauthoryear{{Cargill} \& {Papadopoulos}}{{Cargill} \&
  {Papadopoulos}}{1988}]{1988ApJ...329L..29C}
{Cargill} P.~J.,  {Papadopoulos} K.,  1988, \mn@doi [\apjl L] {10.1086/185170},
  \href {https://ui.adsabs.harvard.edu/abs/1988ApJ...329L..29C} {329, L29}

\bibitem[\protect\citeauthoryear{{Chandra}, {Chevalier}, {Maeda}, {Ray}  \&
  {Nayana}}{{Chandra} et~al.}{2024}]{Chandra_2024ApJ}
{Chandra} P.,  {Chevalier} R.~A.,  {Maeda} K.,  {Ray} A.~K.,   {Nayana} A.~J.,
  2024, \mn@doi [\apjl L] {10.3847/2041-8213/ad275d}, \href
  {https://ui.adsabs.harvard.edu/abs/2024ApJ...963L...4C} {963, L4}

\bibitem[\protect\citeauthoryear{{Chevalier} \& {Fransson}}{{Chevalier} \&
  {Fransson}}{1994}]{1994ApJ...420..268C}
{Chevalier} R.~A.,  {Fransson} C.,  1994, \mn@doi [\apj] {10.1086/173557},
  \href {https://ui.adsabs.harvard.edu/abs/1994ApJ...420..268C} {420, 268}

\bibitem[\protect\citeauthoryear{Chevalier \& Fransson}{Chevalier \&
  Fransson}{2006}]{Chevalier2006}
Chevalier R.~A.,  Fransson C.,  2006, \mn@doi [\apj] {10.1086/507606}, 651, 381

\bibitem[\protect\citeauthoryear{{Chevalier} \& {Fransson}}{{Chevalier} \&
  {Fransson}}{2017}]{2017hsn..book..875C}
{Chevalier} R.~A.,  {Fransson} C.,  2017, in {Alsabti} A.~W.,  {Murdin} P.,
  eds, , Handbook of Supernovae.
p.~875, \mn@doi{10.1007/978-3-319-21846-5_34}

\bibitem[\protect\citeauthoryear{{Chevalier} \& {Irwin}}{{Chevalier} \&
  {Irwin}}{2011}]{2011ApJ...729L...6C}
{Chevalier} R.~A.,  {Irwin} C.~M.,  2011, \mn@doi [\apjl L]
  {10.1088/2041-8205/729/1/L6}, \href
  {https://ui.adsabs.harvard.edu/abs/2011ApJ...729L...6C} {729, L6}

\bibitem[\protect\citeauthoryear{{Chornock}, {Blanchard}, {Gomez},
  {Hosseinzadeh}  \& {Berger}}{{Chornock} et~al.}{2019}]{2019TNSCR1016....1C}
{Chornock} R.,  {Blanchard} P.~K.,  {Gomez} S.,  {Hosseinzadeh} G.,   {Berger}
  E.,  2019, Transient Name Server Classification Report, \href
  {https://ui.adsabs.harvard.edu/abs/2019TNSCR1016....1C} {2019-1016, 1}

\bibitem[\protect\citeauthoryear{{Cosentino}, {Pumo}  \&
  {Cherubini}}{{Cosentino} et~al.}{2024}]{CosPumChe2024}
{Cosentino} S.~P.,  {Pumo} M.~L.,   {Cherubini} S.,  2024, \mn@doi [Il Nuovo
  Cimento C] {10.1393/ncc/i2024-24357-7}, 47

\bibitem[\protect\citeauthoryear{Courant, Isaacson  \& Rees}{Courant
  et~al.}{1952}]{Courant52}
Courant R.,  Isaacson E.,   Rees M.,  1952, \mn@doi [Communications on Pure and
  Applied Mathematics] {https://doi.org/10.1002/cpa.3160050303}, 5, 243

\bibitem[\protect\citeauthoryear{{Davies}, {Plez}  \& {Petrault}}{{Davies}
  et~al.}{2022}]{2022MNRAS.517.1483D}
{Davies} B.,  {Plez} B.,   {Petrault} M.,  2022, \mn@doi [\mnras]
  {10.1093/mnras/stac2427}, \href
  {https://ui.adsabs.harvard.edu/abs/2022MNRAS.517.1483D} {517, 1483}

\bibitem[\protect\citeauthoryear{{Dessart} \& {Jacobson-Gal{\'a}n}}{{Dessart}
  \& {Jacobson-Gal{\'a}n}}{2023}]{2023A&A...677A.105D}
{Dessart} L.,  {Jacobson-Gal{\'a}n} W.~V.,  2023, \mn@doi [\aap]
  {10.1051/0004-6361/202346754}, \href
  {https://ui.adsabs.harvard.edu/abs/2023A&A...677A.105D} {677, A105}

\bibitem[\protect\citeauthoryear{{Dessart}, {Hillier}, {Audit}, {Livne}  \&
  {Waldman}}{{Dessart} et~al.}{2016}]{DessartHiller2016}
{Dessart} L.,  {Hillier} D.~J.,  {Audit} E.,  {Livne} E.,   {Waldman} R.,
  2016, \mn@doi [\mnras] {10.1093/mnras/stw336}, \href
  {https://ui.adsabs.harvard.edu/abs/2016MNRAS.458.2094D} {458, 2094}

\bibitem[\protect\citeauthoryear{{Draine}}{{Draine}}{2011}]{2011piim.book.....D}
{Draine} B.~T.,  2011, {Physics of the Interstellar and Intergalactic Medium}

\bibitem[\protect\citeauthoryear{{Drury}}{{Drury}}{1983}]{1983RPPh...46..973D}
{Drury} L.~O.,  1983, \mn@doi [Reports on Progress in Physics]
  {10.1088/0034-4885/46/8/002}, \href
  {https://ui.adsabs.harvard.edu/abs/1983RPPh...46..973D} {46, 973}

\bibitem[\protect\citeauthoryear{Enberg, Reno  \& Sarcevic}{Enberg
  et~al.}{2008}]{PhysRevD.78.043005}
Enberg R.,  Reno M.~H.,   Sarcevic I.,  2008, \mn@doi [Phys. Rev. D]
  {10.1103/PhysRevD.78.043005}, 78, 043005

\bibitem[\protect\citeauthoryear{{Esteban}, {Gonzalez-Garcia}, {Maltoni},
  {Schwetz}  \& {Zhou}}{{Esteban} et~al.}{2020}]{2020JHEP...09..178E}
{Esteban} I.,  {Gonzalez-Garcia} M.~C.,  {Maltoni} M.,  {Schwetz} T.,   {Zhou}
  A.,  2020, \mn@doi [Journal of High Energy Physics]
  {10.1007/JHEP09(2020)178}, \href
  {https://ui.adsabs.harvard.edu/abs/2020JHEP...09..178E} {2020, 178}

\bibitem[\protect\citeauthoryear{Fang, Metzger, Vurm, Aydi  \& Chomiuk}{Fang
  et~al.}{2020}]{Fang2020}
Fang K.,  Metzger B.~D.,  Vurm I.,  Aydi E.,   Chomiuk L.,  2020, \mn@doi
  [\apj] {10.3847/1538-4357/abbc6e}, 904, 4

\bibitem[\protect\citeauthoryear{{Feldman} \& {Cousins}}{{Feldman} \&
  {Cousins}}{1998}]{1998PhRvD..57.3873F}
{Feldman} G.~J.,  {Cousins} R.~D.,  1998, \mn@doi [\prd]
  {10.1103/PhysRevD.57.3873}, \href
  {https://ui.adsabs.harvard.edu/abs/1998PhRvD..57.3873F} {57, 3873}

\bibitem[\protect\citeauthoryear{{Finke} \& {Dermer}}{{Finke} \&
  {Dermer}}{2012}]{2012ApJ...751...65F}
{Finke} J.~D.,  {Dermer} C.~D.,  2012, \mn@doi [\apj]
  {10.1088/0004-637X/751/1/65}, \href
  {https://ui.adsabs.harvard.edu/abs/2012ApJ...751...65F} {751, 65}

\bibitem[\protect\citeauthoryear{{Flinner}, {Tucker}, {Beacom}  \&
  {Shappee}}{{Flinner} et~al.}{2023}]{Flinner_2023RNAAS}
{Flinner} N.,  {Tucker} M.~A.,  {Beacom} J.~F.,   {Shappee} B.~J.,  2023,
  \mn@doi [Research Notes of the American Astronomical Society]
  {10.3847/2515-5172/acefc4}, \href
  {https://ui.adsabs.harvard.edu/abs/2023RNAAS...7..174F} {7, 174}

\bibitem[\protect\citeauthoryear{Fransson et~al.,}{Fransson
  et~al.}{2014}]{Fransson_2014}
Fransson C.,  et~al., 2014, \mn@doi [\apj] {10.1088/0004-637X/797/2/118}, 797,
  118

\bibitem[\protect\citeauthoryear{{Fuller}}{{Fuller}}{2017}]{Fuller2017}
{Fuller} J.,  2017, \mn@doi [\mnras] {10.1093/mnras/stx1314}, \href
  {https://ui.adsabs.harvard.edu/abs/2017MNRAS.470.1642F} {470, 1642}

\bibitem[\protect\citeauthoryear{Ginzburg \& Balberg}{Ginzburg \&
  Balberg}{2012}]{Ginzburg_2012}
Ginzburg S.,  Balberg S.,  2012, \mn@doi [\apj] {10.1088/0004-637X/757/2/178},
  757, 178

\bibitem[\protect\citeauthoryear{{Gould}}{{Gould}}{1975}]{1975ApJ...196..689G}
{Gould} R.~J.,  1975, \mn@doi [\apj] {10.1086/153457}, \href
  {https://ui.adsabs.harvard.edu/abs/1975ApJ...196..689G} {196, 689}

\bibitem[\protect\citeauthoryear{Grefenstette, Brightman, Earnshaw, Harrison
  \& Margutti}{Grefenstette et~al.}{2023}]{Grefenstette_2023}
Grefenstette B.~W.,  Brightman M.,  Earnshaw H.~P.,  Harrison F.~A.,   Margutti
  R.,  2023, \mn@doi [\apjl L] {10.3847/2041-8213/acdf4e}, 952, L3

\bibitem[\protect\citeauthoryear{Guetta, Langella, Gagliardini  \&
  Valle}{Guetta et~al.}{2023}]{Guetta_2023}
Guetta D.,  Langella A.,  Gagliardini S.,   Valle M.~D.,  2023, \mn@doi [\apjl
  L] {10.3847/2041-8213/acf573}, 955, L9

\bibitem[\protect\citeauthoryear{Hiramatsu et~al.,}{Hiramatsu
  et~al.}{2023}]{Hiramatsu_2023}
Hiramatsu D.,  et~al., 2023, \mn@doi [\apjl L] {10.3847/2041-8213/acf299}, 955,
  L8

\bibitem[\protect\citeauthoryear{Honda, Kajita, Kasahara, Midorikawa  \&
  Sanuki}{Honda et~al.}{2007}]{PhysRevD.75.043006}
Honda M.,  Kajita T.,  Kasahara K.,  Midorikawa S.,   Sanuki T.,  2007, \mn@doi
  [Phys. Rev. D] {10.1103/PhysRevD.75.043006}, 75, 043006

\bibitem[\protect\citeauthoryear{{IceCube Collaboration}}{{IceCube
  Collaboration}}{2013}]{2013Sci...342E...1I}
{IceCube Collaboration} 2013, \mn@doi [Science] {10.1126/science.1242856},
  \href {https://ui.adsabs.harvard.edu/abs/2013Sci...342E...1I} {342, 1242856}

\bibitem[\protect\citeauthoryear{{IceCube Collaboration}}{{IceCube
  Collaboration}}{2020}]{2020GCN.27865....1I}
{IceCube Collaboration} 2020, GRB Coordinates Network, \href
  {https://ui.adsabs.harvard.edu/abs/2020GCN.27865....1I} {27865, 1}

\bibitem[\protect\citeauthoryear{{IceCube Collaboration} et~al.,}{{IceCube
  Collaboration} et~al.}{2018}]{2018Sci...361.1378I}
{IceCube Collaboration} et~al., 2018, \mn@doi [Science]
  {10.1126/science.aat1378}, \href
  {https://ui.adsabs.harvard.edu/abs/2018Sci...361.1378I} {361, eaat1378}

\bibitem[\protect\citeauthoryear{{IceCube Collaboration} et~al.,}{{IceCube
  Collaboration} et~al.}{2021}]{2021arXiv210109836I}
{IceCube Collaboration} et~al., 2021, \mn@doi [arXiv e-prints]
  {10.48550/arXiv.2101.09836}, \href
  {https://ui.adsabs.harvard.edu/abs/2021arXiv210109836I} {p. arXiv:2101.09836}

\bibitem[\protect\citeauthoryear{{IceCube-Gen2 Collaboration}
  et~al.,}{{IceCube-Gen2 Collaboration} et~al.}{2014}]{2014arXiv1412.5106I}
{IceCube-Gen2 Collaboration} et~al., 2014, \mn@doi [arXiv e-prints]
  {10.48550/arXiv.1412.5106}, \href
  {https://ui.adsabs.harvard.edu/abs/2014arXiv1412.5106I} {p. arXiv:1412.5106}

\bibitem[\protect\citeauthoryear{{Inoue}, {Marcowith}, {Giacinti}, {Jan van
  Marle}  \& {Nishino}}{{Inoue} et~al.}{2021}]{2021ApJ...922....7I}
{Inoue} T.,  {Marcowith} A.,  {Giacinti} G.,  {Jan van Marle} A.,   {Nishino}
  S.,  2021, \mn@doi [\apj] {10.3847/1538-4357/ac21ce}, \href
  {https://ui.adsabs.harvard.edu/abs/2021ApJ...922....7I} {922, 7}

\bibitem[\protect\citeauthoryear{{Inserra}}{{Inserra}}{2019}]{Inserra2019}
{Inserra} C.,  2019, \mn@doi [Nature Astronomy] {10.1038/s41550-019-0854-4},
  \href {https://ui.adsabs.harvard.edu/abs/2019NatAs...3..697I} {3, 697}

\bibitem[\protect\citeauthoryear{{Itagaki}}{{Itagaki}}{2023}]{2023TNSTR1158....1I}
{Itagaki} K.,  2023, Transient Name Server Discovery Report, \href
  {https://www.wis-tns.org/object/2023ixf/discovery-cert} {2023-1158, 1}

\bibitem[\protect\citeauthoryear{{Ivezi{\'c}} et~al.,}{{Ivezi{\'c}}
  et~al.}{2019}]{2019ApJ...873..111I}
{Ivezi{\'c}} {\v{Z}}.,  et~al., 2019, \mn@doi [\apj]
  {10.3847/1538-4357/ab042c}, \href
  {https://ui.adsabs.harvard.edu/abs/2019ApJ...873..111I} {873, 111}

\bibitem[\protect\citeauthoryear{Jacobson-Galán et~al.,}{Jacobson-Galán
  et~al.}{2023}]{Jacobson-Galán_2023}
Jacobson-Galán W.~V.,  et~al., 2023, \mn@doi [\apjl L]
  {10.3847/2041-8213/acf2ec}, 954, L42

\bibitem[\protect\citeauthoryear{Janka}{Janka}{2012}]{Janka_12}
Janka H.-T.,  2012, \mn@doi [Annual Review of Nuclear and Particle Science]
  {https://doi.org/10.1146/annurev-nucl-102711-094901}, 62, 407

\bibitem[\protect\citeauthoryear{Jencson et~al.,}{Jencson
  et~al.}{2023}]{Jencson_2023}
Jencson J.~E.,  et~al., 2023, \mn@doi [\apjl L] {10.3847/2041-8213/ace618},
  952, L30

\bibitem[\protect\citeauthoryear{{KM3NeT Collaboration} et~al.,}{{KM3NeT
  Collaboration} et~al.}{2024}]{2024EPJC...84..885K}
{KM3NeT Collaboration} et~al., 2024, \mn@doi [European Physical Journal C]
  {10.1140/epjc/s10052-024-13137-2}, \href
  {https://ui.adsabs.harvard.edu/abs/2024EPJC...84..885K} {84, 885}

\bibitem[\protect\citeauthoryear{{Kasen} \& {Woosley}}{{Kasen} \&
  {Woosley}}{2009}]{2009ApJ...703.2205K}
{Kasen} D.,  {Woosley} S.~E.,  2009, \mn@doi [\apj]
  {10.1088/0004-637X/703/2/2205}, \href
  {https://ui.adsabs.harvard.edu/abs/2009ApJ...703.2205K} {703, 2205}

\bibitem[\protect\citeauthoryear{{Katz}, {Sapir}  \& {Waxman}}{{Katz}
  et~al.}{2012}]{2012IAUS..279..274K}
{Katz} B.,  {Sapir} N.,   {Waxman} E.,  2012, in {Roming} P.,  {Kawai} N.,
  {Pian} E.,  eds,  IAU Symposium Vol. 279, Death of Massive Stars: Supernovae
  and Gamma-Ray Bursts. pp 274--281 (\mn@eprint {arXiv} {1106.1898}),
  \mn@doi{10.1017/S174392131201304X}

\bibitem[\protect\citeauthoryear{Kelner, Aharonian  \& Bugayov}{Kelner
  et~al.}{2006}]{Kelner2006}
Kelner S.~R.,  Aharonian F.~A.,   Bugayov V.~V.,  2006, \mn@doi [Phys. Rev. D]
  {10.1103/PhysRevD.74.034018}, 74, 034018

\bibitem[\protect\citeauthoryear{Khatami \& Kasen}{Khatami \&
  Kasen}{2019}]{Khatami_2019}
Khatami D.~K.,  Kasen D.~N.,  2019, \mn@doi [\apj] {10.3847/1538-4357/ab1f09},
  878, 56

\bibitem[\protect\citeauthoryear{{Khatami} \& {Kasen}}{{Khatami} \&
  {Kasen}}{2024}]{KK_2023}
{Khatami} D.~K.,  {Kasen} D.~N.,  2024, \mn@doi [\apj]
  {10.3847/1538-4357/ad60c0}, \href
  {https://ui.adsabs.harvard.edu/abs/2024ApJ...972..140K} {972, 140}

\bibitem[\protect\citeauthoryear{Kheirandish \& Murase}{Kheirandish \&
  Murase}{2023}]{KheiMura_2023}
Kheirandish A.,  Murase K.,  2023, \mn@doi [\apjl L]
  {10.3847/2041-8213/acf84f}, 956, L8

\bibitem[\protect\citeauthoryear{Kilpatrick et~al.,}{Kilpatrick
  et~al.}{2023}]{Kilpatrick_2023}
Kilpatrick C.~D.,  et~al., 2023, \mn@doi [\apjl L] {10.3847/2041-8213/ace4ca},
  952, L23

\bibitem[\protect\citeauthoryear{{Kumar}, {Dastidar}, {Maund}, {Singleton}  \&
  {Sun}}{{Kumar} et~al.}{2025}]{2025MNRAS.538..659K}
{Kumar} A.,  {Dastidar} R.,  {Maund} J.~R.,  {Singleton} A.~J.,   {Sun} N.-C.,
  2025, \mn@doi [\mnras] {10.1093/mnras/staf312}, \href
  {https://ui.adsabs.harvard.edu/abs/2025MNRAS.538..659K} {538, 659}

\bibitem[\protect\citeauthoryear{Li et~al.,}{Li et~al.}{2024}]{li_shock_2024}
Li G.,  et~al., 2024, \mn@doi [Nature] {10.1038/s41586-023-06843-6}, 627, 754

\bibitem[\protect\citeauthoryear{{Lodders}}{{Lodders}}{2019}]{2019arXiv191200844L}
{Lodders} K.,  2019, \mn@doi [arXiv e-prints] {10.48550/arXiv.1912.00844},
  \href {https://ui.adsabs.harvard.edu/abs/2019arXiv191200844L} {p.
  arXiv:1912.00844}

\bibitem[\protect\citeauthoryear{Margalit, Quataert  \& Ho}{Margalit
  et~al.}{2022}]{Margalit_2022}
Margalit B.,  Quataert E.,   Ho A. Y.~Q.,  2022, \mn@doi [\apj]
  {10.3847/1538-4357/ac53b0}, 928, 122

\bibitem[\protect\citeauthoryear{{Marti-Devesa}}{{Marti-Devesa}}{2023}]{2023ATel16075....1M}
{Marti-Devesa} G.,  2023, The Astronomer's Telegram, \href
  {https://ui.adsabs.harvard.edu/abs/2023ATel16075....1M} {16075, 1}

\bibitem[\protect\citeauthoryear{{Mart{\'\i}-Devesa}, {Cheung}, {Di Lalla},
  {Renaud}, {Principe}, {Omodei}  \& {Acero}}{{Mart{\'\i}-Devesa}
  et~al.}{2024}]{2024A&A...686A.254M}
{Mart{\'\i}-Devesa} G.,  {Cheung} C.~C.,  {Di Lalla} N.,  {Renaud} M.,
  {Principe} G.,  {Omodei} N.,   {Acero} F.,  2024, \mn@doi [\aap]
  {10.1051/0004-6361/202349061}, \href
  {https://ui.adsabs.harvard.edu/abs/2024A&A...686A.254M} {686, A254}

\bibitem[\protect\citeauthoryear{{Martinez}, {Bersten}, {Folatelli}, {Orellana}
   \& {Ertini}}{{Martinez} et~al.}{2024}]{Martinez_AA2024}
{Martinez} L.,  {Bersten} M.~C.,  {Folatelli} G.,  {Orellana} M.,   {Ertini}
  K.,  2024, \mn@doi [\aap] {10.1051/0004-6361/202348142}, \href
  {https://ui.adsabs.harvard.edu/abs/2024A&A...683A.154M} {683, A154}

\bibitem[\protect\citeauthoryear{{Matzner} \& {McKee}}{{Matzner} \&
  {McKee}}{1999}]{1999ApJ...510..379M}
{Matzner} C.~D.,  {McKee} C.~F.,  1999, \mn@doi [\apj] {10.1086/306571}, \href
  {https://ui.adsabs.harvard.edu/abs/1999ApJ...510..379M} {510, 379}

\bibitem[\protect\citeauthoryear{{Moriya} \& {Maeda}}{{Moriya} \&
  {Maeda}}{2012}]{2012ApJ...756L..22M}
{Moriya} T.~J.,  {Maeda} K.,  2012, \mn@doi [\apjl L]
  {10.1088/2041-8205/756/1/L22}, \href
  {https://ui.adsabs.harvard.edu/abs/2012ApJ...756L..22M} {756, L22}

\bibitem[\protect\citeauthoryear{Moriya \& Maeda}{Moriya \&
  Maeda}{2014}]{Moriya_2014}
Moriya T.~J.,  Maeda K.,  2014, \mn@doi [\apjl L]
  {10.1088/2041-8205/790/2/L16}, 790, L16

\bibitem[\protect\citeauthoryear{{Moriya} \& {Singh}}{{Moriya} \&
  {Singh}}{2024}]{Moriya2024}
{Moriya} T.~J.,  {Singh} A.,  2024, \mn@doi [\pasj] {10.1093/pasj/psae070},
  \href {https://ui.adsabs.harvard.edu/abs/2024PASJ...76.1050M} {76, 1050}

\bibitem[\protect\citeauthoryear{Moriya, Tominaga, Blinnikov, Baklanov  \&
  Sorokina}{Moriya et~al.}{2011}]{Moriya_11}
Moriya T.,  Tominaga N.,  Blinnikov S.~I.,  Baklanov P.~V.,   Sorokina E.~I.,
  2011, \mn@doi [\mnras] {10.1111/j.1365-2966.2011.18689.x}, 415, 199

\bibitem[\protect\citeauthoryear{Moriya, Maeda, Taddia, Sollerman, Blinnikov
  \& Sorokina}{Moriya et~al.}{2013}]{Moriya2013}
Moriya T.~J.,  Maeda K.,  Taddia F.,  Sollerman J.,  Blinnikov S.~I.,
  Sorokina E.~I.,  2013, \mn@doi [\mnras] {10.1093/mnras/stt1392}, 435, 1520

\bibitem[\protect\citeauthoryear{{Moriya}, {Yoon}, {Gr{\"a}fener}  \&
  {Blinnikov}}{{Moriya} et~al.}{2017}]{MYGB2017}
{Moriya} T.~J.,  {Yoon} S.-C.,  {Gr{\"a}fener} G.,   {Blinnikov} S.~I.,  2017,
  \mn@doi [\mnras] {10.1093/mnrasl/slx056}, \href
  {https://ui.adsabs.harvard.edu/abs/2017MNRAS.469L.108M} {469, L108}

\bibitem[\protect\citeauthoryear{{Morozova}, {Piro}  \& {Valenti}}{{Morozova}
  et~al.}{2018}]{MorzovaPiroValenti2018}
{Morozova} V.,  {Piro} A.~L.,   {Valenti} S.,  2018, \mn@doi [\apj]
  {10.3847/1538-4357/aab9a6}, \href
  {https://ui.adsabs.harvard.edu/abs/2018ApJ...858...15M} {858, 15}

\bibitem[\protect\citeauthoryear{{Muller}, {Heijboer}  \& {van Eeden}}{{Muller}
  et~al.}{2023}]{Muller:2023xO}
{Muller} R.,  {Heijboer} A.,   {van Eeden} T.,  2023, \mn@doi [PoS]
  {10.22323/1.444.1018}, ICRC2023, 1018

\bibitem[\protect\citeauthoryear{Murase}{Murase}{2018}]{Murase18}
Murase K.,  2018, \mn@doi [Phys. Rev. D] {10.1103/PhysRevD.97.081301}, 97,
  081301

\bibitem[\protect\citeauthoryear{{Murase}}{{Murase}}{2024}]{Murase24}
{Murase} K.,  2024, \mn@doi [\prd] {10.1103/PhysRevD.109.103020}, \href
  {https://ui.adsabs.harvard.edu/abs/2024PhRvD.109j3020M} {109, 103020}

\bibitem[\protect\citeauthoryear{Murase \& Waxman}{Murase \&
  Waxman}{2016}]{Murase16}
Murase K.,  Waxman E.,  2016, \mn@doi [Phys. Rev. D]
  {10.1103/PhysRevD.94.103006}, 94, 103006

\bibitem[\protect\citeauthoryear{Murase, Thompson, Lacki  \& Beacom}{Murase
  et~al.}{2011}]{Murase11}
Murase K.,  Thompson T.~A.,  Lacki B.~C.,   Beacom J.~F.,  2011, \mn@doi [Phys.
  Rev. D] {10.1103/PhysRevD.84.043003}, 84, 043003

\bibitem[\protect\citeauthoryear{Murase, Thompson  \& Ofek}{Murase
  et~al.}{2014}]{Murase14}
Murase K.,  Thompson T.~A.,   Ofek E.~O.,  2014, \mn@doi [\mnras]
  {10.1093/mnras/stu384}, 440, 2528

\bibitem[\protect\citeauthoryear{Murase, Franckowiak, Maeda, Margutti  \&
  Beacom}{Murase et~al.}{2019}]{Murase_2019}
Murase K.,  Franckowiak A.,  Maeda K.,  Margutti R.,   Beacom J.~F.,  2019,
  \mn@doi [\apj] {10.3847/1538-4357/ab0422}, 874, 80

\bibitem[\protect\citeauthoryear{Müller, Prieto, Pejcha  \&
  Clocchiatti}{Müller et~al.}{2017}]{Müller_2017}
Müller T.,  Prieto J.~L.,  Pejcha O.,   Clocchiatti A.,  2017, \mn@doi [\apj]
  {10.3847/1538-4357/aa72f1}, 841, 127

\bibitem[\protect\citeauthoryear{Nicholl}{Nicholl}{2018}]{Nicholl_2018}
Nicholl M.,  2018, \mn@doi [Research Notes of the AAS]
  {10.3847/2515-5172/aaf799}, 2, 230

\bibitem[\protect\citeauthoryear{Niu, Sun, Maund, Zhang, Zhao  \& Liu}{Niu
  et~al.}{2023}]{Niu_2023}
Niu Z.,  Sun N.-C.,  Maund J.~R.,  Zhang Y.,  Zhao R.,   Liu J.,  2023, \mn@doi
  [\apjl L] {10.3847/2041-8213/acf4e3}, 955, L15

\bibitem[\protect\citeauthoryear{Ofek et~al.,}{Ofek et~al.}{2014}]{Ofek_2014}
Ofek E.~O.,  et~al., 2014, \mn@doi [\apj] {10.1088/0004-637X/788/2/154}, 788,
  154

\bibitem[\protect\citeauthoryear{{Ohira}, {Murase}  \& {Yamazaki}}{{Ohira}
  et~al.}{2010}]{2010A&A...513A..17O}
{Ohira} Y.,  {Murase} K.,   {Yamazaki} R.,  2010, \mn@doi [\aap]
  {10.1051/0004-6361/200913495}, \href
  {https://ui.adsabs.harvard.edu/abs/2010A&A...513A..17O} {513, A17}

\bibitem[\protect\citeauthoryear{{Pan}, {Patnaude}  \& {Loeb}}{{Pan}
  et~al.}{2013}]{2013MNRAS.433..838P}
{Pan} T.,  {Patnaude} D.,   {Loeb} A.,  2013, \mn@doi [\mnras]
  {10.1093/mnras/stt780}, \href
  {https://ui.adsabs.harvard.edu/abs/2013MNRAS.433..838P} {433, 838}

\bibitem[\protect\citeauthoryear{Perley et~al.,}{Perley
  et~al.}{2020}]{Perley_2020}
Perley D.~A.,  et~al., 2020, \mn@doi [\apj] {10.3847/1538-4357/abbd98}, 904, 35

\bibitem[\protect\citeauthoryear{{Petropoulou}, {Kamble}  \&
  {Sironi}}{{Petropoulou} et~al.}{2016}]{2016MNRAS.460...44P}
{Petropoulou} M.,  {Kamble} A.,   {Sironi} L.,  2016, \mn@doi [\mnras]
  {10.1093/mnras/stw920}, \href
  {https://ui.adsabs.harvard.edu/abs/2016MNRAS.460...44P} {460, 44}

\bibitem[\protect\citeauthoryear{Petropoulou, Coenders, Vasilopoulos, Kamble
  \& Sironi}{Petropoulou et~al.}{2017}]{Petropoulou17}
Petropoulou M.,  Coenders S.,  Vasilopoulos G.,  Kamble A.,   Sironi L.,  2017,
  \mn@doi [\mnras] {10.1093/mnras/stx1251}, 470, 1881

\bibitem[\protect\citeauthoryear{Piro, Haynie  \& Yao}{Piro
  et~al.}{2021}]{Piro_2021}
Piro A.~L.,  Haynie A.,   Yao Y.,  2021, \mn@doi [\apj]
  {10.3847/1538-4357/abe2b1}, 909, 209

\bibitem[\protect\citeauthoryear{Pitik, Tamborra, Angus  \& Auchettl}{Pitik
  et~al.}{2022}]{Pitik2022}
Pitik T.,  Tamborra I.,  Angus C.~R.,   Auchettl K.,  2022, \mn@doi [\apj]
  {10.3847/1538-4357/ac5ab1}, 929, 163

\bibitem[\protect\citeauthoryear{{Pitik}, {Tamborra}, {Lincetto}  \&
  {Franckowiak}}{{Pitik} et~al.}{2023}]{2023MNRAS.524.3366P}
{Pitik} T.,  {Tamborra} I.,  {Lincetto} M.,   {Franckowiak} A.,  2023, \mn@doi
  [\mnras] {10.1093/mnras/stad2025}, \href
  {https://ui.adsabs.harvard.edu/abs/2023MNRAS.524.3366P} {524, 3366}

\bibitem[\protect\citeauthoryear{{Planck Collaboration} et~al.,}{{Planck
  Collaboration} et~al.}{2020}]{2020A&A...641A...6P}
{Planck Collaboration} et~al., 2020, \mn@doi [\aap]
  {10.1051/0004-6361/201833910}, \href
  {https://ui.adsabs.harvard.edu/abs/2020A&A...641A...6P} {641, A6}

\bibitem[\protect\citeauthoryear{Pledger \& Shara}{Pledger \&
  Shara}{2023}]{Pledger_2023}
Pledger J.~L.,  Shara M.~M.,  2023, \mn@doi [\apjl L]
  {10.3847/2041-8213/ace88b}, 953, L14

\bibitem[\protect\citeauthoryear{{Popov}}{{Popov}}{1993}]{1993ApJ...414..712P}
{Popov} D.~V.,  1993, \mn@doi [\apj] {10.1086/173117}, \href
  {https://ui.adsabs.harvard.edu/abs/1993ApJ...414..712P} {414, 712}

\bibitem[\protect\citeauthoryear{{Protheroe} \& {Clay}}{{Protheroe} \&
  {Clay}}{2004}]{2004PASA...21....1P}
{Protheroe} R.~J.,  {Clay} R.~W.,  2004, \mn@doi [\pasa] {10.1071/AS03047},
  \href {https://ui.adsabs.harvard.edu/abs/2004PASA...21....1P} {21, 1}

\bibitem[\protect\citeauthoryear{{Pumo} \& {Cosentino}}{{Pumo} \&
  {Cosentino}}{2025}]{PC2025}
{Pumo} M.~L.,  {Cosentino} S.~P.,  2025, \mn@doi [\mnras]
  {10.1093/mnras/staf288}, \href
  {https://ui.adsabs.harvard.edu/abs/2025arXiv250209752P} {p. arXiv:2502.09752
  (PC25)}

\bibitem[\protect\citeauthoryear{{Pumo} \& {Zampieri}}{{Pumo} \&
  {Zampieri}}{2011}]{PZ2011}
{Pumo} M.~L.,  {Zampieri} L.,  2011, \mn@doi [\apj]
  {10.1088/0004-637X/741/1/41}, \href
  {https://ui.adsabs.harvard.edu/abs/2011ApJ...741...41P} {741, 41}

\bibitem[\protect\citeauthoryear{{Pumo} \& {Zampieri}}{{Pumo} \&
  {Zampieri}}{2013}]{2013MNRAS.434.3445P}
{Pumo} M.~L.,  {Zampieri} L.,  2013, \mn@doi [\mnras] {10.1093/mnras/stt1256},
  \href {https://ui.adsabs.harvard.edu/abs/2013MNRAS.434.3445P} {434, 3445}

\bibitem[\protect\citeauthoryear{{Pumo} et~al.,}{{Pumo}
  et~al.}{2009}]{2009ApJ...705L.138P}
{Pumo} M.~L.,  et~al., 2009, \mn@doi [\apjl L] {10.1088/0004-637X/705/2/L138},
  \href {https://ui.adsabs.harvard.edu/abs/2009ApJ...705L.138P} {705, L138}

\bibitem[\protect\citeauthoryear{{Pumo}, {Zampieri}, {Spiro}, {Pastorello},
  {Benetti}, {Cappellaro}, {Manic{\`o}}  \& {Turatto}}{{Pumo}
  et~al.}{2017}]{2017MNRAS.464.3013P}
{Pumo} M.~L.,  {Zampieri} L.,  {Spiro} S.,  {Pastorello} A.,  {Benetti} S.,
  {Cappellaro} E.,  {Manic{\`o}} G.,   {Turatto} M.,  2017, \mn@doi [\mnras]
  {10.1093/mnras/stw2625}, \href
  {https://ui.adsabs.harvard.edu/abs/2017MNRAS.464.3013P} {464, 3013}

\bibitem[\protect\citeauthoryear{{Pumo}, {Cosentino}, {Pastorello}, {Benetti},
  {Cherubini}, {Manic{\`o}}  \& {Zampieri}}{{Pumo}
  et~al.}{2023}]{2023MNRAS.521.4801P}
{Pumo} M.~L.,  {Cosentino} S.~P.,  {Pastorello} A.,  {Benetti} S.,  {Cherubini}
  S.,  {Manic{\`o}} G.,   {Zampieri} L.,  2023, \mn@doi [\mnras]
  {10.1093/mnras/stad861}, \href
  {https://ui.adsabs.harvard.edu/abs/2023MNRAS.521.4801P} {521, 4801}

\bibitem[\protect\citeauthoryear{{Qin} et~al.,}{{Qin} et~al.}{2024}]{Qin_2023}
{Qin} Y.-J.,  et~al., 2024, \mn@doi [\mnras] {10.1093/mnras/stae2012}, \href
  {https://ui.adsabs.harvard.edu/abs/2024MNRAS.534..271Q} {534, 271}

\bibitem[\protect\citeauthoryear{Riess et~al.,}{Riess
  et~al.}{2022}]{Riess_2022}
Riess A.~G.,  et~al., 2022, \mn@doi [\apjl L] {10.3847/2041-8213/ac5c5b}, 934,
  L7

\bibitem[\protect\citeauthoryear{{Rodr{\'\i}guez}, {Meza}, {Pineda-Garc{\'\i}a}
   \& {Ramirez}}{{Rodr{\'\i}guez} et~al.}{2021}]{2021MNRAS.505.1742R}
{Rodr{\'\i}guez} {\'O}.,  {Meza} N.,  {Pineda-Garc{\'\i}a} J.,   {Ramirez} M.,
  2021, \mn@doi [\mnras] {10.1093/mnras/stab1335}, \href
  {https://ui.adsabs.harvard.edu/abs/2021MNRAS.505.1742R} {505, 1742}

\bibitem[\protect\citeauthoryear{{Rozwadowska}, {Vissani}  \&
  {Cappellaro}}{{Rozwadowska} et~al.}{2021}]{2021NewA...8301498R}
{Rozwadowska} K.,  {Vissani} F.,   {Cappellaro} E.,  2021, \mn@doi [\na]
  {10.1016/j.newast.2020.101498}, \href
  {https://ui.adsabs.harvard.edu/abs/2021NewA...8301498R} {83, 101498}

\bibitem[\protect\citeauthoryear{{Salmaso} et~al.,}{{Salmaso}
  et~al.}{2025}]{Salmaso24}
{Salmaso} I.,  et~al., 2025, \mn@doi [\aap] {10.1051/0004-6361/202451764},
  \href {https://ui.adsabs.harvard.edu/abs/2025A&A...695A..29S} {695, A29}

\bibitem[\protect\citeauthoryear{{Sarmah}}{{Sarmah}}{2024}]{2024JCAP...04..083S}
{Sarmah} P.,  2024, \mn@doi [\jcap] {10.1088/1475-7516/2024/04/083}, \href
  {https://ui.adsabs.harvard.edu/abs/2024JCAP...04..083S} {2024, 083}

\bibitem[\protect\citeauthoryear{Sarmah, Chakraborty, Tamborra  \&
  Auchettl}{Sarmah et~al.}{2022}]{Sarmah_2022}
Sarmah P.,  Chakraborty S.,  Tamborra I.,   Auchettl K.,  2022, \mn@doi
  [Journal of Cosmology and Astroparticle Physics]
  {10.1088/1475-7516/2022/08/011}, 2022, 011

\bibitem[\protect\citeauthoryear{{Singh}, {Kumar}, {Moriya}, {Anupama}, {Sahu},
  {Brown}, {Andrews}  \& {Smith}}{{Singh}
  et~al.}{2019}]{Singh2019ApJ...882...68S}
{Singh} A.,  {Kumar} B.,  {Moriya} T.~J.,  {Anupama} G.~C.,  {Sahu} D.~K.,
  {Brown} P.~J.,  {Andrews} J.~E.,   {Smith} N.,  2019, \mn@doi [\apj]
  {10.3847/1538-4357/ab3050}, \href
  {https://ui.adsabs.harvard.edu/abs/2019ApJ...882...68S} {882, 68}

\bibitem[\protect\citeauthoryear{Singh et~al.,}{Singh
  et~al.}{2024}]{Singh_2024}
Singh A.,  et~al., 2024, \mn@doi [The Astrophysical Journal]
  {10.3847/1538-4357/ad7955}, 975, 132

\bibitem[\protect\citeauthoryear{{Smith}}{{Smith}}{2014}]{2014ARA&A..52..487S}
{Smith} N.,  2014, \mn@doi [\araa] {10.1146/annurev-astro-081913-040025}, \href
  {https://ui.adsabs.harvard.edu/abs/2014ARA&A..52..487S} {52, 487}

\bibitem[\protect\citeauthoryear{Smith}{Smith}{2017}]{Smith2017}
Smith N.,  2017, Interacting Supernovae: Types IIn and Ibn.
Springer International Publishing, Cham, pp 403--429,
  \mn@doi{10.1007/978-3-319-21846-5_38}

\bibitem[\protect\citeauthoryear{Soker}{Soker}{2023}]{Soker2023}
Soker N.,  2023, \mn@doi [Research in Astronomy and Astrophysics]
  {10.1088/1674-4527/ace51f}, 23, 081002

\bibitem[\protect\citeauthoryear{Soraisam et~al.,}{Soraisam
  et~al.}{2023}]{Soraisam_2023}
Soraisam M.~D.,  et~al., 2023, \mn@doi [\apj] {10.3847/1538-4357/acef22}, 957,
  64

\bibitem[\protect\citeauthoryear{Strotjohann et~al.,}{Strotjohann
  et~al.}{2021}]{Strotjohann_2021}
Strotjohann N.~L.,  et~al., 2021, \mn@doi [\apj] {10.3847/1538-4357/abd032},
  907, 99

\bibitem[\protect\citeauthoryear{Sturner, Skibo, Dermer  \& Mattox}{Sturner
  et~al.}{1997}]{Sturner_1997}
Sturner S.~J.,  Skibo J.~G.,  Dermer C.~D.,   Mattox J.~R.,  1997, \mn@doi
  [\apj] {10.1086/304894}, 490, 619

\bibitem[\protect\citeauthoryear{Suzuki, Moriya  \& Takiwaki}{Suzuki
  et~al.}{2020}]{Suzuki_2020}
Suzuki A.,  Moriya T.~J.,   Takiwaki T.,  2020, \mn@doi [\apj]
  {10.3847/1538-4357/aba0ba}, 899, 56

\bibitem[\protect\citeauthoryear{Teja et~al.,}{Teja et~al.}{2023}]{Teja_2023}
Teja R.~S.,  et~al., 2023, \mn@doi [\apjl L] {10.3847/2041-8213/acef20}, 954,
  L12

\bibitem[\protect\citeauthoryear{{Thwaites}, {Vandenbroucke}, {Santander}  \&
  {IceCube Collaboration}}{{Thwaites} et~al.}{2023}]{2023ATel16043....1T}
{Thwaites} J.,  {Vandenbroucke} J.,  {Santander} M.,   {IceCube Collaboration}
  2023, The Astronomer's Telegram, \href
  {https://ui.adsabs.harvard.edu/abs/2023ATel16043....1T} {16043, 1}

\bibitem[\protect\citeauthoryear{Trovato \& for~the
  KM3NeT~Collaboration}{Trovato \& for~the
  KM3NeT~Collaboration}{2017}]{Trovato_2017}
Trovato A.,  for~the KM3NeT~Collaboration 2017, \mn@doi [Journal of Physics:
  Conference Series] {10.1088/1742-6596/888/1/012108}, 888, 012108

\bibitem[\protect\citeauthoryear{Tsuna, Kashiyama  \& Shigeyama}{Tsuna
  et~al.}{2019}]{Tsuna_2019}
Tsuna D.,  Kashiyama K.,   Shigeyama T.,  2019, \mn@doi [\apj]
  {10.3847/1538-4357/ab40ba}, 884, 87

\bibitem[\protect\citeauthoryear{Tsuna, Murase  \& Moriya}{Tsuna
  et~al.}{2023}]{Tsuna_2023}
Tsuna D.,  Murase K.,   Moriya T.~J.,  2023, \mn@doi [\apj]
  {10.3847/1538-4357/acdb71}, 952, 115

\bibitem[\protect\citeauthoryear{Tsvetkov et~al.,}{Tsvetkov
  et~al.}{2024}]{Tsvetkov_2024}
Tsvetkov D.~Y.,  et~al., 2024, \mn@doi [Astronomische Nachrichten]
  {https://doi.org/10.1002/asna.20230166}, 345, e230166

\bibitem[\protect\citeauthoryear{{Utrobin} \& {Chugai}}{{Utrobin} \&
  {Chugai}}{2011}]{2011A&A...532A.100U}
{Utrobin} V.~P.,  {Chugai} N.~N.,  2011, \mn@doi [\aap]
  {10.1051/0004-6361/201117137}, \href
  {https://ui.adsabs.harvard.edu/abs/2011A&A...532A.100U} {532, A100}

\bibitem[\protect\citeauthoryear{{Van Dyk} et~al.,}{{Van Dyk}
  et~al.}{2024}]{VanDyk_2023}
{Van Dyk} S.~D.,  et~al., 2024, \mn@doi [\apj] {10.3847/1538-4357/ad414b},
  \href {https://ui.adsabs.harvard.edu/abs/2024ApJ...968...27V} {968, 27}

\bibitem[\protect\citeauthoryear{Vasylyev et~al.,}{Vasylyev
  et~al.}{2023}]{Vasylyev_2023}
Vasylyev S.~S.,  et~al., 2023, \mn@doi [\apjl L] {10.3847/2041-8213/acf1a3},
  955, L37

\bibitem[\protect\citeauthoryear{Waxman \& Loeb}{Waxman \&
  Loeb}{2001}]{PhysRevLett.87.071101}
Waxman E.,  Loeb A.,  2001, \mn@doi [Phys. Rev. Lett.]
  {10.1103/PhysRevLett.87.071101}, 87, 071101

\bibitem[\protect\citeauthoryear{{Weaver}}{{Weaver}}{1976}]{1976ApJS...32..233W}
{Weaver} T.~A.,  1976, \mn@doi [\apjs] {10.1086/190398}, \href
  {https://ui.adsabs.harvard.edu/abs/1976ApJS...32..233W} {32, 233}

\bibitem[\protect\citeauthoryear{Woosley, Heger  \& Weaver}{Woosley
  et~al.}{2002}]{RevModPhys.74.1015}
Woosley S.~E.,  Heger A.,   Weaver T.~A.,  2002, \mn@doi [Rev. Mod. Phys.]
  {10.1103/RevModPhys.74.1015}, 74, 1015

\bibitem[\protect\citeauthoryear{Xiang, Mo, Wang, Wang, Zhang, Lin  \&
  Wang}{Xiang et~al.}{2023}]{xiang_dusty_2023}
Xiang D.,  Mo J.,  Wang L.,  Wang X.,  Zhang J.,  Lin H.,   Wang L.,  2023,
  \mn@doi [Science China Physics, Mechanics \& Astronomy]
  {10.1007/s11433-023-2267-0}, 67, 219514

\bibitem[\protect\citeauthoryear{{Yaron} et~al.,}{{Yaron}
  et~al.}{2017}]{2017NatPh..13..510Y}
{Yaron} O.,  et~al., 2017, \mn@doi [Nature Physics] {10.1038/nphys4025}, \href
  {https://ui.adsabs.harvard.edu/abs/2017NatPh..13..510Y} {13, 510}

\bibitem[\protect\citeauthoryear{Zhang et~al.,}{Zhang
  et~al.}{2023}]{ZHANG20232548}
Zhang J.,  et~al., 2023, \mn@doi [Science Bulletin]
  {https://doi.org/10.1016/j.scib.2023.09.015}, 68, 2548

\bibitem[\protect\citeauthoryear{Zimmerman et~al.,}{Zimmerman
  et~al.}{2024}]{zimmerman_complex_2024}
Zimmerman E.~A.,  et~al., 2024, \mn@doi [Nature] {10.1038/s41586-024-07116-6},
  627, 759

\bibitem[\protect\citeauthoryear{{Zirakashvili} \& {Aharonian}}{{Zirakashvili}
  \& {Aharonian}}{2007}]{2007A&A...465..695Z}
{Zirakashvili} V.~N.,  {Aharonian} F.,  2007, \mn@doi [\aap]
  {10.1051/0004-6361:20066494}, \href
  {https://ui.adsabs.harvard.edu/abs/2007A&A...465..695Z} {465, 695}

\bibitem[\protect\citeauthoryear{{de Vaucouleurs}, {de Vaucouleurs}, {Corwin},
  {Buta}, {Paturel}  \& {Fouque}}{{de Vaucouleurs}
  et~al.}{1991}]{1991rc3..book.....D}
{de Vaucouleurs} G.,  {de Vaucouleurs} A.,  {Corwin} Herold~G. J.,  {Buta}
  R.~J.,  {Paturel} G.,   {Fouque} P.,  1991, {Third Reference Catalogue of
  Bright Galaxies}

\makeatother
\end{thebibliography}






\bsp	
\label{lastpage}
\end{document}